\documentclass[12pt]{article}
\usepackage{amssymb}
\usepackage{graphics}
\usepackage{epsfig}
\usepackage{a4wide}
\usepackage[numbers,square,comma,sort&compress]{natbib}

\begin{document}
\newcommand{\ba}{\begin{array}}
\newcommand{\ea}{\end{array}}
\newcommand{\gdtsl}{$gd_k \to t\slep_i^-$ }
\newcommand{\pptsl}{$pp \to t\slep_i+X~$}
\newcommand{\realgluon}{$gd_k \to t\slep_i+g$}
\newcommand{\tabincell}[2]{\begin{tabular}{@{}#1@{}}#2\end{tabular}}

\def\rp{R\!\!\!/ _p}
\def\slep{\tilde{\ell}}
\def\snu{\tilde{\nu}}
\def\smu{\tilde{\mu}}
\def\stau{\tilde{\tau}}
\def\sel{\tilde{e}}

\title{ Single slepton production associated with a top quark at LHC in NLO QCD }
\author{ Li Xiao-Peng$^1$, Guo Lei$^1$, Ma Wen-Gan$^1$, Han Liang$^1$, \\
         Zhang Ren-You$^1$, and Wang Shao-Ming$^{1,2}$ \\
{\small $^1$ Department of Modern Physics, University of Science and Technology}  \\
{\small of China (USTC), Hefei, Anhui 230026, P.R.China} \\
{\small $^2$ Department of Physics, Chongqing University, Chongqing
401331, P.R. China   }}

\date{}
\maketitle \vskip 15mm
\begin{abstract}
Single slepton production in association with a top quark at the
CERN Large Hadron Collider (LHC) is one of the important processes
in probing the R-parity violation couplings. We calculate the QCD
next-to-leading order (NLO) corrections to the $pp \to
t\slep^{-}(\bar{t}\slep^{+})+X$ process at the LHC and discuss the
impacts of the QCD corrections on kinematic distributions. We
investigate the dependence of the leading order (LO) and the NLO QCD
corrected integrated cross section on the
factorization/renormalization energy scale, slepton, stop-quark and
gluino masses. We find that the uncertainty of the LO cross section
due to the energy scale is obviously improved by the NLO QCD
corrections, and the exclusive jet event selection scheme keeps the
convergence of the perturbative series better than the inclusive scheme.
The results show that the polarization asymmetry of the top-quark
will be reduced by the NLO QCD corrections, and the QCD corrections
generally increase with the increment of the $\tilde{t}_1$ or $\tilde{g}$ mass
value.
\end{abstract}


\vfill \eject

\baselineskip=0.32in

\renewcommand{\theequation}{\arabic{section}.\arabic{equation}}
\renewcommand{\thesection}{\Roman{section}.}
\newcommand{\nb}{\nonumber}

\newcommand{\Dir}{\kern -6.4pt\Big{/}}
\newcommand{\Dirin}{\kern -10.4pt\Big{/}\kern 4.4pt}
\newcommand{\DDir}{\kern -7.6pt\Big{/}}
\newcommand{\DGir}{\kern -6.0pt\Big{/}}

\makeatletter      
\@addtoreset{equation}{section}
\makeatother       

\section{Introduction}
\par
Supersymmetry (SUSY) is one of the most appealing theories as an
extension of the standard model (SM). It provides an elegant way to
solve the gauge hierarchy problem and cancel the quadratic
divergences of the radiative correction to the Higgs boson mass. The
minimal supersymmetric standard model (MSSM) contains SM particles,
their superpartners and an additional Higgs doublet. In order to
avoid the rapid proton decay, a discrete $R$-parity
symmetry\cite{Rv_1,Rv_2} is introduced in the MSSM, which implies a
conserved quantum number, $R_p=(-1)^{3B+L+2S}$, with $B$, $L$ and $S$
being baryon number, lepton number and spin of the particle,
respectively. For all the SM particles $R_p$ is equal to $+1$,
while for all the superpartners $R_p$ is $-1$. So this symmetry
leads to the result that the superpartners can only be produced in a pair and the
lightest SUSY particle (LSP) is stable. However, such a stringent
symmetry appears to be a theoretical basis; especially, we
know that a stable proton can survive by imposing either $L$- or
$B$-conservation \cite{LBviolation}. At the same time, there is not
enough experimental evidence for $R$-parity conservation, so
$R$-parity conservation is not necessary in the MSSM. Moreover,
non-zero $R$-parity violating (RPV) couplings could provide small
neutrino masses, which could explain the phenomena of neutrino
oscillation experiments. Thus, there are strong theoretical and
phenomenological motivations to introduce partial $R$-parity
violation into the most general representation of superpotential
with the respect to the renormalizability and preservation of the
gauge symmetries of SM and supersymmetry, which can be written as
\cite{superpotential_1,superpotential_2,superpotential_3}
\begin{equation}\label{Lagrangian-1}
{\cal W}_{\rlap/R_{p}} = \frac{1}{2}\epsilon_{ab}
\lambda_{ijk}\hat{L}_{i}^a \hat{L}_{j}^b \hat{E}_{k} +
\epsilon_{ab}\lambda^{'}_{ijk} \hat{L}_{i}^a \hat{Q}_{j}^b
\hat{D}_{k} +
\frac{1}{2}\epsilon_{\alpha\beta\gamma}\lambda^{''}_{ijk}
   \hat{U}_{i}^{\alpha} \hat{D}_{j}^{\beta} \hat{D}_{k}^{\gamma} +
\epsilon_{ab}\mu_{i} \hat{L}_{i}^a \hat{H}_{2}^b,
\end{equation}
where $i$, $j$, $k$ denote generation indices, $a,b~(= 1,2)$ are
SU(2) isospin indices, and $\alpha,\beta,\gamma$ are SU(3) color
indices. $\hat{L}_i$ ($\hat{Q}_i$) are the left-handed leptons
(quarks) SU(2)-doublet chiral superfields, and $\hat{E}_i$
($\hat{U}_i$, $\hat{D}_i$) are the right-handed leptons (up- and
down-type quarks) SU(2)-singlet chiral superfields. $H_{1,2}$ are
the Higgs chiral superfields. The $\lambda_{ijk}$,
$\lambda^{\prime}_{ijk}$ and $\mu_i$ are the dimensionless
$L$-violating coupling coefficients, and
$\lambda^{\prime\prime}_{ijk}$ are the $B$-violating dimensionless
coupling constants. As mentioned above, $B$ and $L$ cannot be
violated at the same time. In this paper we concentrate ourselves
only on the $L$-violating couplings, where the coefficients
$\lambda^\prime_{ijk}$ may be assumed to be non-zero.

\par
The LHC can be used as a top-factory, and it is advantageous to study
the production and decay of top quark. We suspect that the processes
related with top quark can be used to probe the new physics effects
since the top mass is close to the weak scale \cite{top},
particularly in the single top-quark production process the chiral
structure of the interaction of the top quark may impact the
polarization observable of the final top quark \cite{toppol}.
Therefore, the predictions including higher order corrections to
single top-quark production within and beyond the SM are very
important in exploring the new physics.

\par
The single charged slepton production in association with a top
quark at hadron collider is induced by the non-zero term
$\lambda^\prime_{i3k} \hat{L}_{i} \hat{Q}_{3} \hat{D}_{k}$ in
Eq.(\ref{Lagrangian-1}). The $pp \to t\slep_i^-(\bar{t}\slep_i^+)+X$
process at the LHC receives the contributions from the partonic
processes $gd_k(g\bar{d}_k) \to t\slep_i^-(\bar{t}\slep_i^+)$, where
$i$ and $k$ are the generation indices. The second term in
Eq.(\ref{Lagrangian-1}) is related to the Born process $pp \to t
\slep_i^-(\bar{t} \slep^+_i) + X$, which can be expressed as
\begin{equation}\label{Lagrangian-2}
\ba{rcl} {\cal L}_{LQD} & = & \lambda^\prime_{ijk} \bigg[
   \snu_{iL}     \bar{d}_{kR}         d_{jL}
+  \tilde{d}_{jL}    \bar{d}_{kR}         \nu_{iL}
+ \tilde{d}_{kR}^* \overline{\nu_{iL}^c} d_{jL} \bigg. \\
&& \bigg. -  \slep_{iL}            \bar{d}_{kR}            u_{jL} -
\tilde{u}_{jL}     \bar{d}_{kR}          \ell_{iL} -
\tilde{d}_{kR}^* \overline{\ell_{iL}^c} u_{jL} \bigg] + ~
\textrm{h.c}. \ea
\end{equation}
For the NLO QCD calculations, the interaction vertices of two
squarks and a  slepton may be involved, which can be extracted from
the general soft SUSY-breaking Lagrangian \cite{Barbier}
\begin{equation}\label{softlagrangian}
\mathcal{L}_{\tilde\ell \tilde q \tilde q}^{soft} =
 -\lambda^{\prime}_{ijk} A\left(
 \tilde{\nu}_i \,\tilde{d}_{jL}\,\tilde{d}_{kR}^*
 - \tilde{\ell}_{Li}\,\tilde{u}_{jL}\,\tilde{d}_{kR}^*
 \right)+ ~\textrm{h.c}.
\end{equation}
In the above equation it is assumed that the soft breaking terms have a
universal dimensionful parameter $A$ and are proportional to the
dimensionless coupling constant $\lambda^{\prime}_{ijk}$. In this
work we take the SUSY-breaking parameter $A=0,~\pm1~TeV$,
separately.

\par
The single top-quark production processes at colliders in the
$R$-parity violating MSSM has been studied in several references
\cite{top,singletopRPV_1,singletopRPV_2,singletopRPV_3,singletopRPV_4,singletopRPV_5,singletop_1,singletop_2}. The top-quark production in
association with a slepton $\slep_i$ at hadron collider has been
studied at leading order (LO) in Refs.\cite{topspin,LO}, there the
authors performed the signal analysis, and found that the final
states in the $t\slep_i^-(\bar{t}\slep_i^+)$ production at the LHC
have distinct kinematic signatures, which can be distinguished from
the backgrounds.

\par
In this paper, we present the calculations of the next-to-leading
order (NLO) QCD corrections to this process. The paper is organized
as follows: In section II, we present the calculations for the
relevant partonic processes and parent process $pp \to
t\slep^-_i(\bar{t}\slep^+_i)+X$ at the LO and QCD NLO. In section
III, we give some numerical results and discussions. Finally, a
short summary is given.

\vskip 5mm
\section{ Calculations }
\par
{\bf A. LO calculation }
\par
In both the LO and NLO calculations, we apply FeynArts3.4 and
FormCalc5.3 packages \cite{feynarts, formcalc} to generate Feynman
diagrams, their corresponding amplitudes, and to simplify the
amplitudes, separately. In Table \ref{tab-1} the
upper $2\sigma$ bounds on $\lambda^\prime_{i3k}$ originating from
Refs.\cite{LO,bounds} are listed. There the coefficients
$\lambda^\prime_{i33}<{\cal O}(10^{-4})$ stem from assuming
$m_{\nu} < 1eV$ and left-right mixing in the sbottom sector. Since we have
the strong constraints on the $\lambda'_{i33}$ coupling shown in
Table \ref{tab-1} and the low (anti)bottom luminosity in parton
distribution function (PDF) of the proton, which indicates there cannot
be any significant production rates via $g b(g\bar b) \to
t\slep_i^-(\bar t\slep_i^+)$ partonic processes at the LHC,
we ignore their contributions in the following calculation.
\begin{table}[!hbp]
\begin{center}
\begin{tabular}{|c|c||c|c|}
\hline
$\lambda^\prime_{131}$ & $0.019\times(m_{\tilde t_{L}}/100GeV)$ &
$\lambda^\prime_{132}$ & $0.28\times (m_{\tilde t_{L}}/100GeV)$ \\
\hline
$\lambda^\prime_{231}$ & $0.18\times(m_{\tilde b_{L}}/100GeV)$ &
$\lambda^\prime_{232}$ & $0.45 ~(m_{\tilde s_{R}}=100GeV)$ \\
\hline
$\lambda^\prime_{331}$ & $0.45 ~(m_{\tilde q}=100GeV)$ &
$\lambda^\prime_{332}$ & $0.45 ~(m_{\tilde q}=100GeV)$ \\
\hline
$\lambda^\prime_{i33}$ &  $\mathcal{O}(10^{-4})$ &&\\
\hline
\end{tabular}
\end{center}
\caption{ \label{tab-1} Upper $2\sigma$ bounds on
$\lambda^\prime_{i3k}$, where $m_{\tilde{q}_{L(R)}}$ is the mass of
the left (right) handed squark $\tilde{q}_{L(R)}$. }
\end{table}

\par
\begin{figure*}
\begin{center}
\includegraphics[ scale = 1 ]{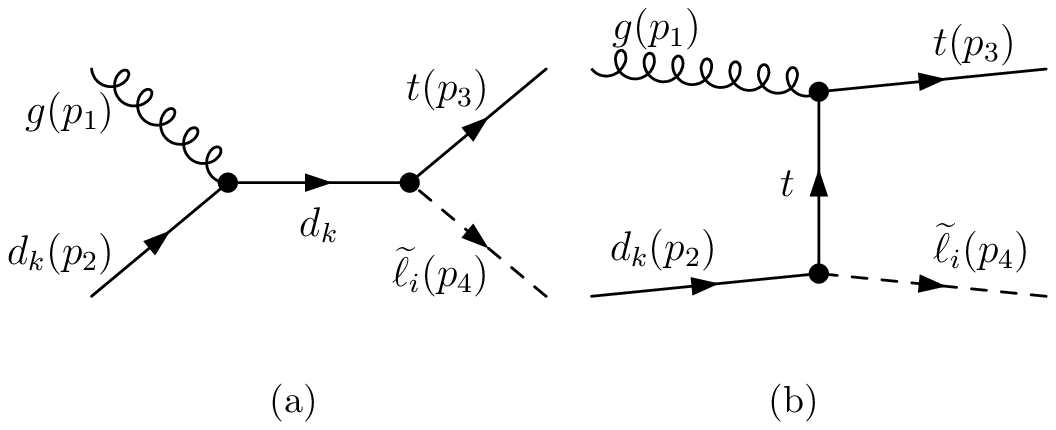}
\caption{\label{fig1} The tree-level Feynman diagrams of the \gdtsl
partonic process. }
\end{center}
\end{figure*}

Due to the CP-conservation the production cross section for $g d_k \to
t\slep_i^-$ ($i=1,2,3,k=1,2$) subprocess is the same as that for the $g \bar{d}_k \to \bar
t\slep_i^+$ ($i=1,2,3,k=1,2$) subprocess. In this section we present only the
calculations of the former subprocess. There are two tree-level
Feynman diagrams contributing to the partonic process of \gdtsl
$(i=1,2,3,k=1,2)$ as shown in Fig.\ref{fig1}(a) (for the s-channel) and
Fig.\ref{fig1}(b) (for the t-channel).

\par
The expression of the LO cross section for the partonic process
$g(p_1) d_k(p_2) \to t(p_3) \slep_i^-(p_4)$ has the form 
\begin{eqnarray}\label{eq-1}
\hat{\sigma}_{0}(gd_k \to t \slep_i^-)&=&
\frac{1}{4}\frac{1}{24}\frac{(2 \pi )^4}{2\hat{s}}\int
\sum_{spin}^{color} |{\cal M}_{LO}(gd_k \to t \slep_i^-)|^2
d\Omega_{2},
\end{eqnarray}
where the factors $\frac{1}{4}$ and $\frac{1}{24}$ come from the
averaging over the spins and colors of the initial partons, 
respectively, $\hat{s}$ is the partonic center-of-mass energy
squared, ${\cal M}_{LO}(gd_k \to t \slep_i^-)$ is the amplitude
of the tree-level Feynman diagrams shown in Fig.\ref{fig1}. The
summation in Eq.(\ref{eq-1}) is taken over the spins and colors of
all the relevant initial and final particles. The phase-space
element $d\Omega_2$ is expressed as
\begin{equation}
{d\Omega_{2}}=\delta^{(4)} \left( p_1+p_2-p_3-p_4 \right)
\prod_{i=3,4} \frac{d^3 \textbf{\textsl{p}}_i}{(2 \pi)^3 2 E_i}.
\end{equation}

\par
The LO cross-section for the parent process $pp \to gd_k \to
t\slep_i^-+X$ at the LHC can be obtained by performing the following
integrations:
\begin{equation} \label{LOsigma}
\sigma_{LO} = \sum_{d_k=d,s}\int_{0}^{1} dx_{1} \int_{0}^{1} dx_{2}
\hat{\sigma}_{0}(gd_k \to t \slep_i^-) \left[
G_{g/P_1}(x_1,\mu_f) G_{d_k /P_2}(x_2,\mu_f)+(x_1 \leftrightarrow
x_2, P_1 \leftrightarrow P_2)\right],
\end{equation}
where $G_{j/A}(x,\mu_f)$ is the PDF of parton $j~(=g$ or $d_k)$ in proton
$A~(=P_1 ,P_2)$ which describes the probability to find a parton $j$
with momentum $xP_A$ in proton $A$, $\mu_f$ is the factorization
energy scale. We adopt the CTEQ6L1 PDFs in the LO calculations.

\par
{\bf B. Real and virtual corrections }
\par
\begin{figure*}
\begin{center}
\includegraphics[ scale = 0.75 ]{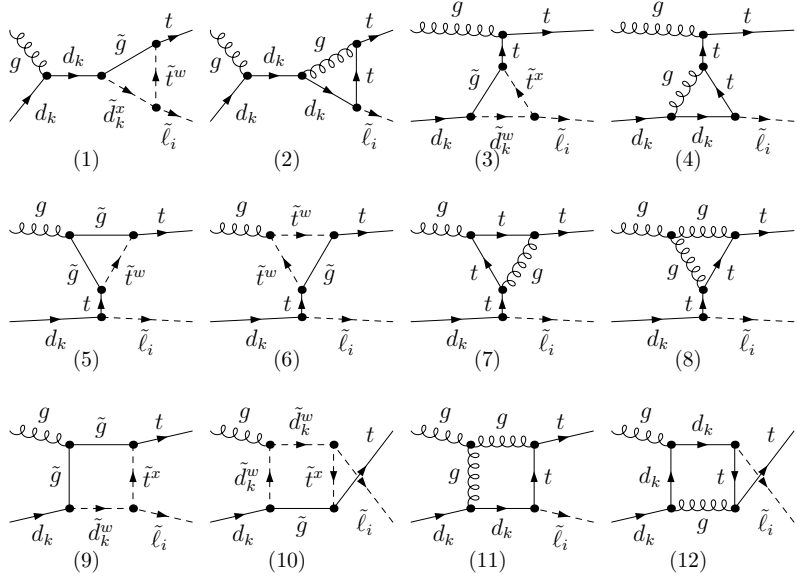}
\caption{\label{fig2} Some of the QCD one-loop Feynman diagrams for
the \gdtsl partonic process. The upper indices $x,w=1,2$, and the lower
indexes $i$ and $k$ run from the first generation to the third generation. }
\end{center}
\end{figure*}

In the NLO calculations we use the dimensional regularization method
in $D=4-2\epsilon$ dimensions to isolate the UV and IR
singularities. Some of the virtual QCD one-loop diagrams are shown
in Fig.\ref{fig2}.

\par
The NLO QCD corrections to the $gd_k \to t\slep_i^-$ partonic
process can be divided into two components: the virtual correction
and the real radiation correction. There exist ultraviolet (UV) and
infrared (IR) singularities in the virtual correction and only IR
singularity in the real radiation correction. The IR singularity
includes soft divergence and collinear divergence. In the virtual
correction component, the UV divergence vanishes by performing
renormalization procedure, and the soft IR divergency can be
completely eliminated by adding the contribution from the real gluon
emission partonic process $gd_k \to t \slep_i +g$. The collinear
divergence in the virtual correction can be partially canceled by
the collinear divergences in real gluon/light-(anti)quark emission
processes, but there still exists residual collinear divergence
which will be absorbed by the redefinitions of the PDFs.

\par
The one-loop diagrams can be divided into two independent parts. One
is the SM-like component arising from the diagrams including
gluon/quark loop, another is the pure supersymmetric (pSUSY) QCD
part where each diagram includes a gluino/squark loop.
Correspondingly we divide the counterterms also into the SM-like QCD
and the pSUSY QCD parts. The definitions of the relevant
counterterms are adopted as
\begin{eqnarray}\label{eq2-4}
q_{L/R}&& \to \left (1+\frac{1}{2}\delta Z^q_{L/R}\right )q_{L/R}=
\left [1+\frac{1}{2}\left (\delta Z^{q,(SM-like)}_{L/R} +
\delta Z^{q,(pSUSY)}_{L/R}\right )\right]q_{L/R} \nb \\
&& G^a_\mu \to \left (1+\frac{1}{2}\delta Z_{g}\right )G^a_\mu=
\left [1+\frac{1}{2}\left (\delta Z^{(SM-like)}_{g}+
\delta Z^{(pSUSY)}_{g}\right )\right ]G^a_\mu  \nb  \\
&& m_t \to m_t+\delta m_t=m_t+\delta m_t^{(SM-like)}+
\delta m^{(pSUSY)}_t   \nb \\
&& g_s\to g_s+\delta g_s=g_s+\delta g^{(SM-like)}_s+
\delta g^{(pSUSY)}_s   \nb \\
&& \lambda^\prime_{i3k}\to
\lambda^\prime_{i3k}+\delta\lambda^\prime_{i3k},
\end{eqnarray}
where $g_s$ is the strong coupling constant, $q_{L/R}$ denote the
fields of top and $d_k(=d,s)$ quarks, and $G^a_\mu$ represents the
gluon field. We renormalize the relevant fields and top-quark mass
in the on-shell scheme \cite{counterterm}. For the renormalization of
the QCD strong coupling constant $g_s$ and the $R$-parity violating
coupling coefficient $\lambda^{\prime}_{i3k}$, we use the
$\overline{MS}$ scheme \cite{MSbar_1,MSbar_2}. The counterterm of the $g_s$
can be expressed as
\begin{eqnarray}\label{eq2-5}
\frac{\delta g^{(SM-like)}_s}{g_s}&=&
-\frac{\alpha_s(\mu_r)}{4\pi}
\left[\frac{\beta^{(SM-like)}_0}{2}\frac{1}{\bar{\epsilon}}
+\frac{1}{3}\ln\frac{m_{t}^2} {\mu_{r}^2}+\frac{1}{3}\ln\frac{m_{b}^2} {\mu_{r}^2}\right], \nb \\
\frac{\delta g^{(pSUSY)}_s}{g_s}&=&  -\frac{\alpha_s(\mu_r)}{4\pi}
\left[\frac{\beta^{(pSUSY)}_1}{2}\frac{1}{\bar{\epsilon}}
+\frac{N}{3}\ln\frac{m_{\tilde{g}}^2} {\mu_{r}^2}
+\sum_{U=u,c,t}^{i=1,2}\frac{1}{12}\ln\frac{m_{\tilde{U}_i}^2}{\mu_{r}^2}
+\sum_{D=d,s,b}^{j=1,2}\frac{1}{12}\ln\frac{m_{\tilde{D}_j}^2}{\mu_{r}^2}\right ], \nb \\
\end{eqnarray}
where $1/\bar\epsilon=1/\epsilon_{UV}-\gamma_E+ln(4\pi)$,
$\beta^{(SM-like)}_0=\frac{11}{3}N-\frac{2}{3}n_{lf}-\frac{4}{3}$
and $\beta^{(pSUSY)}_1=-\frac{2}{3}N-\frac{1}{3}(n_{lf}+2)$ with
$N=3$ and $n_{lf}=4$.
\par
With the Lagrangian shown in Eq.(\ref{Lagrangian-2}), the
counterterm of the $\slep_i - \bar d_k - t$ vertex is expressed as
below:
\begin{equation}
\delta V_{\slep_i \bar d_k
t}=-\lambda^\prime_{i3k}\left(\frac{\delta\lambda^\prime_{i3k}}{\lambda^\prime_{i3k}}
+\frac{1}{2}\delta Z^{d_k}_{R}+\frac{1}{2}\delta Z^t_L\right).
\end{equation}
By using the $\overline{MS}$ scheme to renormalize the
$\slep_i-\bar d_k-t$ coupling, we get
\begin{equation}
\delta\lambda^\prime_{i3k}=\lambda^\prime_{i3k}\left(-\frac{C_F}{2}\right)
\frac{\alpha_s}{\pi\bar{\epsilon}},
\end{equation}
where $C_F=4/3$. After the renormalization we get a UV-finite
virtual correction to the partonic process \gdtsl.

\par
\begin{figure*}
\begin{center}
\includegraphics[ scale = 1 ]{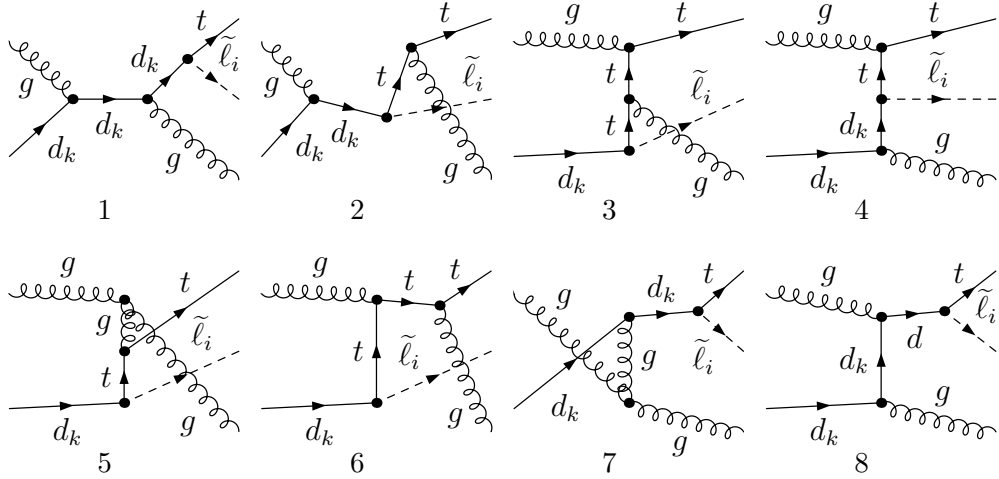}
\caption{\label{fig3} The tree-level Feynman diagrams for the real
gluon emission process $gd_k\to t \slep_{i}^-+g$. }
\end{center}
\end{figure*}

The real radiation correction includes the contributions from the
gluon and light-(anti)-quark emission processes. The contribution of
real radiation processes is at the same $\alpha_s$ order as the
virtual correction to the partonic process $gd_k \to t\slep_i^-$ in
perturbation theory according to the Kinoshita--Lee--Nauenberg(KLN)
theorem \cite{KLN_1,KLN_2}. The tree-level Feynman diagrams of the real
gluon emission partonic process $g(p_1)d_k(p_2)\to t(p_3)
\slep_{i}^-(p_4)g(p_5)$ are depicted in Fig.\ref{fig3}. We adopt the
two cutoff phase-space slicing (TCPSS) method \cite{TCPSS} to
isolate the IR singularities for the real emission subprocesses by
introducing two cutoff parameters $\delta_{s}$ and $\delta_{c}$. The
arbitrary small soft cutoff $\delta_{s}$ separates the three-body
final state phase space of real emission subprocess into two
regions: the soft region ($E_{5}\leq \delta_{s}\sqrt{\hat{s}}/2$)
and the hard region ($E_{5}>\delta_{s}\sqrt{\hat{s}}/2$). The
collinear cutoff $\delta_{c}$ separates hard region into the hard
collinear ($HC$) region and hard non-collinear ($\overline{HC}$)
region. The region for real hard gluon/light-(anti)quark emission
with $\hat{s}_{15}$ (or $\hat{s}_{25}$) $< \delta_{c}\hat{s}$ (where
$\hat{s}_{ij}=(p_{i}+p_{j})^{2}$) is called the $HC$ region.
Otherwise it is called the $\overline{HC}$ region. Then the cross
section of the real gluon emission partonic process can be written
as
\begin{equation}\label{colinear-1}
\hat \sigma^{R}_{g} ( gd_k \to t\slep_ig ) = \hat
\sigma^{S}_{g}+\hat \sigma^{H}_{g} = \hat \sigma^{S}_{g}+ \hat
\sigma^{HC}_{g}+\hat \sigma^{\overline {HC}}_{g}
\end{equation}
where $\hat \sigma^{S}_{g}$, $\hat \sigma^{HC}_{g}$ and
$\sigma^{\overline {HC}}_{g}$ are the cross sections in the soft
gluon region, hard collinear region and hard non-collinear region,
respectively.

\par
\begin{figure*}
\begin{center}
\includegraphics[ scale = 1 ]{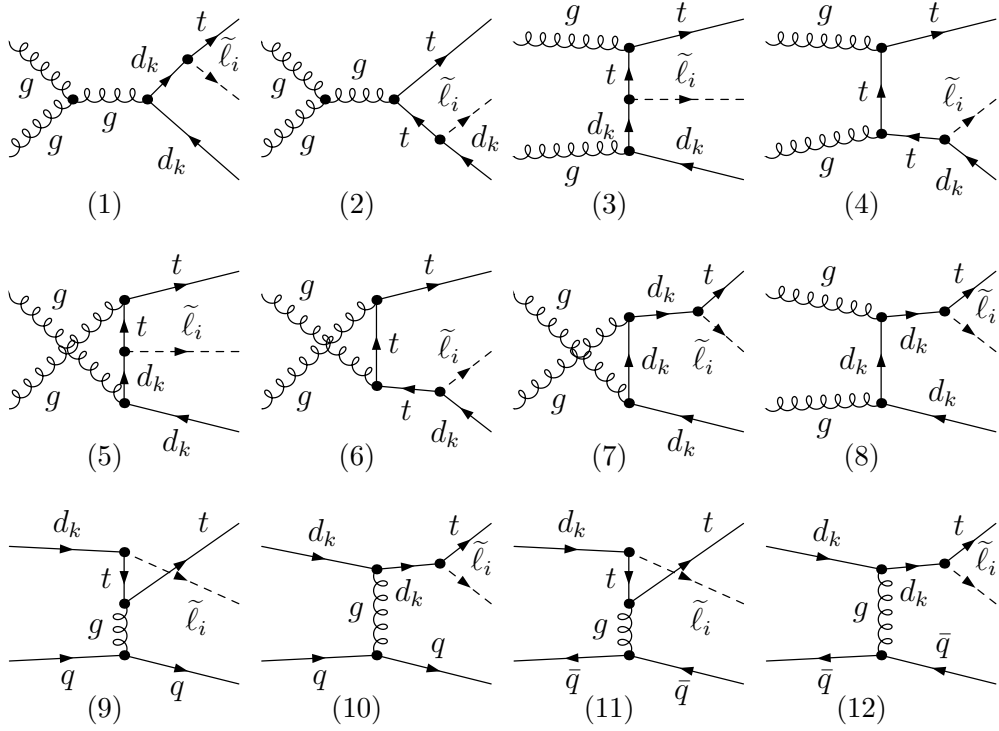}
\caption{\label{fig4} The tree-level Feynman diagrams for the real
light-(anti)quark emission processes. (1--8), (9--10) and (11--12) are
for the $gg \to t\slep_i d_k^-$, $qd_k \to t\slep_i^- q$ and $\bar
qd_k \to t\slep_i^- \bar q$ $(q=u,d,s,c)$ partonic processes,
respectively. }
\end{center}
\end{figure*}

The light-(anti)quark emission contribution at the QCD NLO to the
$pp \to t\slep_i^-+X$ process includes the partonic channels: (1)
$q(\bar q)d_k \to t\slep_i^- + q(\bar q)$, (2) $gg \to t\slep_i^-
+d_k$. The corresponding Feynman diagrams of these partonic
processes at the tree-level are shown in Fig.\ref{fig4}. There the
diagrams by exchanging the identical incoming quarks in
Figs.\ref{fig4}(9)-(10) for the partonic process $qd_k \to
t\slep_i^- q$, i.e., $q=d_k(=d,s)$ are not drawn.

\par
Again we use the TCPSS method to splitting the three-body phase
space into collinear ($C$) and non-collinear ($\overline{C}$)
regions, and get the cross sections for the subprocesses $gg\to
t\slep_i^- \bar d_k$ and $q(\bar q)d_k \to t\slep_i^- q(\bar q)$ at
the tree level expressed as
\begin{eqnarray}\label{colinear-2}
& \hat\sigma^R (gg\to t\slep_i \bar d_k)=\hat\sigma^{R}_{\bar
d_k}=\hat\sigma^{C}_{\bar d_k}+\hat\sigma^{\overline C}_{\bar
d_k},\\ \label{colinear-3}& \hat\sigma^R (q(\bar q)d_k\to t\slep_i
q(\bar q))=\hat\sigma^{R}_{q(\bar q)}=\hat\sigma^{C}_{q(\bar
q)}+\hat\sigma^{\overline C}_{q(\bar q)}.
\end{eqnarray}

\par
The cross-sections in the non-collinear region,
$\hat\sigma^{\overline C}_{\bar d_k}$, $\hat\sigma^{\overline
C}_{q(\bar q)}$(in Eqs.(\ref{colinear-2}), (\ref{colinear-3})) and
$\hat \sigma^{\overline {HC}}_{g}$ (in Eq.(\ref{colinear-1})), are
finite and can be evaluated in four dimensions by using the general
Monte Carlo method. After summing all the contributions mentioned
above there still exists the remaineding collinear divergence,
which will be absorbed by the redefinition of the PDFs at the NLO.

\par
{\bf C. Total NLO QCD correction  }
\par
Given the NLO correction components to the cross sections of
the subprocesses above, the full NLO QCD correction to the cross section for the $pp\to
t\slep_i^-+X$ process at the LHC is formally given by the QCD factorization formula as
\begin{eqnarray}
&& \sigma_{NLO}(pp\to t\slep_i^-+X)  \nonumber \\
&=& \sum_{(jk)} \int_0^1 dx_{1} \int_0^1 dx_{2}
\left\{\left[ G_{j/P_1}(x_1,\mu_f) G_{k /P_2}(x_2,\mu_f)
\hat{\sigma}_{(jk)}(x_1x_2s) \right]+(x_1 \leftrightarrow x_2, P_1
\leftrightarrow P_2)\right\},   \nonumber \\
\end{eqnarray}
where $j$ and $k$ sum over all possible types of initial partons
contributing to the subprocesses up to the QCD NLO, i.e., $(jk)$
represents $(gd),~(gs),~(gg),~(qd),~(qs), ~(\bar{q}d)$ and
$(\bar{q}s)$ (where $q=u,d,c,s$). We consistently adopt the CTEQ6m
PDFs\cite{cteq_1,cteq_2} for $G_{j/P_1(P_2)}(x_{1},\mu_f)$ and
$G_{k/P_2(P_1)}(x_{2},\mu_f)$. The total NLO QCD correction $\Delta
\sigma_{NLO}$ to the $pp\to t\slep_i^-+X$ process can be divided
into two-body term and three-body term, i.e., $\Delta
\sigma_{NLO}=\Delta \sigma^{(2)}+\Delta\sigma^{(3)}$. The two-body
term consists of the virtual correction and the cross section for
the real gluon/light-(anti)quark emission process in the soft and
hard collinear phase-space region. The three-body term consists of
the cross section for the real gluon/light-(anti)quark emission
process in the hard non-collinear phase space. Finally, the QCD
corrected total cross section for the $pp \to t\slep_i^- + X$
process is
\begin{eqnarray}
\sigma_{NLO} = \sigma_{LO} + \sigma^{(2)} + \sigma^{(3)}.
\end{eqnarray}

\vskip 5mm
\section{Numerical results and discussion}
\par
In this section we present and discuss the numerical results for the
LO and NLO QCD corrected cross sections for the $pp \to
t\slep_i^-(\bar{t}\slep_i^+)+X$ process at the early
($\sqrt{s}=7~TeV$) and future ($\sqrt{s}=14~TeV$) LHC. In order to
check the correctness of our LO calculations, we list our LO
numerical results for $pp \to gd_k\to t\slep_i^-+X$ process in Table
\ref{tab2}, and compare them with those presented in Table
\ref{tab2} and Fig.3 of Ref.\cite{LO}. Our results are obtained by
employing the same input parameters and PDFs as used in previous
work \cite{LO}. We can see that they are in agreement. But we should
say that most of the data taken from Ref.\cite{LO} are read from
the figure with large errors.
\begin{table}
\begin{center}
\begin{tabular}{|c|c|c|c|c|}
\hline {$m_{\slep_i}(GeV)$} &
\multicolumn{2}{|c|}{$pp \to
t\slep_i^-+X$ via $\lambda^{\prime}_{i31}=0.1$} &
\multicolumn{2}{|c|}{$pp \to t\slep_i^-+X$ via $\lambda^{\prime}_{i32}=0.1$} \\
\cline{2-5} & our $\sigma_{LO}$(fb) & $\sigma_{LO}$(fb) in
Ref.\cite{LO}& our $\sigma_{LO}$(fb)
& $\sigma_{LO}$(fb) in Ref.\cite{LO} \\
\hline 200    & 377.8(5) & 378   & 73.5(4)  & 73.5 \\
\hline 500    & 51.80(7) & 52    & 7.64(4)  & 7.7  \\
\hline 800    & 11.58(8) & 11.6  & 1.370(7) & 1.4  \\
\hline
\end{tabular}
\end{center}
\caption{ \label{tab2} The comparison of our LO numerical results
for the $pp \to gd_k\to t\slep_i^-+X$ process at the
$\sqrt{s}=14~TeV$ LHC with those in Ref.\cite{LO}. The relevant
parameters and the PDFs being the same as used in Ref.\cite{LO}.  }
\end{table}

\par
In the following numerical calculation, we take one-loop running
$\alpha_s$ and two-loop running $\alpha_s$ in the LO and NLO
calculations, respectively \cite{Amsler}. The number of active
flavors is taken as $N_f=4$, and the QCD parameter
$\Lambda_4^{LO}=215~MeV$ and the CTEQ6L1 PDFs are adopted in the LO
calculation, while $\Lambda_4^{\overline{MS}}=326~MeV$ and the
CTEQ6M PDFs are used for the NLO calculation \cite{cteq_1,cteq_2}. We set
factorization scale and renormalization scale equal and take
$\mu=\mu_0\equiv(m_t + m_{\slep_i})/2$ in default. The CKM matrix is
set to the unit matrix. We ignore the masses of electron and u-, d-,
s-, c-quarks, and take $m_t=172.0~GeV$ and $m_b=4.2~GeV$ in the
numerical calculation \cite{Amsler}.

\par
In the calculation of the process $pp \to t\slep_i^-+X$ we assume that
the masses of the sleptons of three generations are degenerated with
the values of $m_{\slep_i} = m_{\tilde{e}_L},(i=1,2,3)$, and take
the $\rp$ coupling parameters as (1) $\lambda^{\prime}_{i31}=0.1$
and the other $\lambda^{\prime}=0$; (2) $\lambda^{\prime}_{i32}=0.1$ and
the other $\lambda^{\prime}=0$. We follow the SPA benchmark point SPS1a'
\cite{SPA}, where the input parameters are $m_0 = 70~GeV$, $m_{1/2}
= 250~GeV$, $A_0 = -300~GeV$, $sign(\mu) = +1$, $\tan\beta = 10$,
and $m_t =172.0~GeV$, but we assume that there is no left- and
right-squark mixing in the first two generations and the degenerat
sleptons having the masses $m_{\slep_i} = 189.9~GeV~(i=1,2,3)$.
Then we get the left--right stop (sbottom) mixing angle $\theta_t =
57.026^{\circ}$ ($\theta_b = 19.75^{\circ}$) by adopting the ISAJET
program\cite{isajet}, and the other relevant SUSY parameters with
the values:
\begin{eqnarray}\label{SUSY-mass}
m_{\tilde g} &=& 607.1~GeV,~~ m_{\tilde t_1}=366.5~GeV,~~
m_{\tilde t_2}=585.5~GeV,  \nb \\
m_{\tilde b_1} &=& 506.3~GeV,~~ m_{\tilde b_2}=545.7~GeV,~~
m_{\tilde u_1}=m_{\tilde c_1}=m_{\tilde u_R}=547.2~GeV,    \nb \\
m_{\tilde u_2} &=& m_{\tilde c_2}=m_{\tilde u_L}=564.7~GeV,~~
m_{\tilde d_1} = m_{\tilde s_1}=m_{\tilde d_R}=546.9~GeV,  \nb \\
m_{\tilde d_2} &=& m_{\tilde s_2}=m_{\tilde d_L}=570.1~GeV,~~
m_{\tilde{\chi}_{1}^{0}} = 97.7~GeV,~~\mu=396~GeV.
\end{eqnarray}
The SM parameters used in the calculation are taken as follows
\cite{Amsler}
\begin{eqnarray}\label{other-parameter}
\alpha_{ew} &=& 1/137.036,~~\sin^2\theta_w = 0.23119,~~m_W = 80.398~GeV,  \nb \\
m_Z&=&91.1876~GeV,~~m_{\mu} = 0.1057~GeV.
\end{eqnarray}
In the following calculations we take the parameters stated above if
there is no other statement.

\par
\begin{figure}[htbp]
\includegraphics[width=3.2in,height=3.2in]{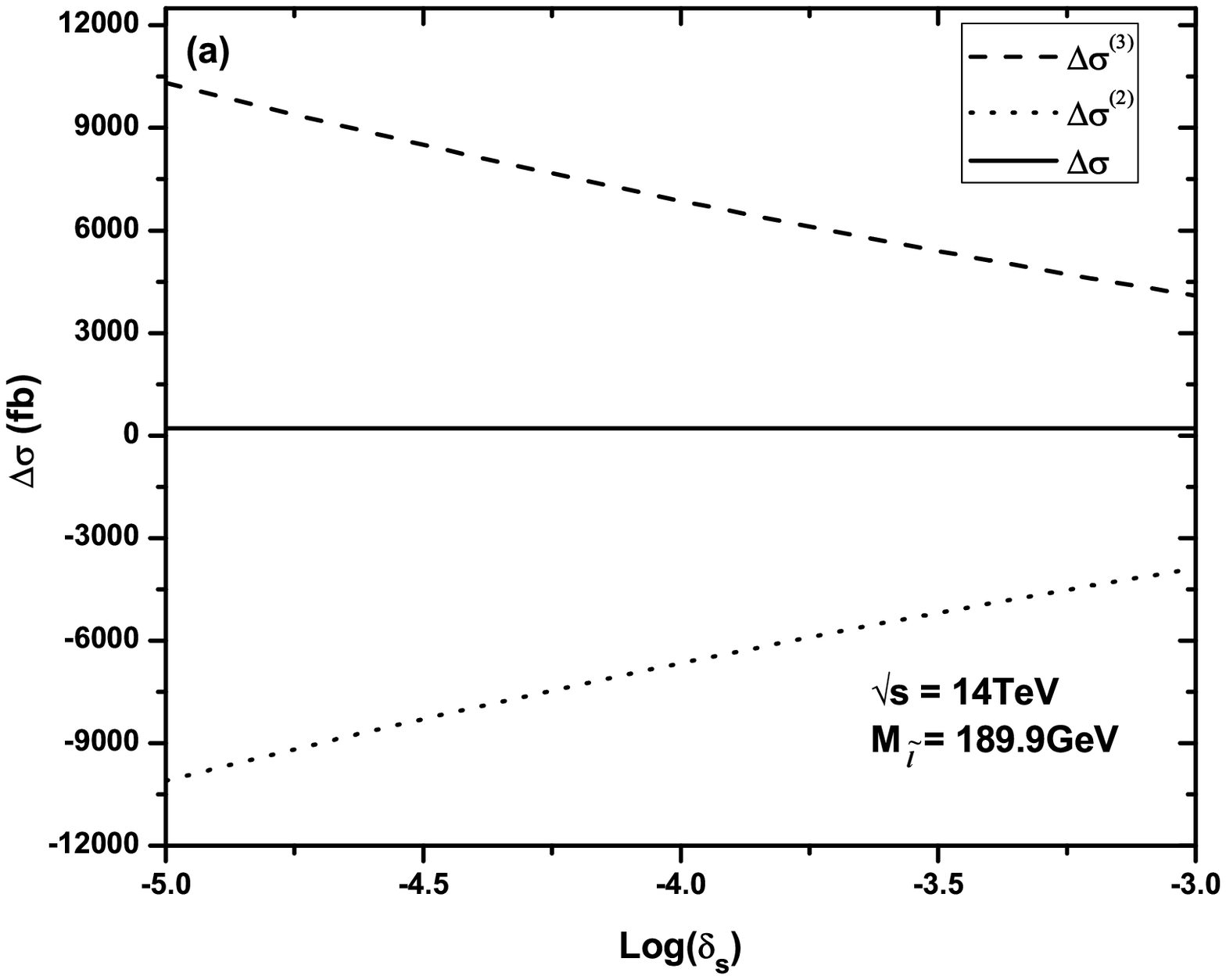}%
\hspace{0in}%
\includegraphics[width=3.2in,height=3.2in]{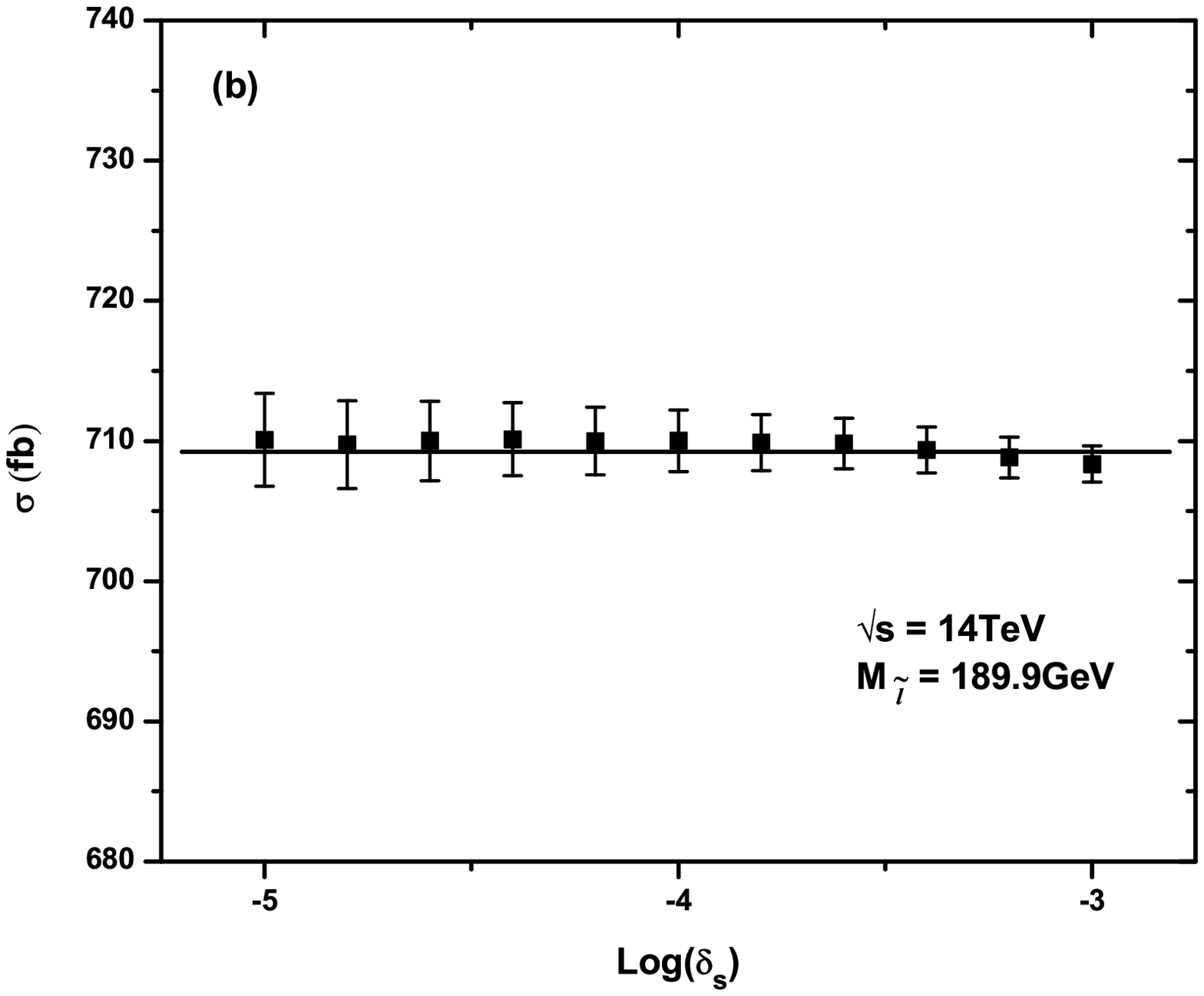}%
\hspace{0in}%
\caption{\label{fig5} (a) The dependence of NLO QCD corrections
to the $pp\to gd\to t\slep_i^-+X$ process on the soft cutoff
$\delta_s$ at the $\sqrt{s}=14~TeV$ LHC with $A=0~TeV$, $m_{\slep_i}=189.9~GeV$
and $\delta_c = \delta_s/50$. (b) The NLO QCD corrected total cross
section $\sigma_{NLO}$ with calculation errors as the function of
$\delta_s$. }
\end{figure}

The independence of the full NLO QCD correction on the two cutoffs,
$\delta_s$ and $\delta_c$, is confirmed numerically.
Figs.\ref{fig5}(a,b) demonstrate that the total NLO QCD correction
to the $pp\to gd \to t\slep_i^- +X$ process with $A=0~TeV$ at the $\sqrt{s}=14~TeV$
LHC does not depend on the arbitrarily chosen values of the
$\delta_s$ and $\delta_c$ within the calculation errors, where we
take $\lambda^{\prime}_{i31}=0.1$ and other $\lambda^{\prime}=0$. In
Fig.5(a), the two-body correction $\Delta\sigma^{(2)}$, three-body
correction $\Delta\sigma^{(3)}$ and the total QCD correction
($\Delta\sigma_{NLO}$) for the $pp\to gd\to t\slep_i^- +X$ process
are depicted as functions of the soft cutoff $\delta_s$ running
from $1\times 10^{-5}$ to $1\times 10^{-3}$ with
$\delta_c=\delta_s/50$. In Fig.\ref{fig5}(b), the amplified curve
for $\Delta\sigma_{NLO}$ is depicted. The results in these two
figures demonstrate that the total QCD correction to the $pp\to
gd\to t\slep_i^- +X$ process is independent of $\delta_s$ and
$\delta_c$. It verifies the cancelation of the soft/collinear IR
divergence in the total QCD correction to $pp\to gd\to
t\slep_i^- +X$. In further numerical calculations, we fix
$\delta_s=5 \times 10^{-5}$ and $\delta_c=\delta_s/50$.

\par
For the gluon/light-(anti)quark jet event selection we adopt two
selection schemes: (1) Inclusive jet event selection scheme. With
this scheme we accept all the real gluon/light-(anti)quark emission
events. (2) Exclusive jet event selection scheme, by which we accept
the real gluon/light-(anti)quark emission event satisfying the
restriction of either $p^{jet}_T <p^{cut}_{T,jet}$ on the jet
transverse momentum or $\eta^{jet}>\eta^{cut}_{jet}$ on the jet
rapidity. In further calculations with exclusive jet event selection
scheme, we take the cut parameters for gluon/light-(anti)quark jet
as $p^{cut}_{T,jet}=50~GeV$ and $\eta^{cut}_{jet}=3$.

\par
\begin{figure}[htbp]
\includegraphics[width=3.2in,height=3.2in]{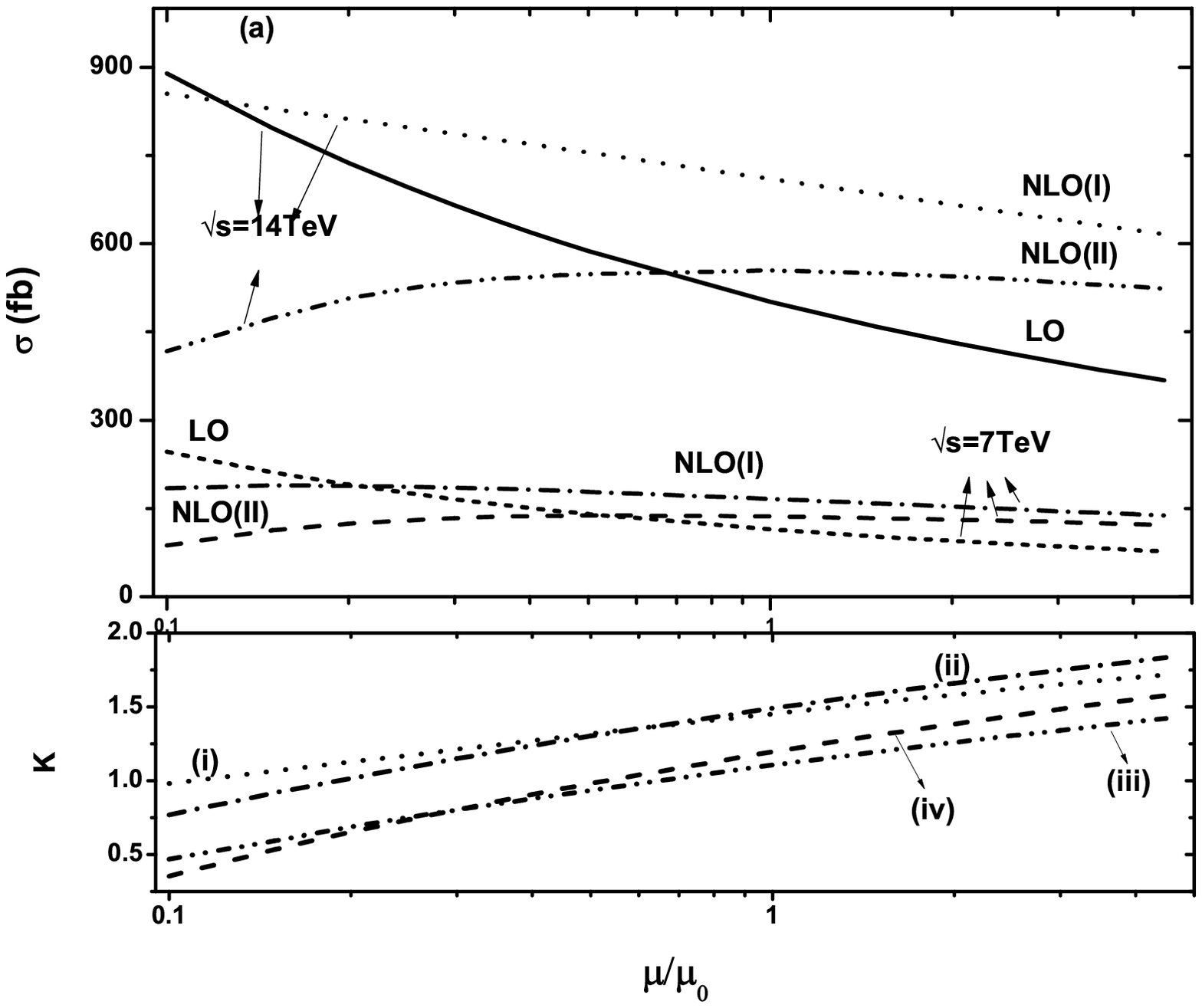}%
\hspace{0in}%
\includegraphics[width=3.2in,height=3.2in]{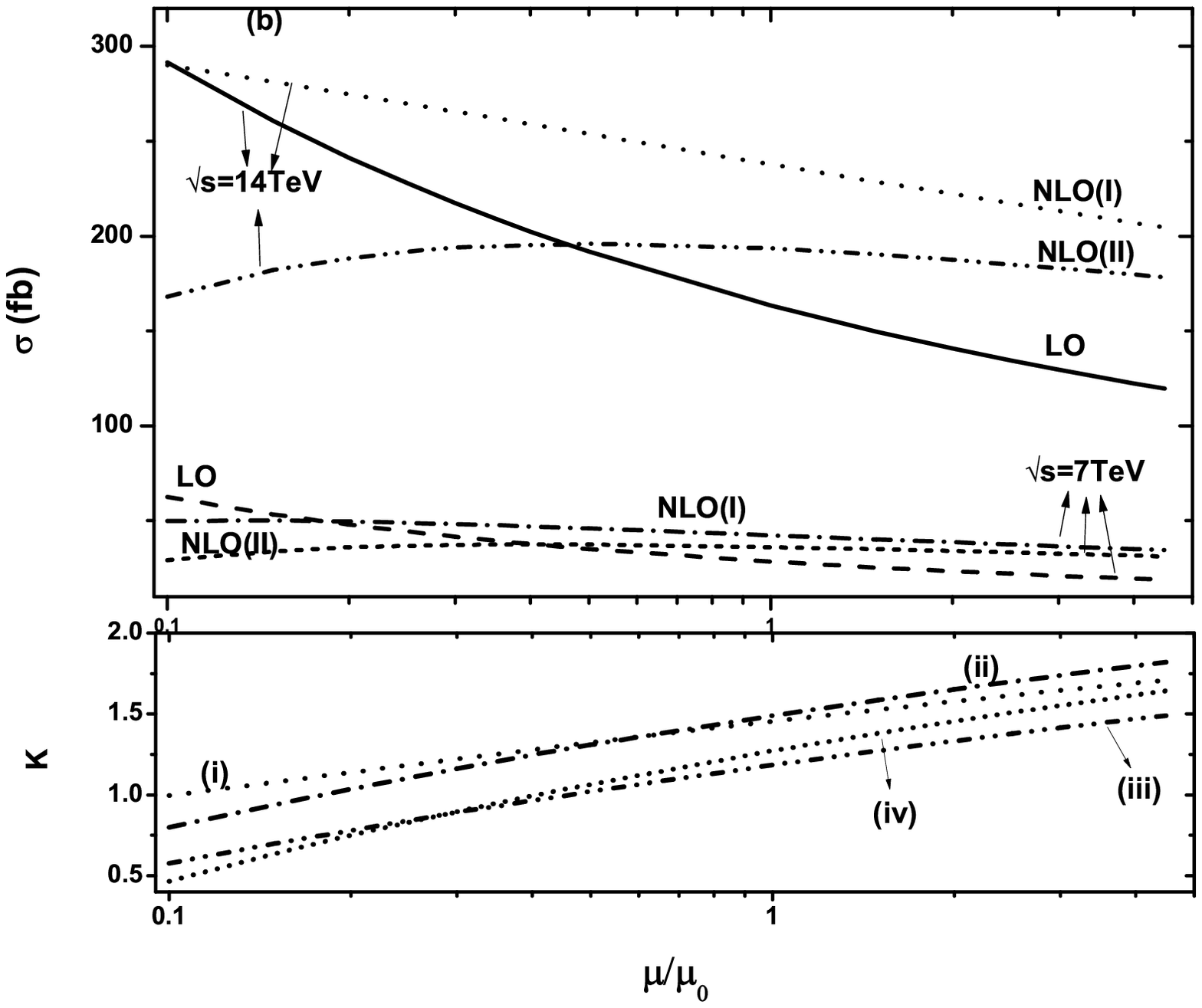}%
\hspace{0in}%
\includegraphics[width=3.2in,height=3.2in]{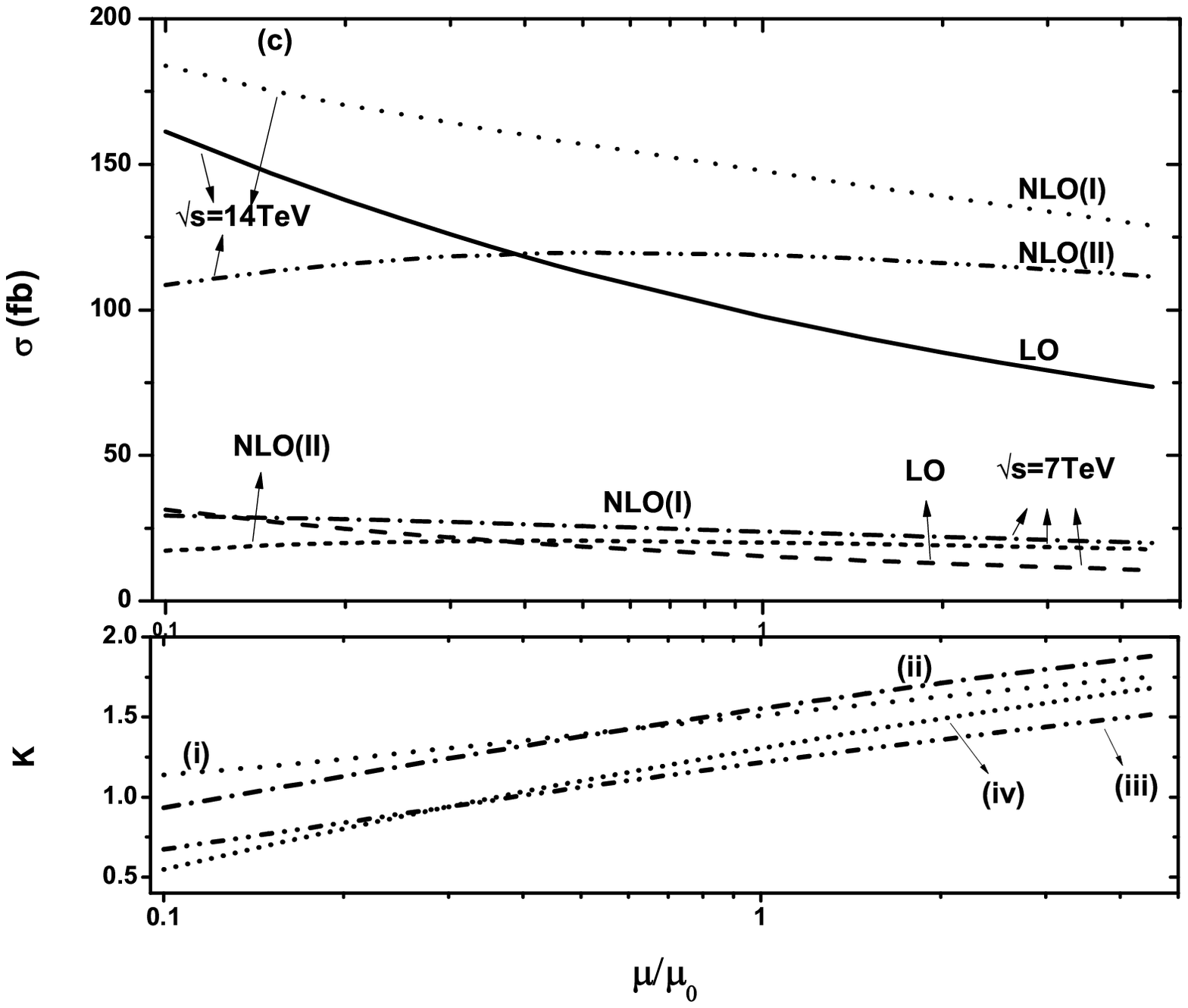}%
\caption{\label{fig6} The LO, NLO QCD corrected cross sections and
the corresponding K-factors versus the factorization/renormalization
scale ($\mu/\mu_0$) with $A=0~TeV$ at the LHC. (a) For the $pp\to gd \to
t\slep^-_i+X$ process. (b) For the $pp\to g \bar{d} \to
\bar{t}\slep^+_i+X$ process. (c) For the $pp\to gs \to t\slep^-_i+X$
process. The K-factor curves labeled (i), (ii), (iii) and (iv)
are for (i) the inclusive scheme with $\sqrt s=14~TeV$, (ii) the
inclusive scheme with $\sqrt s=7~TeV$, (iii) the exclusive scheme
with $\sqrt s=14~TeV$, (iv) the exclusive scheme with $\sqrt
s=7~TeV$, respectively. }
\end{figure}

One of the main reasons to base the LHC analyses on higher order
predictions is the stabilization of the dependence on the unphysical
renormalization and the factorization scales. In the upper figures
of Figs.\ref{fig6}(a,b,c), we show the dependence of the LO and NLO
QCD corrected cross sections for the processes $pp\to gd \to
t\slep^-_i+X$, $pp\to g\bar d \to \bar t\slep^+_i+X$, and $pp\to gs
\to t\slep^-_i+X$ on the factorization/renormalization scale
($\mu/\mu_0$) with $A=0~TeV$ at the $\sqrt{s}=7~TeV$ and
$\sqrt{s}=14~TeV$ LHC, respectively. The corresponding K-factor
$[K\equiv\sigma_{NLO}/\sigma_{LO}]$ are shown in the lower figures
of Figs.\ref{fig6}(a,b,c). Since the luminosities of the $s$- and
$\bar{s}$-quark in proton are the same, the observables for the
process $pp\to g\bar{s} \to \bar{t}\slep^+_i+X$ are equal to those
for the process $pp\to gs \to t\slep^-_i+X$, we present only the
plots for the process $pp\to gs \to t\slep^-_i+X$. There we assume
$\mu\equiv\mu_r=\mu_f$ for simplicity, and set
$\lambda^{\prime}_{i31}=0.1$ and other $\lambda^{\prime}=0$ in
Figs.\ref{fig6}(a,b), while $\lambda^{\prime}_{i32}=0.1$ and other
$\lambda^{\prime}=0$ in Fig.\ref{fig6}(c). The curves labeled
NLO(I) (NLO(II)) denotes the NLO QCD corrected cross section with
the inclusive (exclusive) jet events selection scheme. The curves in
the lower figures of Figs.\ref{fig6}(a,b,c) labeled (i), (ii),
(iii) and (iv) correspond to the K-factors for (i) the inclusive
scheme with $\sqrt s=14~TeV$, (ii) the inclusive scheme with $\sqrt
s=7~TeV$, (iii) the exclusive scheme with $\sqrt s=14~TeV$, (iv) the
exclusive scheme with $\sqrt s=7~TeV$, respectively. These notations
are also adopted in the following figures. We can see from
Figs.\ref{fig6}(a,b,c) when the scale $\mu$ runs from $0.1\mu_0$ to
$4\mu_0$, the curves for the NLO QCD corrected cross section becomes
more stable in comparison with the corresponding curves for the LO.
It demonstrates that the NLO QCD corrections can improve the scale
uncertainty apparently, and the exclusive scheme keeps the
convergence of the perturbative series better than the inclusive
scheme in the plotted $\mu/\mu_0$ range.

\par
\begin{figure}[htbp]
\includegraphics[width=3.2in,height=3.2in]{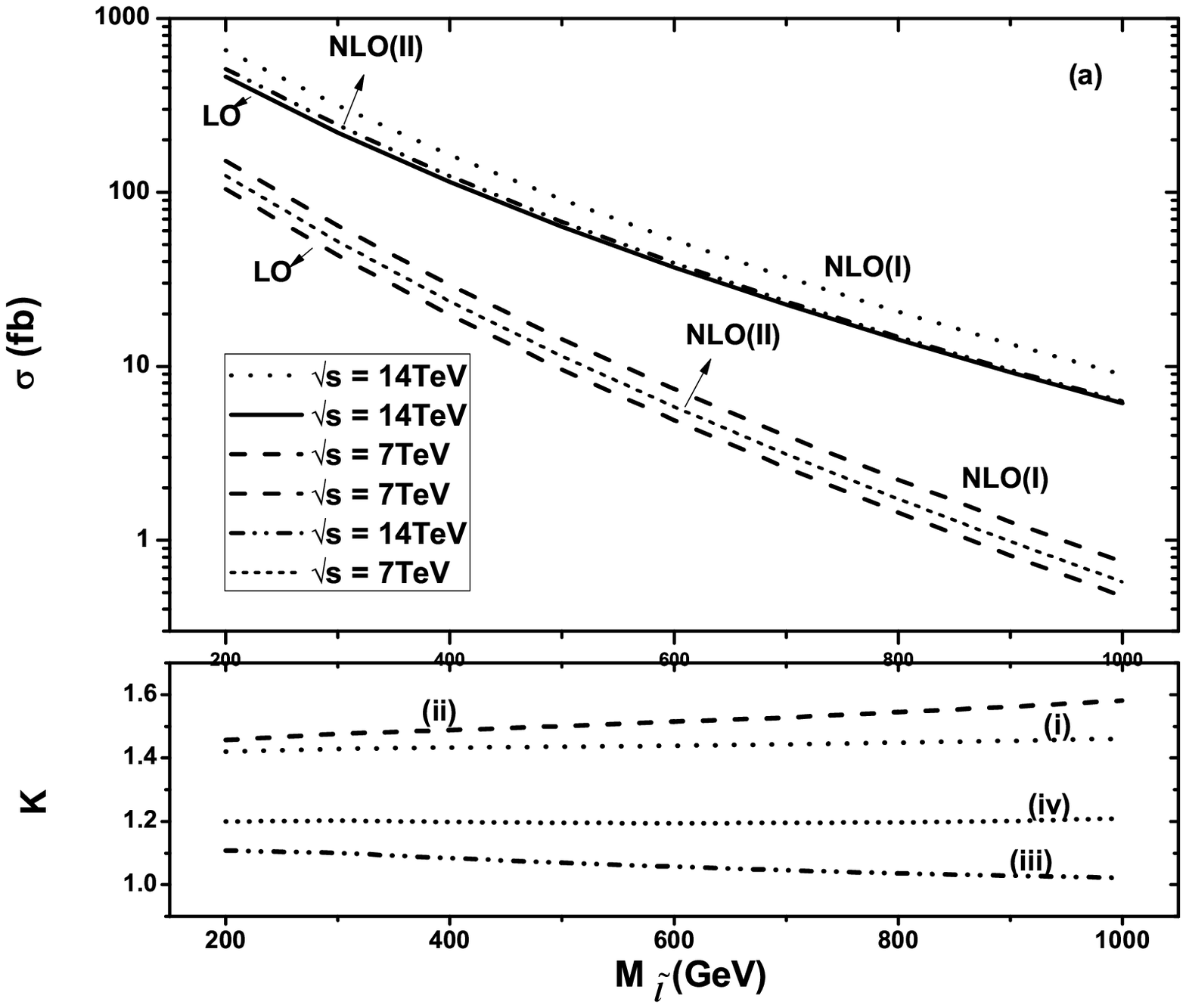}%
\hspace{0in}%
\includegraphics[width=3.2in,height=3.2in]{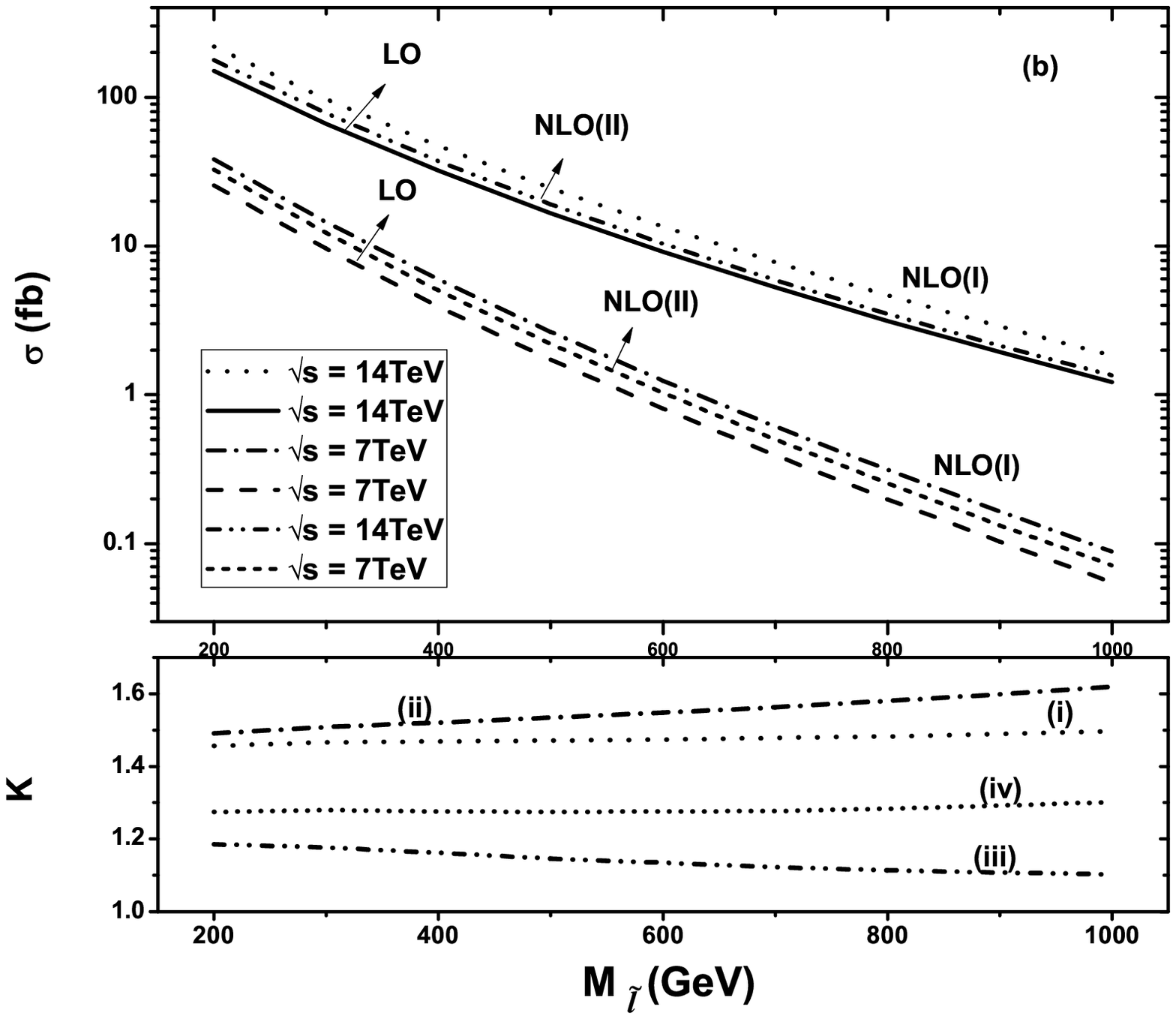}%
\hspace{0in}%
\includegraphics[width=3.2in,height=3.2in]{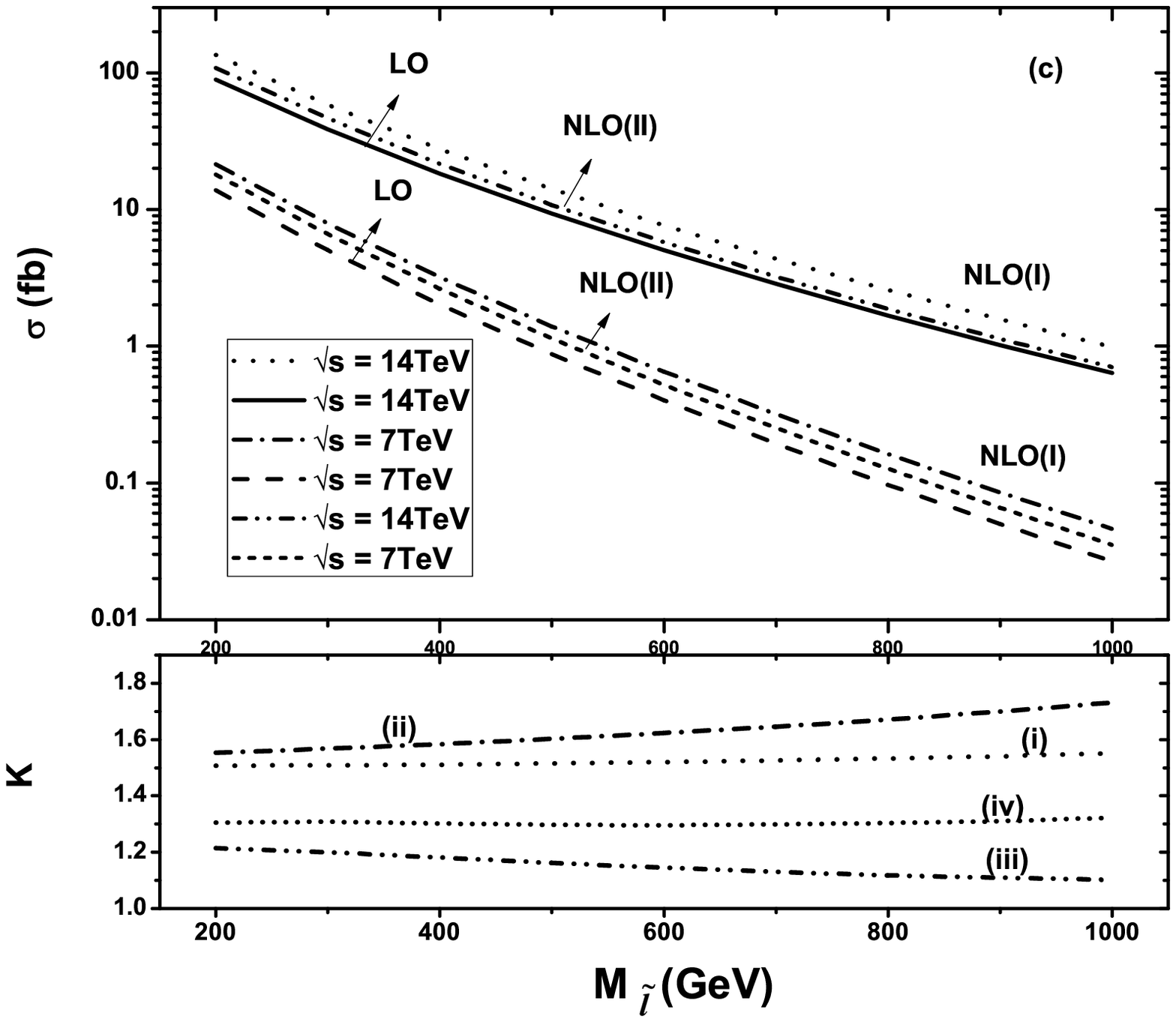}%
\caption{\label{fig7} The LO, NLO QCD corrected cross sections and
the corresponding K-factors versus the mass of slepton $m_{\slep_i}$
with $A=0~TeV$ at the LHC. (a) For the $pp\to gd \to t\slep^-_i+X$ process.
(b) For the $pp\to g \bar{d} \to \bar{t}\slep^+_i+X$ process. (c) For the
$pp\to gs \to t\slep^-_i+X$ process. The descriptions for the
K-factor curves labeled (i), (ii), (iii) and (iv) are the same
as in Figs.\ref{fig6}. }
\end{figure}

In Figs.\ref{fig7}(a,b,c) we plot the LO, NLO QCD corrected cross
sections and the corresponding K-factors as the functions of the
final slepton mass $m_{\slep_i}$ with $A=0~TeV$ for the $pp\to gd \to
t\slep^-_i+X$, $pp\to g\bar d \to \bar t\slep^+_i+X$, and $pp\to gs
\to t\slep^-_i+X$ processes, respectively. In each figure the curves
for the $\sqrt{s}=7~TeV$ and $\sqrt{s}=14~TeV$ LHC are depicted. In
Figs.\ref{fig7}(a,b) we set the $R$-parity violating coupling
coefficients as $\lambda^{\prime}_{i31}=0.1$ and the other
$\lambda^{\prime}=0$, while in Fig.\ref{fig7}(c) we have
$\lambda^{\prime}_{i32}=0.1$ and the other $\lambda^{\prime}=0$. We can
see that both the LO and NLO QCD corrected cross sections decrease
with the increment of the value of $m_{\slep_i}$. The curves in
these figures show that the cross sections for $t\slep^-_i$ and
$\bar{t}\slep^+_i$ production processes are different. Unlike the
luminosities of $s$- and $\bar{s}$-quark in proton, being equal, the
initial d-quark has a higher luminosity than the $\bar{d}$-quark. This
induces the cross-section of the process $pp\to gd \to t\slep^-_i
+X$ to always be larger than the process $pp\to g\bar{d} \to\bar t
\slep^+_i +X$.

\par
\begin{figure}[htbp]
\includegraphics[width=3.2in,height=3.2in]{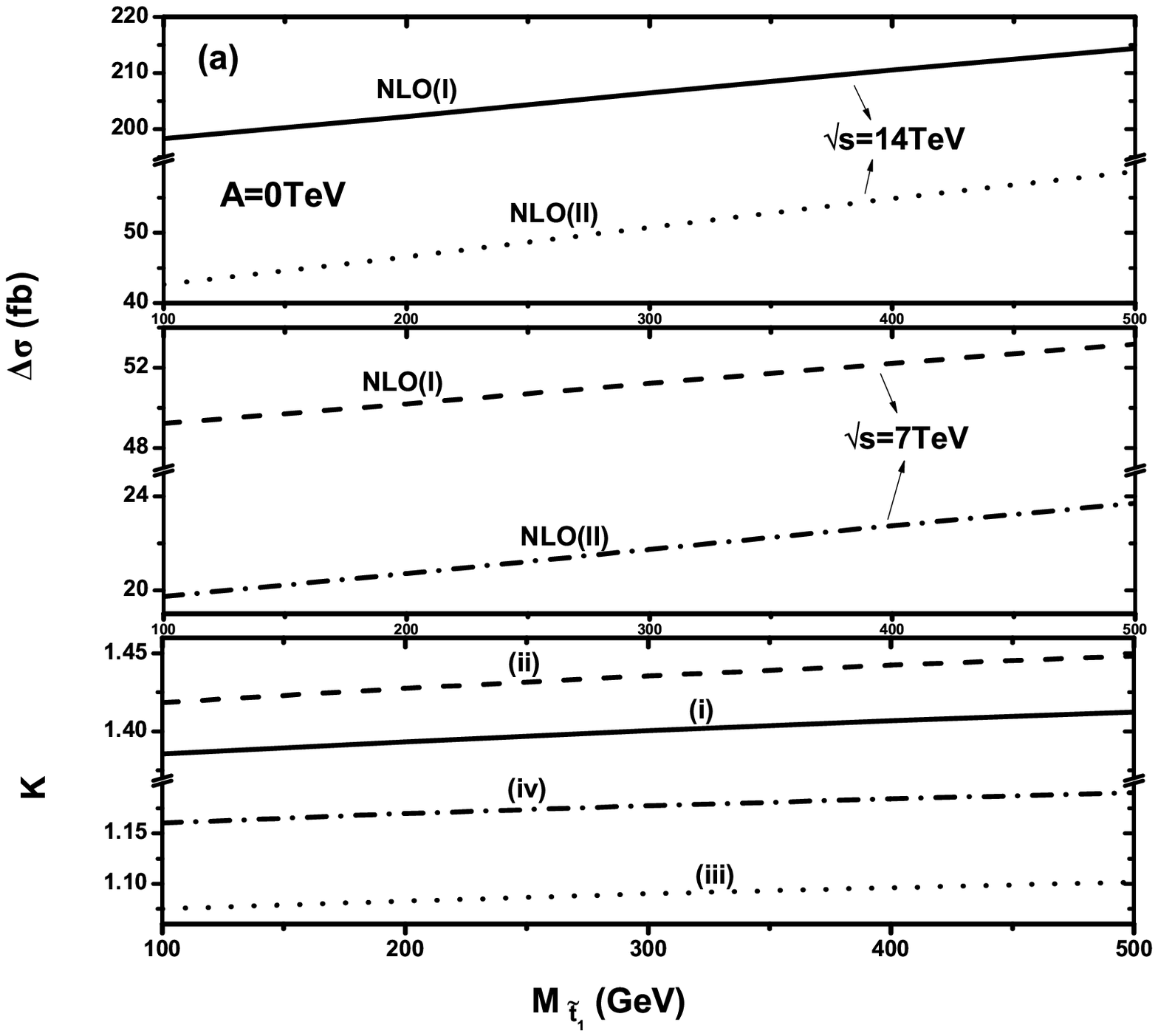}%
\hspace{0in}%
\includegraphics[width=3.2in,height=3.2in]{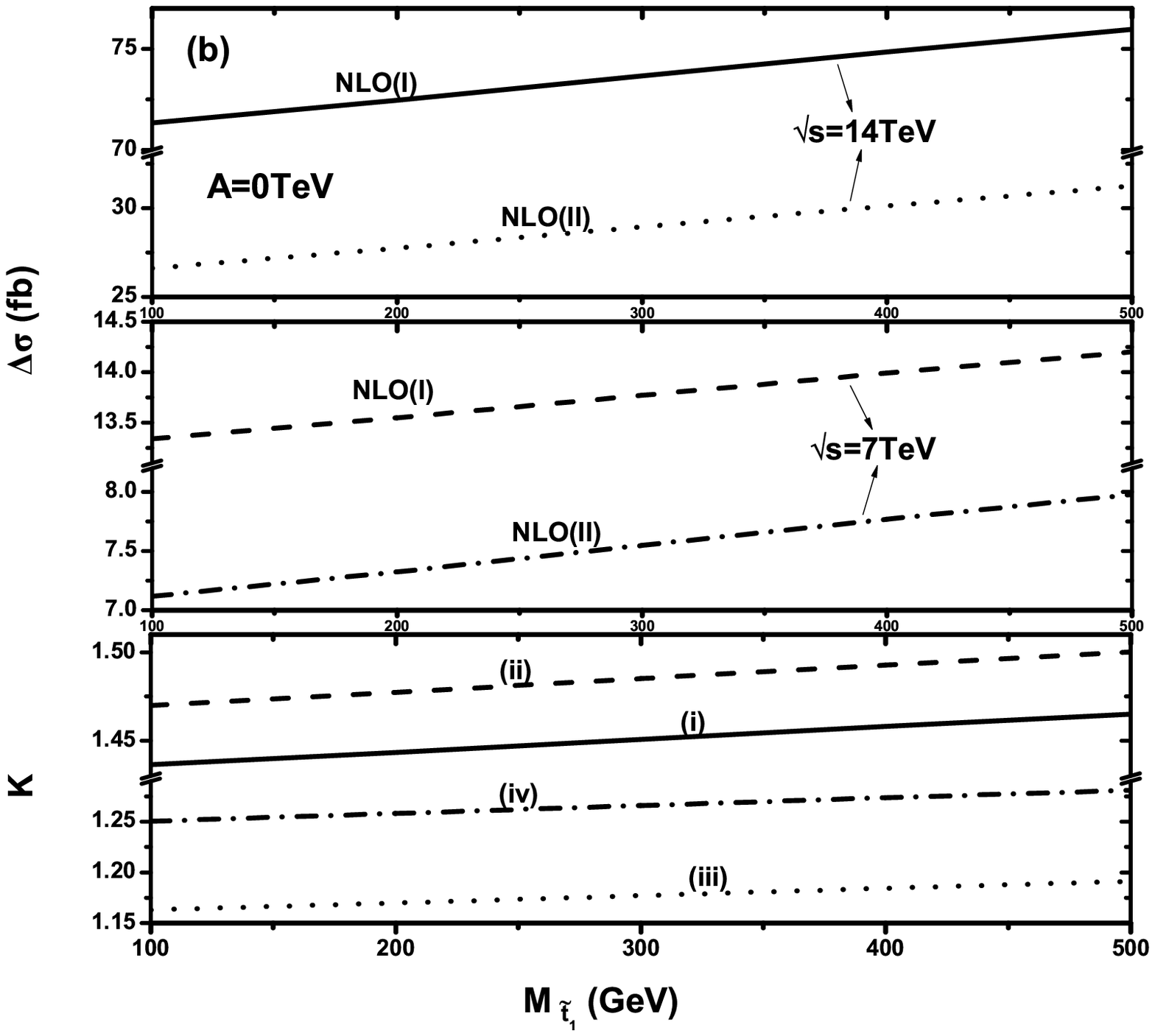}%
\hspace{0in}%
\includegraphics[width=3.2in,height=3.2in]{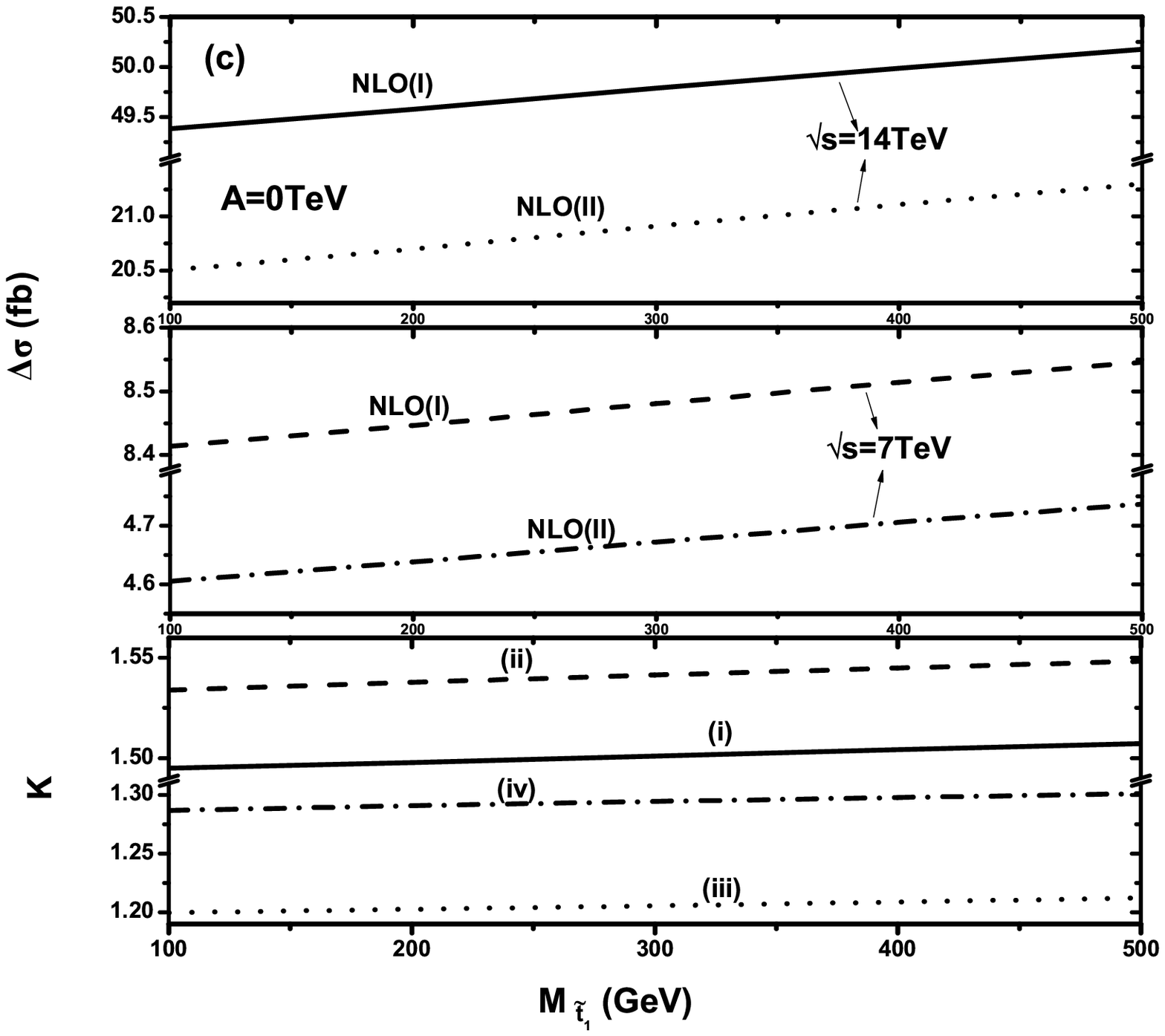}%
\caption{\label{fig8-1} The NLO QCD correction to the LO cross
sections $(\Delta \sigma_{NLO})$ and the corresponding K-factors
versus $m_{\tilde{t}_1}$ with $A=0~TeV$ at the $\sqrt{s}=7~TeV$ and
$\sqrt{s}=14~TeV$ LHC. (a) For the $pp\to gd \to t\slep^-_i+X$
process. (b) For the $pp\to g \bar{d} \to \bar{t}\slep^+_i+X$
process. (c) For the $pp\to gs \to t\slep^-_i+X$ process. The
descriptions for the K-factor curves labeled (i), (ii), (iii)
and (iv) are the same as in Figs.\ref{fig6}. }
\end{figure}
\begin{figure}[htbp]
\includegraphics[width=3.2in,height=3.2in]{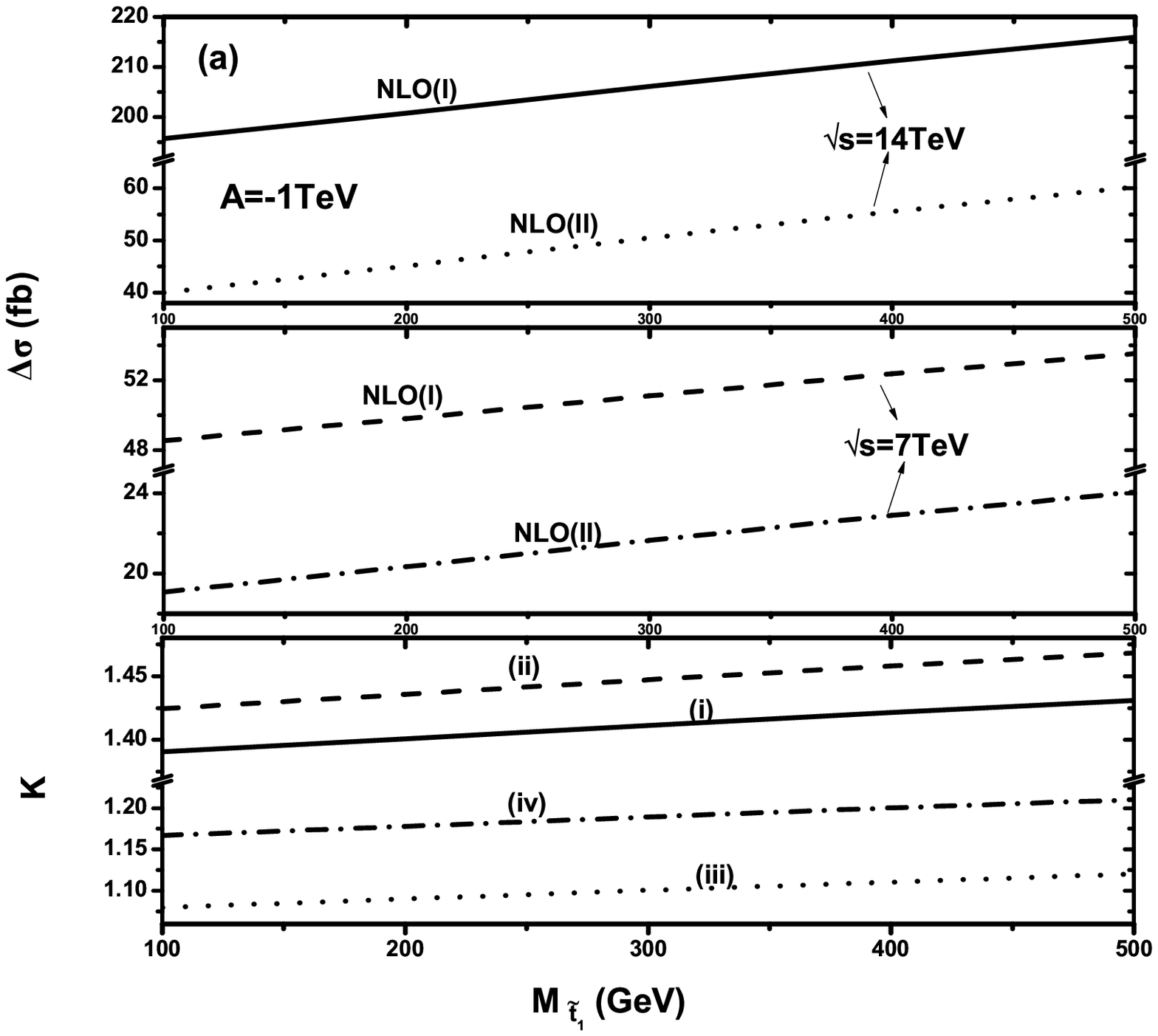}%
\hspace{0in}%
\includegraphics[width=3.2in,height=3.2in]{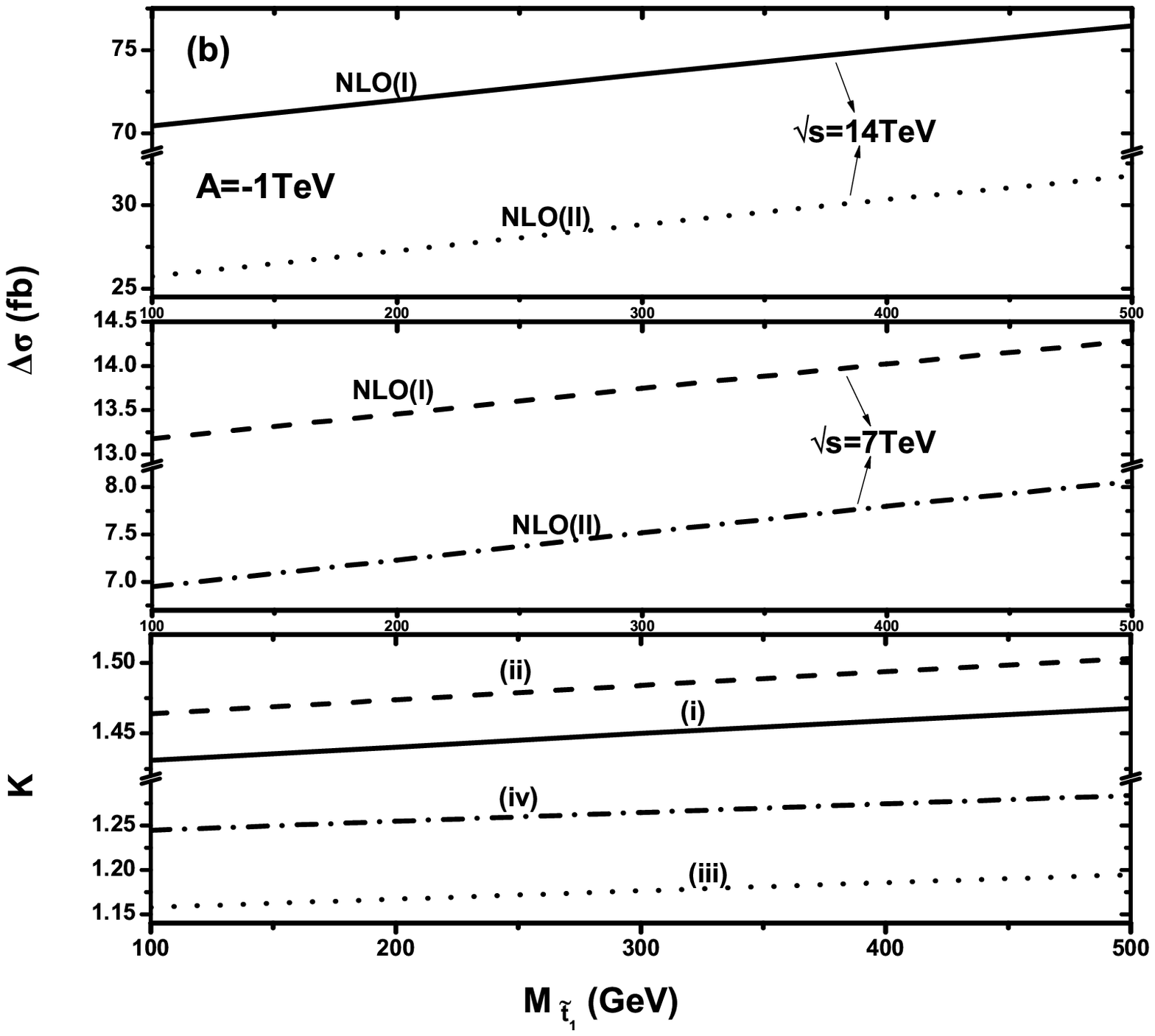}%
\hspace{0in}%
\includegraphics[width=3.2in,height=3.2in]{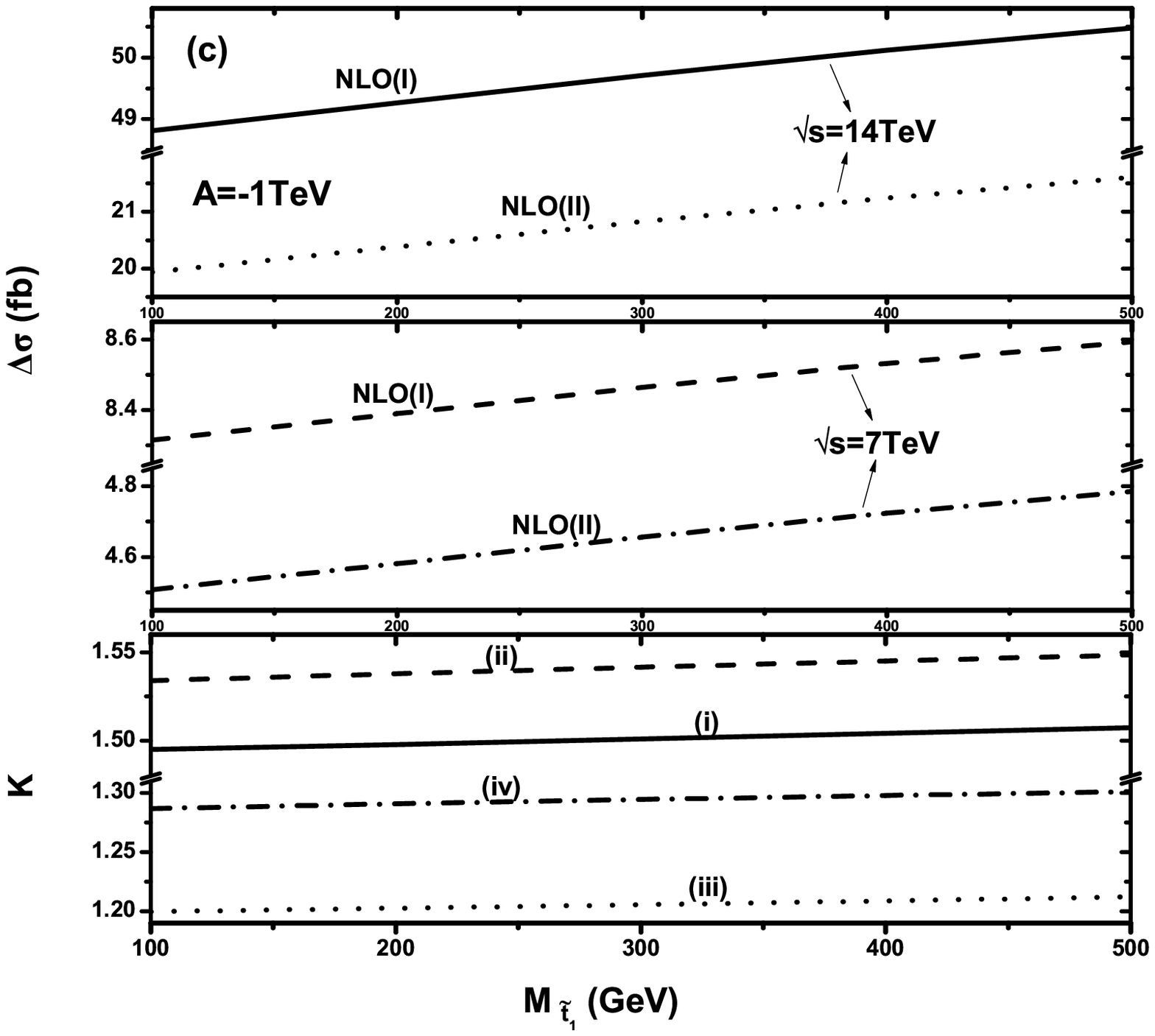}%
\caption{\label{fig8-2} The NLO QCD correction to the LO cross
sections $(\Delta \sigma_{NLO})$ and the corresponding K-factors
versus $m_{\tilde{t}_1}$ with $A=-1~TeV$ at the $\sqrt{s}=7~TeV$ and
$\sqrt{s}=14~TeV$ LHC. (a) For the $pp\to gd \to t\slep^-_i+X$
process. (b) For the $pp\to g \bar{d} \to \bar{t}\slep^+_i+X$
process. (c) For the $pp\to gs \to t\slep^-_i+X$ process. The
descriptions for the K-factor curves labeled (i), (ii), (iii)
and (iv) are the same as in Figs.\ref{fig6}. }
\end{figure}
\begin{figure}[htbp]
\includegraphics[width=3.2in,height=3.2in]{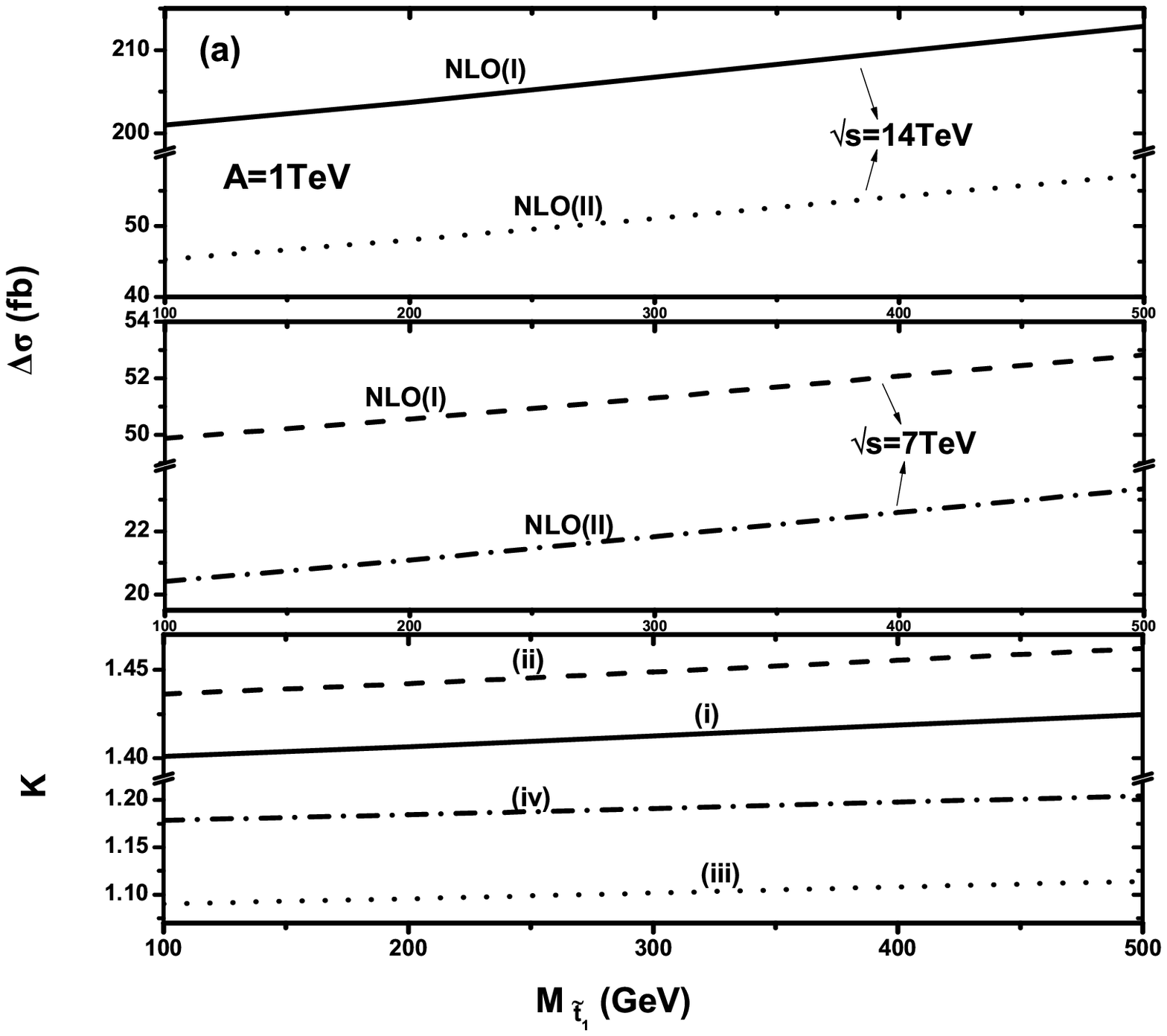}%
\hspace{0in}%
\includegraphics[width=3.2in,height=3.2in]{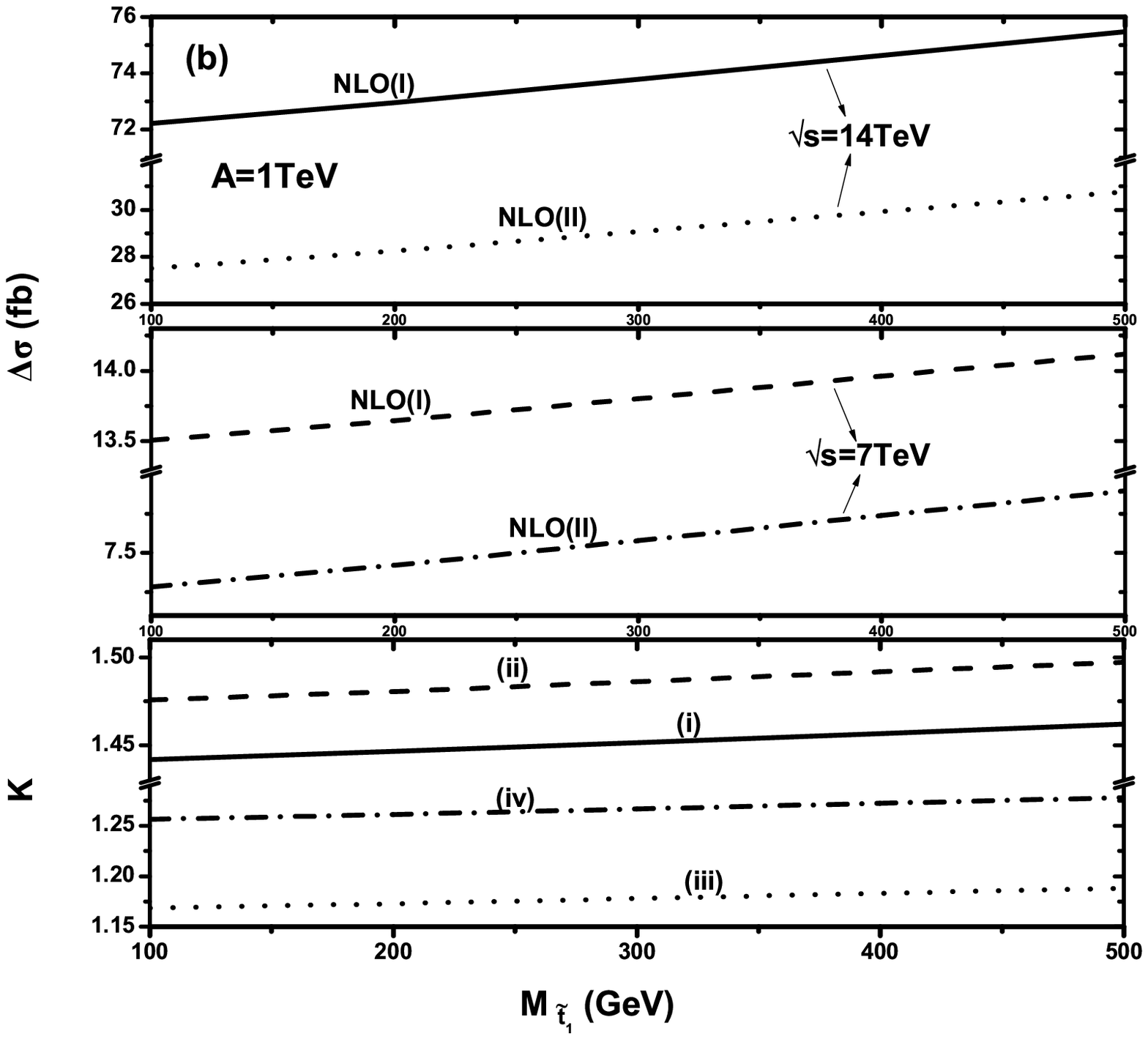}%
\hspace{0in}%
\includegraphics[width=3.2in,height=3.2in]{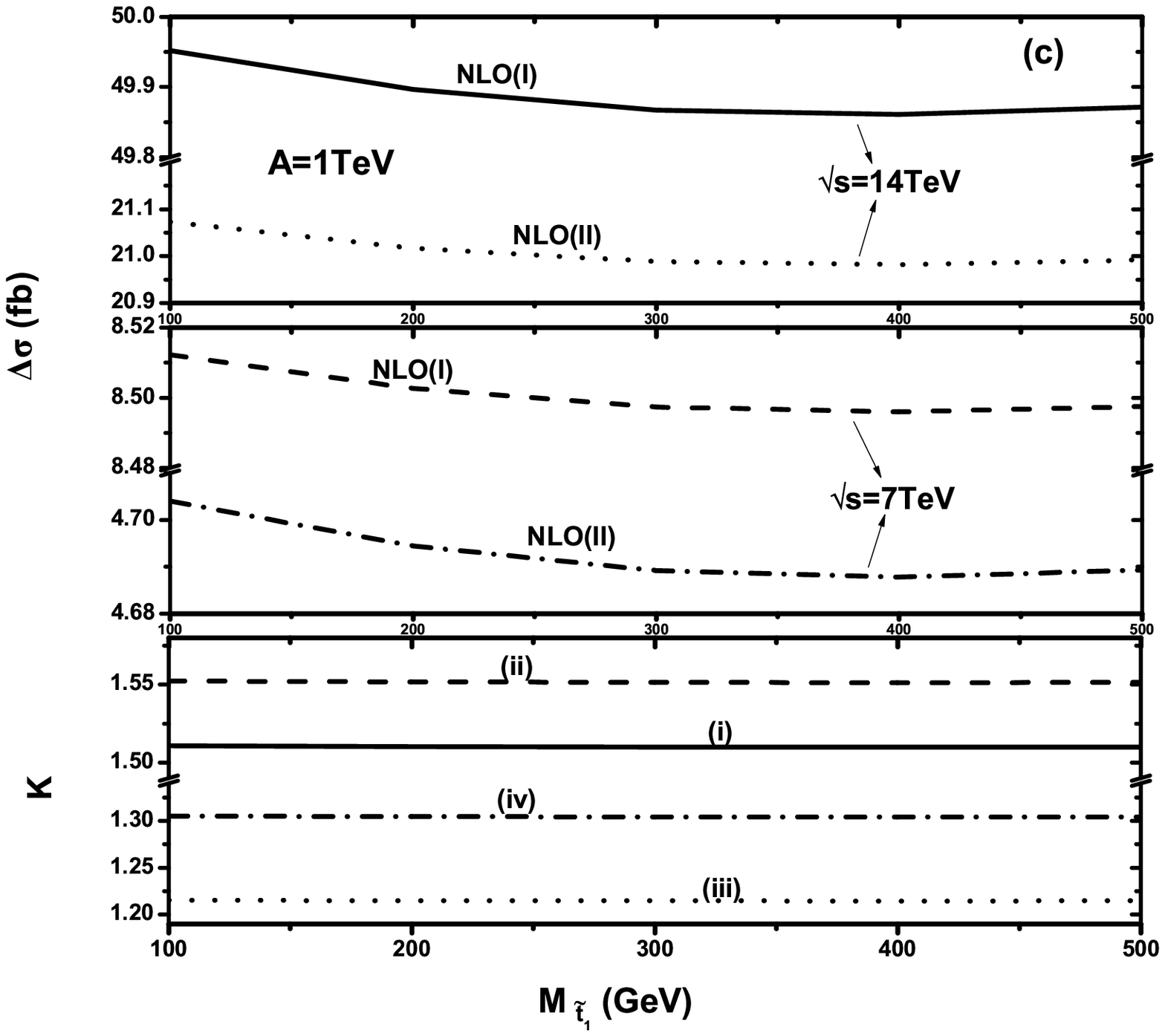}%
\caption{\label{fig8-3} The NLO QCD correction to the LO cross
sections $(\Delta \sigma_{NLO})$ and the corresponding K-factors
versus $m_{\tilde{t}_1}$ with $A=1~TeV$ at the $\sqrt{s}=7~TeV$ and
$\sqrt{s}=14~TeV$ LHC. (a) For the $pp\to gd \to t\slep^-_i+X$
process. (b) For the $pp\to g \bar{d} \to \bar{t}\slep^+_i+X$
process. (c) For the $pp\to gs \to t\slep^-_i+X$ process. The
descriptions for the K-factor curves labeled (i), (ii), (iii)
and (iv) are the same as in Figs.\ref{fig6}. }
\end{figure}
\begin{figure}[htbp]
\includegraphics[width=3.2in,height=3.2in]{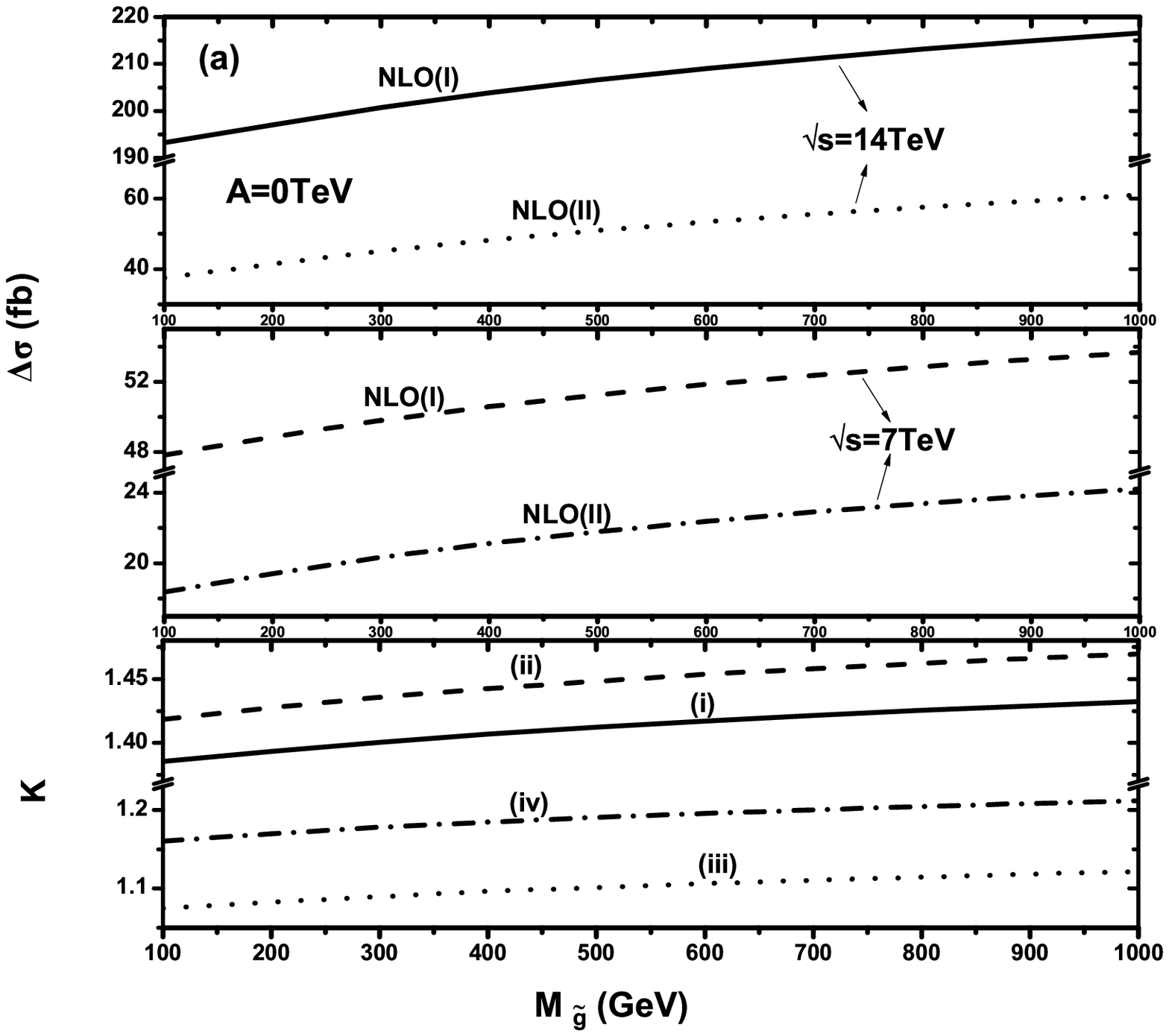}%
\hspace{0in}%
\includegraphics[width=3.2in,height=3.2in]{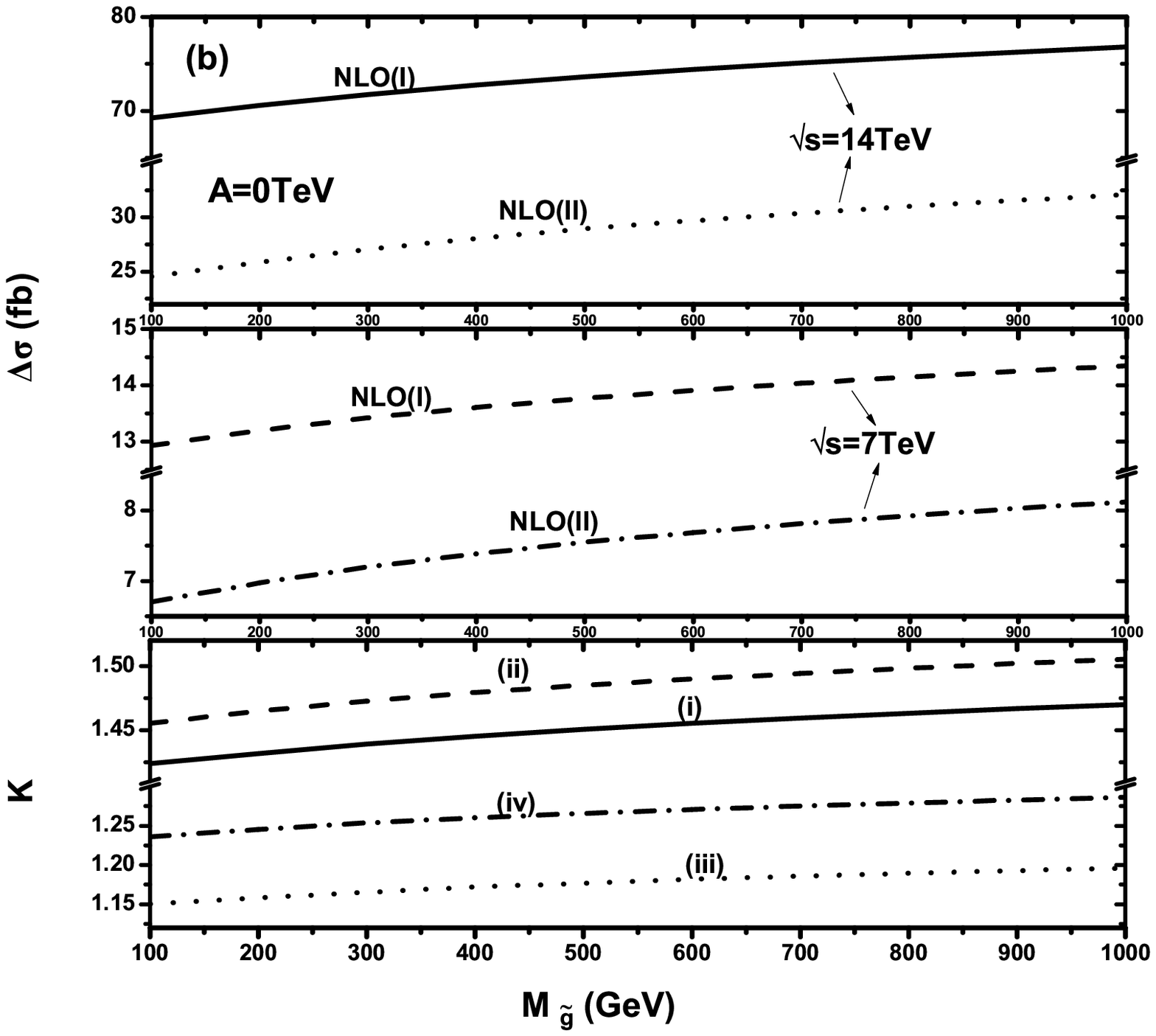}%
\hspace{0in}%
\includegraphics[width=3.2in,height=3.2in]{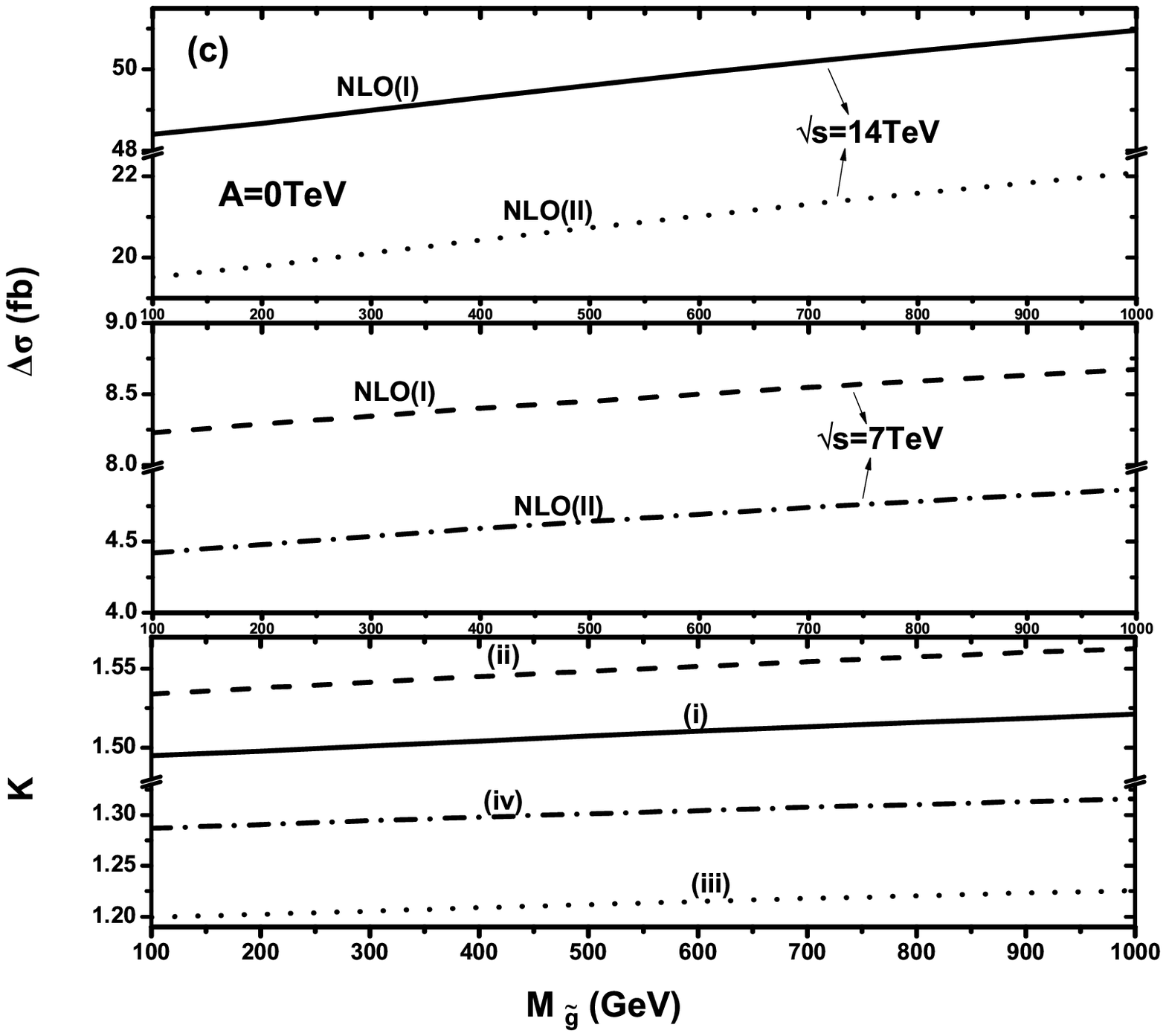}%
\caption{\label{fig9-1} The NLO QCD correction to the LO cross
sections $(\Delta \sigma_{NLO})$ and the corresponding K-factors
versus $m_{\tilde{g}}$ with $A=0~TeV$ at the $\sqrt{s}=7~TeV$ and $\sqrt{s}=14~TeV$
LHC. (a) For the $pp\to gd \to t\slep^-_i+X$ process. (b) For the
$pp\to g \bar{d} \to \bar{t}\slep^+_i+X$ process. (c) For the $pp\to
gs \to t\slep^-_i+X$ process. The descriptions for the K-factor
curves labeled (i), (ii), (iii) and (iv) are the same as in
Figs.\ref{fig6}. }
\end{figure}
\begin{figure}[htbp]
\includegraphics[width=3.2in,height=3.2in]{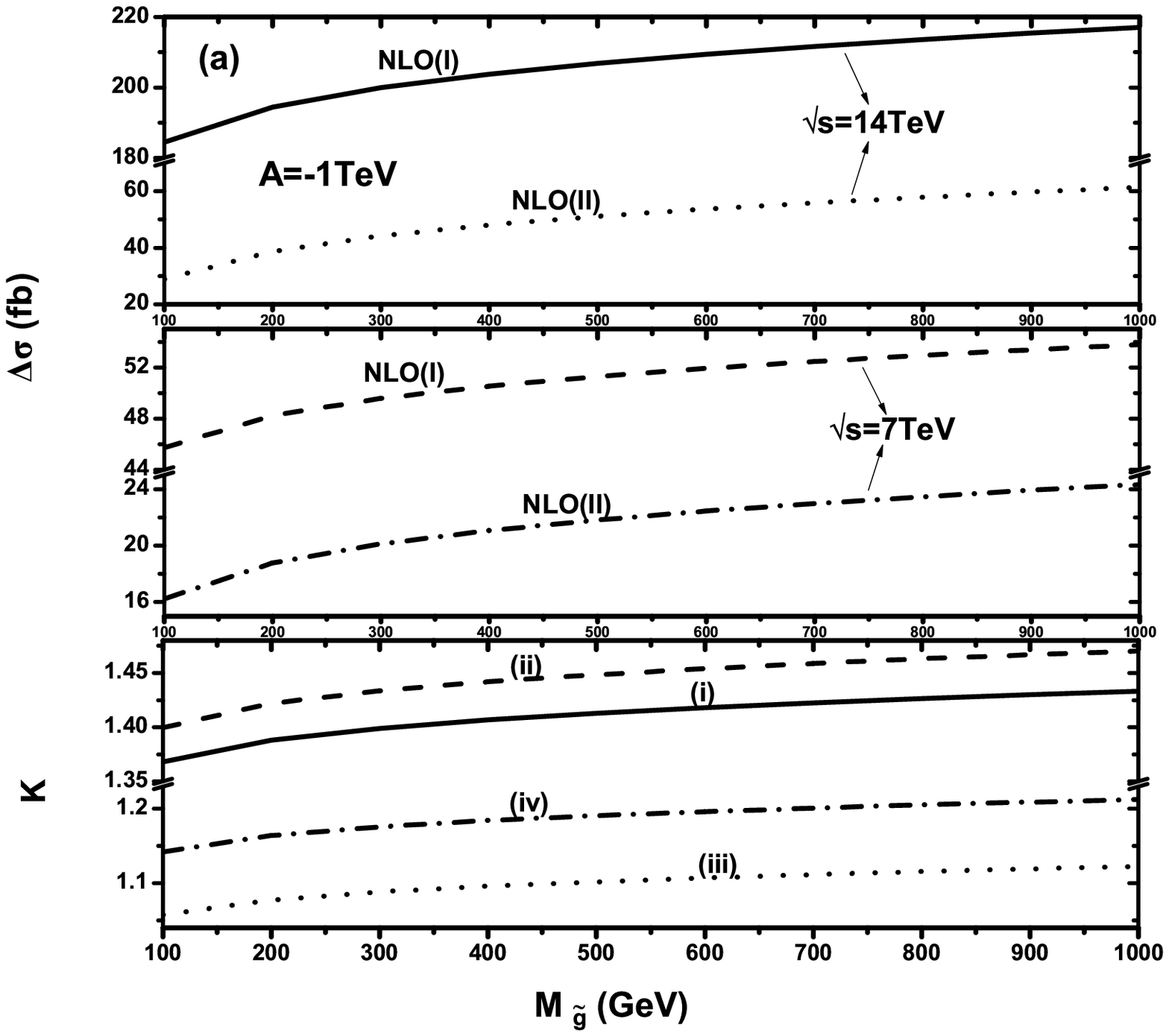}%
\hspace{0in}%
\includegraphics[width=3.2in,height=3.2in]{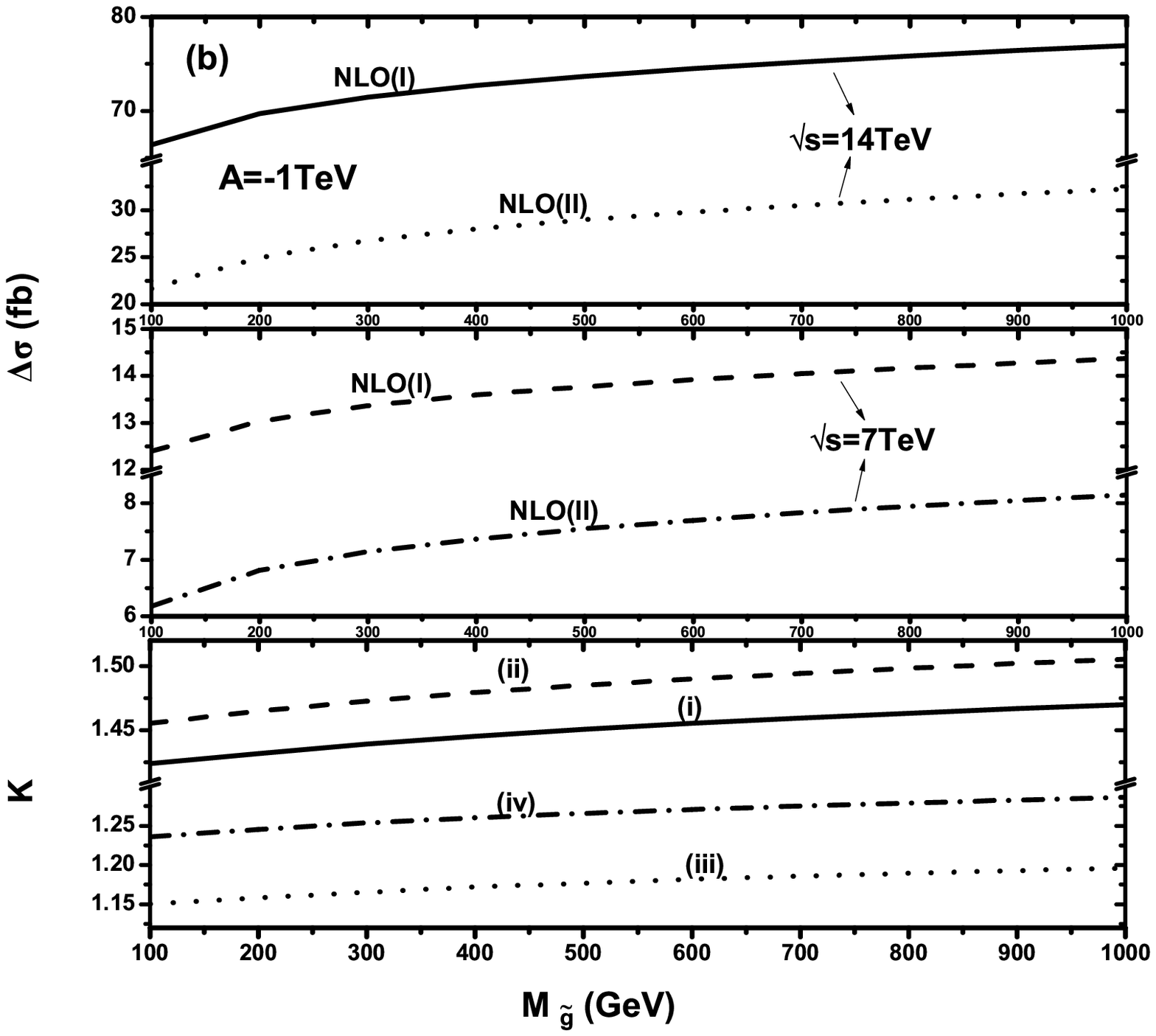}%
\hspace{0in}%
\includegraphics[width=3.2in,height=3.2in]{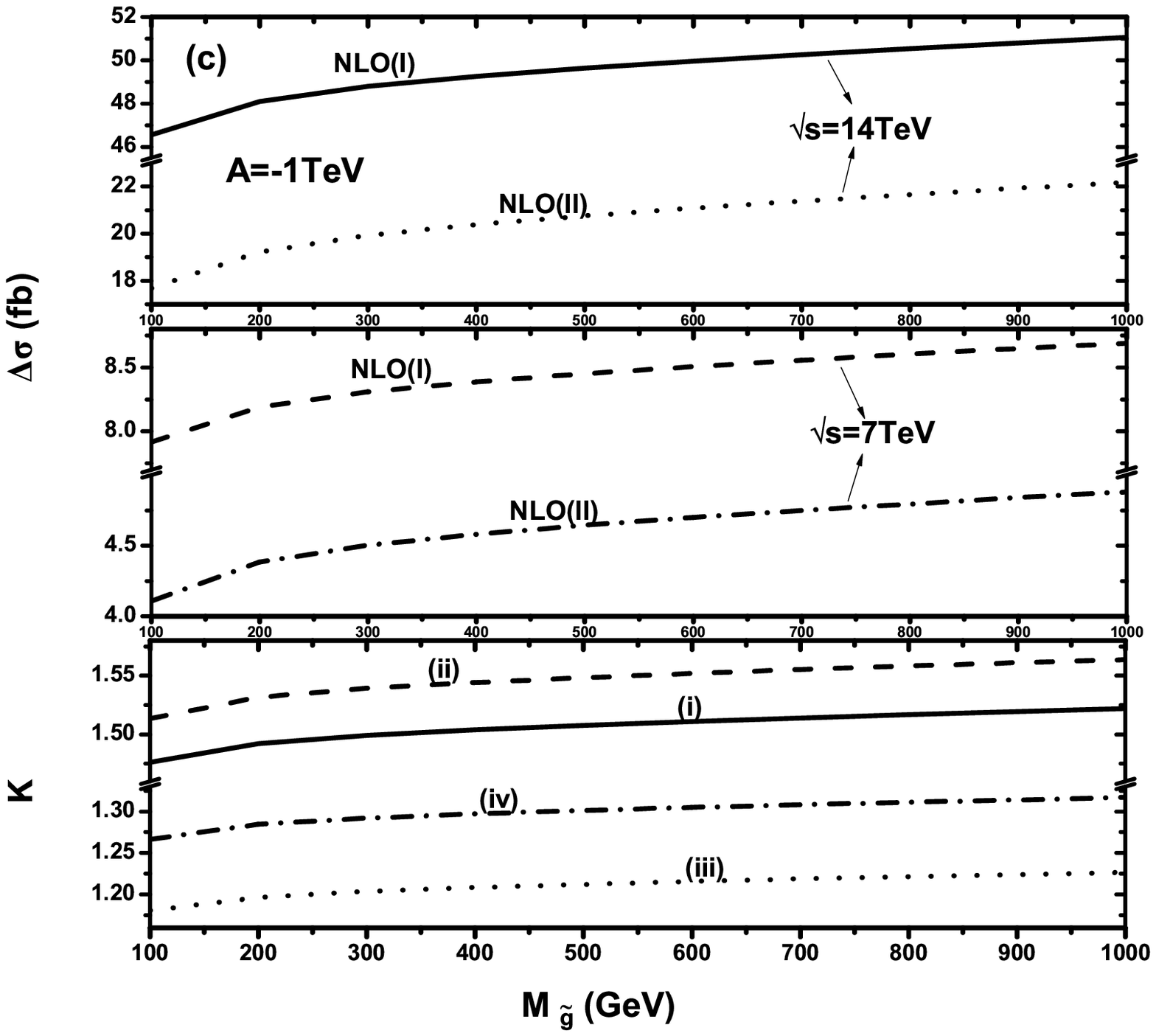}%
\caption{\label{fig9-2} The NLO QCD correction to the LO cross
sections $(\Delta \sigma_{NLO})$ and the corresponding K-factors
versus $m_{\tilde{g}}$ with $A=-1~TeV$ at the $\sqrt{s}=7~TeV$ and $\sqrt{s}=14~TeV$
LHC. (a) For the $pp\to gd \to t\slep^-_i+X$ process. (b) For the
$pp\to g \bar{d} \to \bar{t}\slep^+_i+X$ process. (c) For the $pp\to
gs \to t\slep^-_i+X$ process. The descriptions for the K-factor
curves labeled (i), (ii), (iii) and (iv) are the same as in
Figs.\ref{fig6}. }
\end{figure}
\begin{figure}[htbp]
\includegraphics[width=3.2in,height=3.2in]{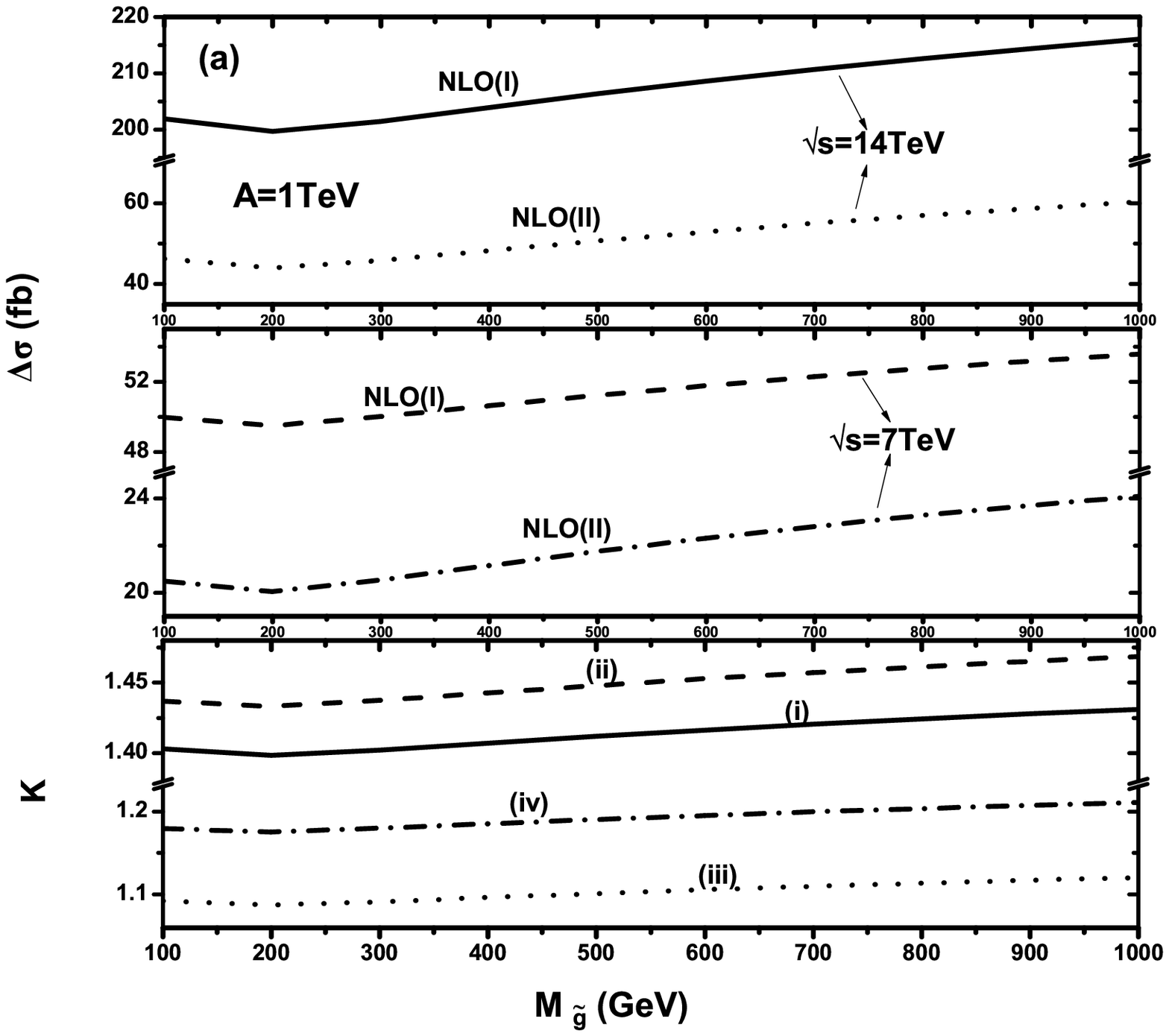}%
\hspace{0in}%
\includegraphics[width=3.2in,height=3.2in]{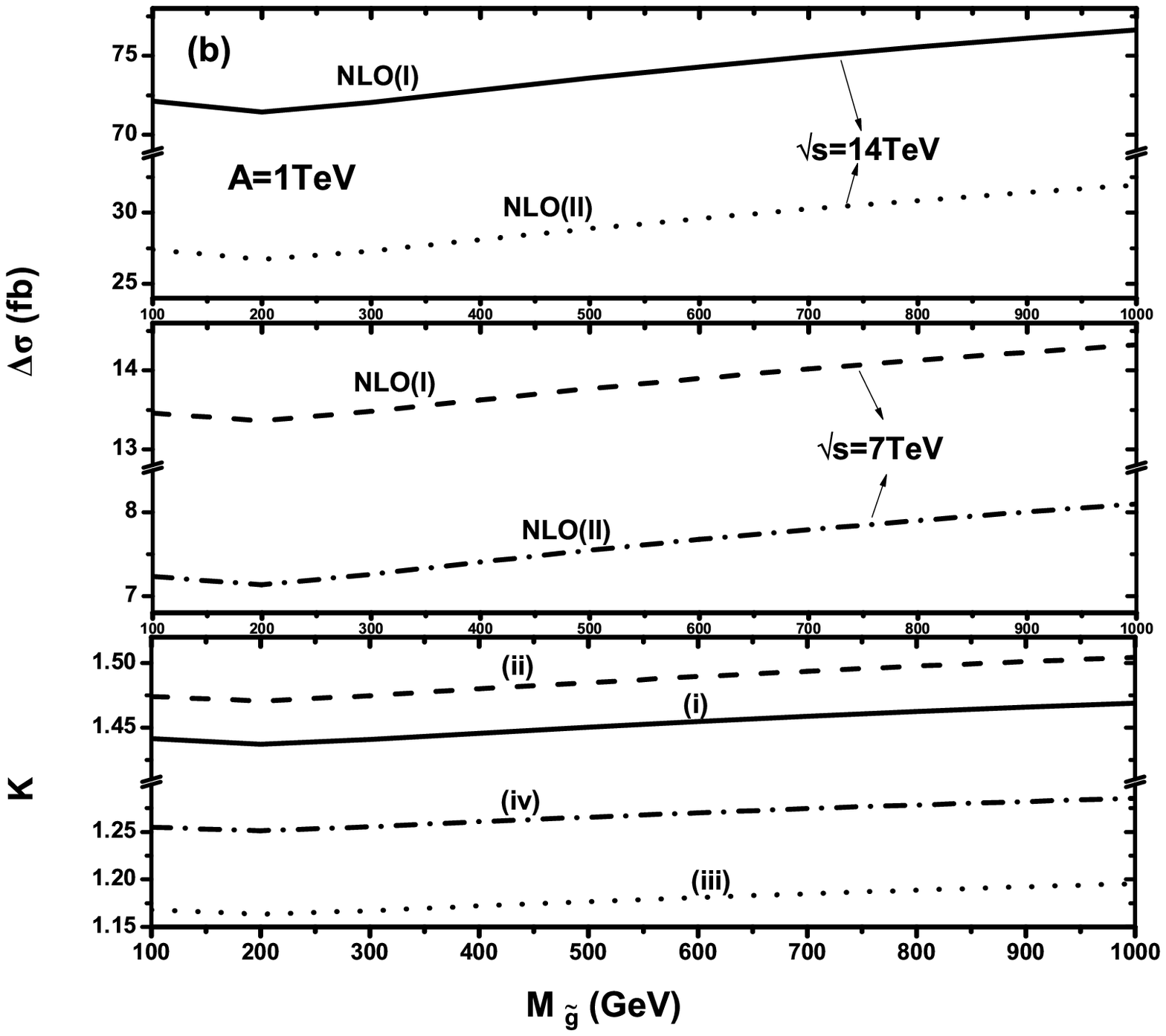}%
\hspace{0in}%
\includegraphics[width=3.2in,height=3.2in]{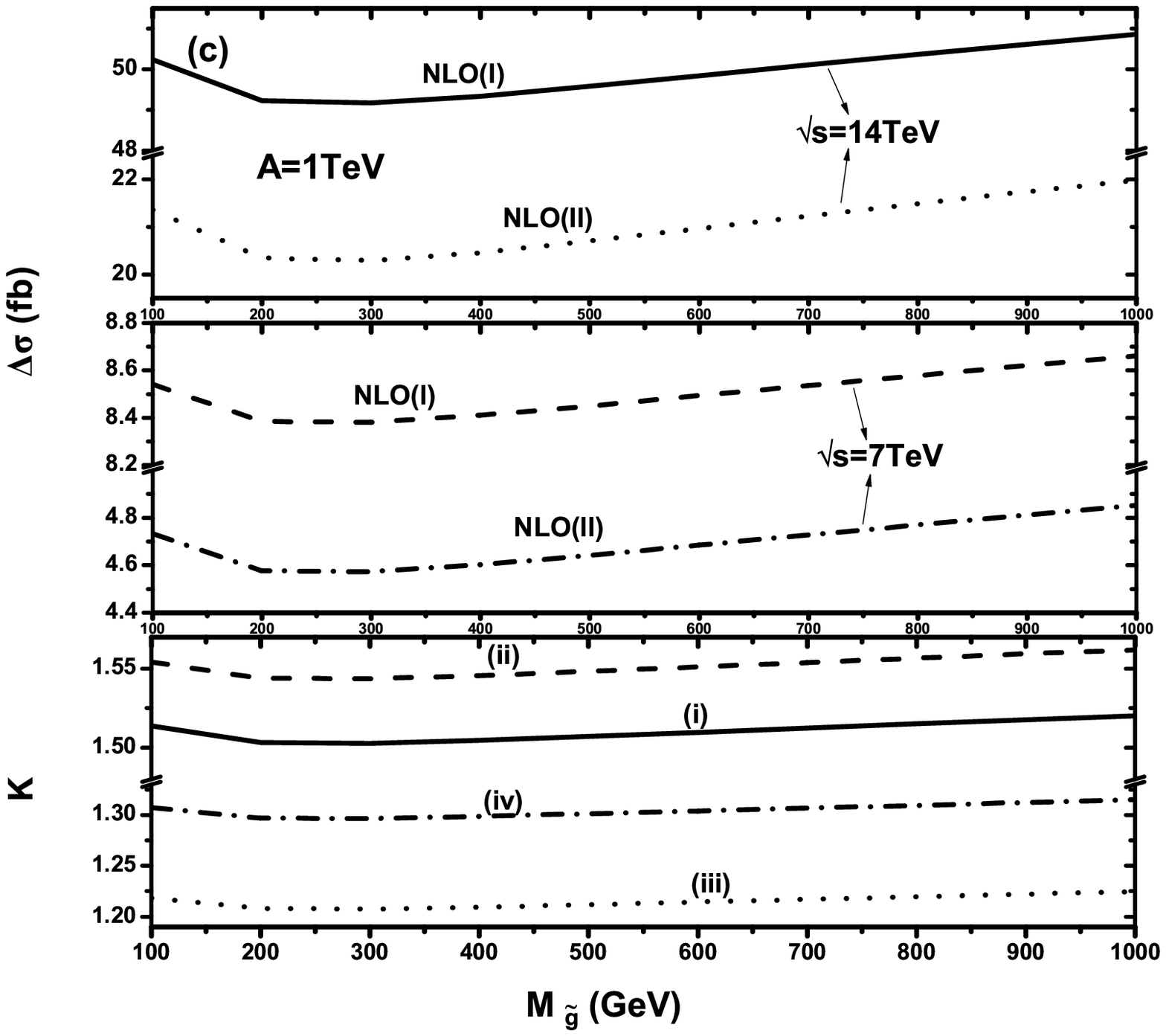}%
\caption{\label{fig9-3} The NLO QCD correction to the LO cross
sections $(\Delta \sigma_{NLO})$ and the corresponding K-factors
versus $m_{\tilde{g}}$ with $A=1~TeV$ at the $\sqrt{s}=7~TeV$ and
$\sqrt{s}=14~TeV$ LHC. (a) For the $pp\to gd \to t\slep^-_i+X$ process.
(b) For the $pp\to g \bar{d} \to \bar{t}\slep^+_i+X$ process.
(c) For the $pp\to gs \to t\slep^-_i+X$ process. The descriptions for
the K-factor curves labeled (i), (ii), (iii) and (iv) are the
same as in Figs.\ref{fig6}. }
\end{figure}

The SUSY partners $\tilde{t}_1$ and $\tilde{g}$ appear only at
one-loop level (shown in Fig.\ref{fig2}). We show the virtual
influences of $m_{\tilde{t}_1}$ and
$m_{\tilde{g}}$ on the NLO QCD corrections by taking three different
values of the SUSY-breaking parameter A (i.e., $A=0~TeV$ and $\pm
1~TeV$) in Figs.\ref{fig8-1}--\ref{fig8-3} and Figs.\ref{fig9-1}
--\ref{fig9-3}, separately. Figs.\ref{fig8-1}--\ref{fig8-3}(a,b,c)
(Figs.\ref{fig9-1}--\ref{fig9-3}(a,b,c)) show the NLO QCD
corrections ($\Delta \sigma_{NLO}$) and the
corresponding K-factors as functions of $m_{\tilde{t}_1}$
($m_{\tilde{g}}$) for the $pp\to gd \to t\slep^-_i+X$, $pp\to g\bar
d \to \bar t\slep^+_i+X$, and $pp\to gs \to t\slep^-_i+X$ processes
with $A=0~TeV$ and $\pm 1~TeV$ at the early and future
LHC, respectively. There we adopt both the inclusive and exclusive
schemes in the NLO calculations. In Figs.\ref{fig8-1}--\ref{fig9-3}(a,b)
we set $\lambda^{\prime}_{i31}=0.1$ and the other $\lambda^{\prime}=0$,
while in Fig.\ref{fig8-1}--\ref{fig9-3}(c) we take $\lambda^{\prime}_{i32}=0.1$
and the other $\lambda^{\prime}=0$. In those figures most of the curves for
the NLO QCD corrections and the corresponding K-factors always
increase with the increment of $m_{\tilde{t}_1}$ ($m_{\tilde{g}}$)
except those $\Delta\sigma_{NLO}$ curves in Fig.\ref{fig8-3}(c)
and the curves in the region of
$m_{\tilde{g}}<200~GeV$ in Figs.\ref{fig9-3}(a,b,c). It shows that
the NLO SUSY QCD corrections to these processes generally increase
in the large $m_{\tilde{t}_1}$ or $m_{\tilde{g}}$
region. That is because the SUSY QCD correction coming from
the renormalization counterterm of the amplitude $\delta {\cal M}$
involves the logarithm terms of $\ln \left(
\frac{m_{\tilde{t}_1}^2}{\mu_r^2} \right)$ and $\ln \left(
\frac{m_{\tilde{g}}^2}{\mu_r^2} \right)$ at the SUSY QCD one-loop
level. The logarithm term contributions can be absorbed by a
redefinition of the coupling $\lambda^{\prime}$ \cite{Dreiner}. In
this work we do not need to adopt this decoupling scheme, since in
our consideration we take the squark and gluino mass
quantitatively at the same order as the scale $\mu$.

\par
\begin{figure}[htbp]
\includegraphics[width=3.2in,height=3.2in]{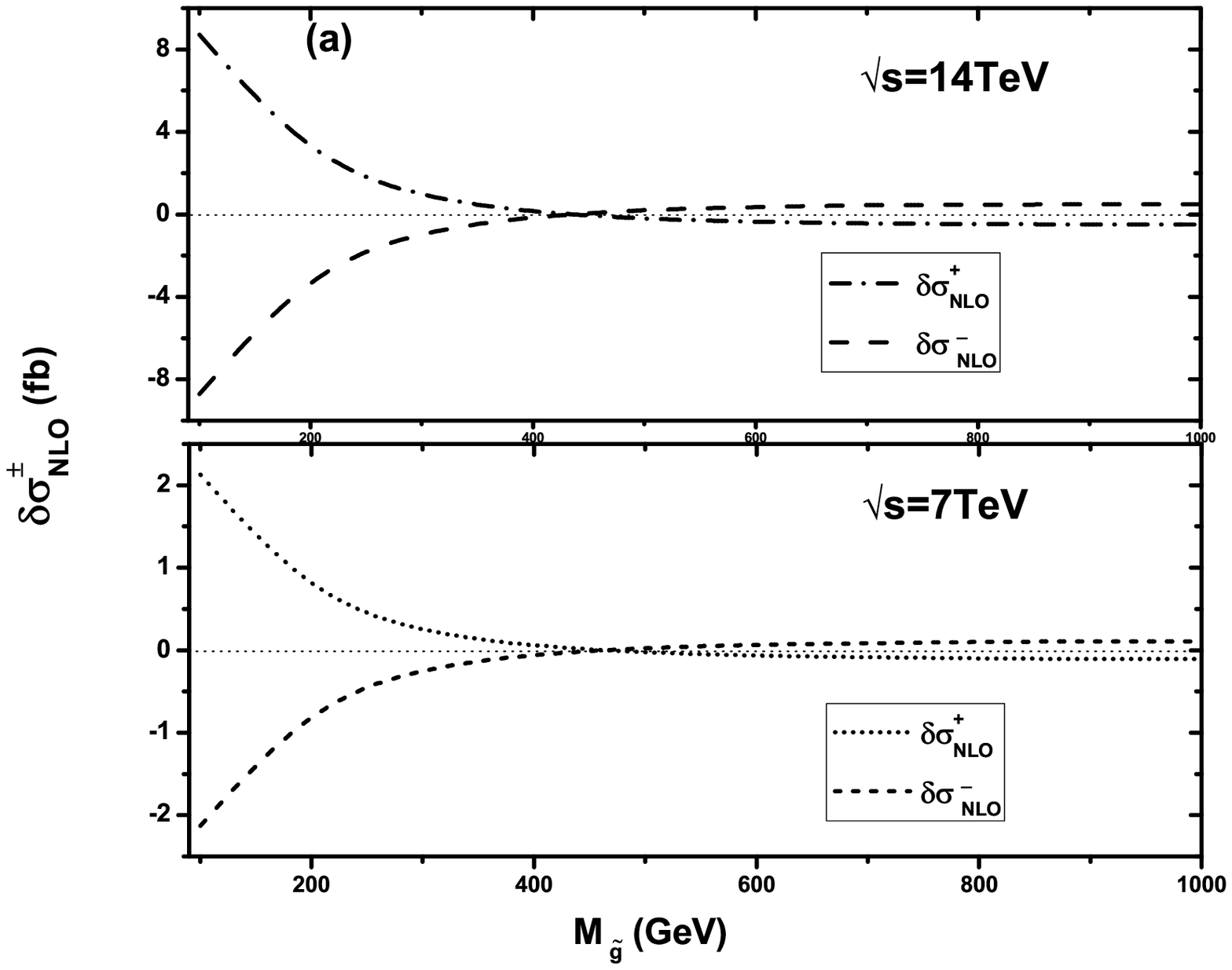}%
\hspace{0in}%
\includegraphics[width=3.2in,height=3.2in]{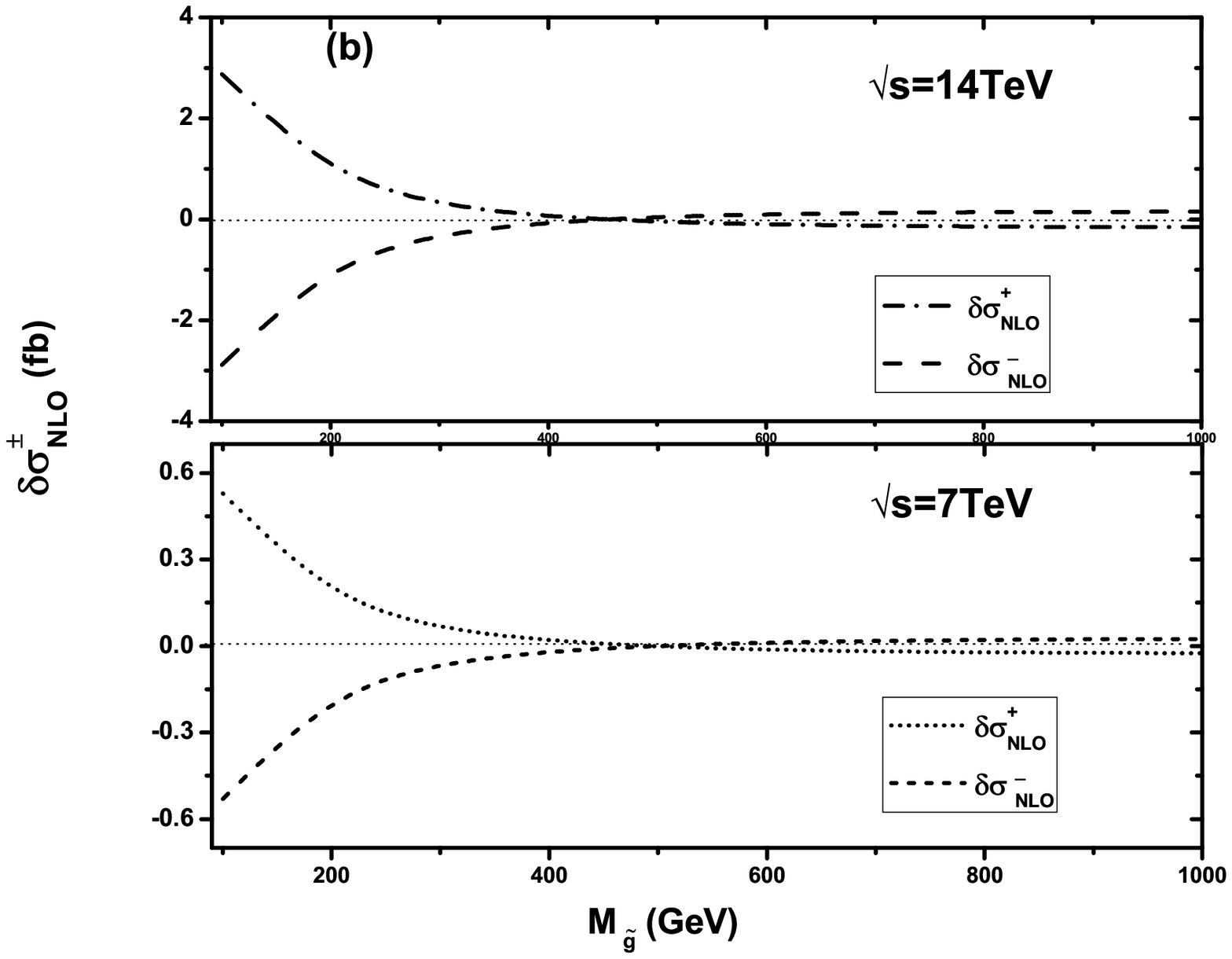}%
\hspace{0in}%
\includegraphics[width=3.2in,height=3.2in]{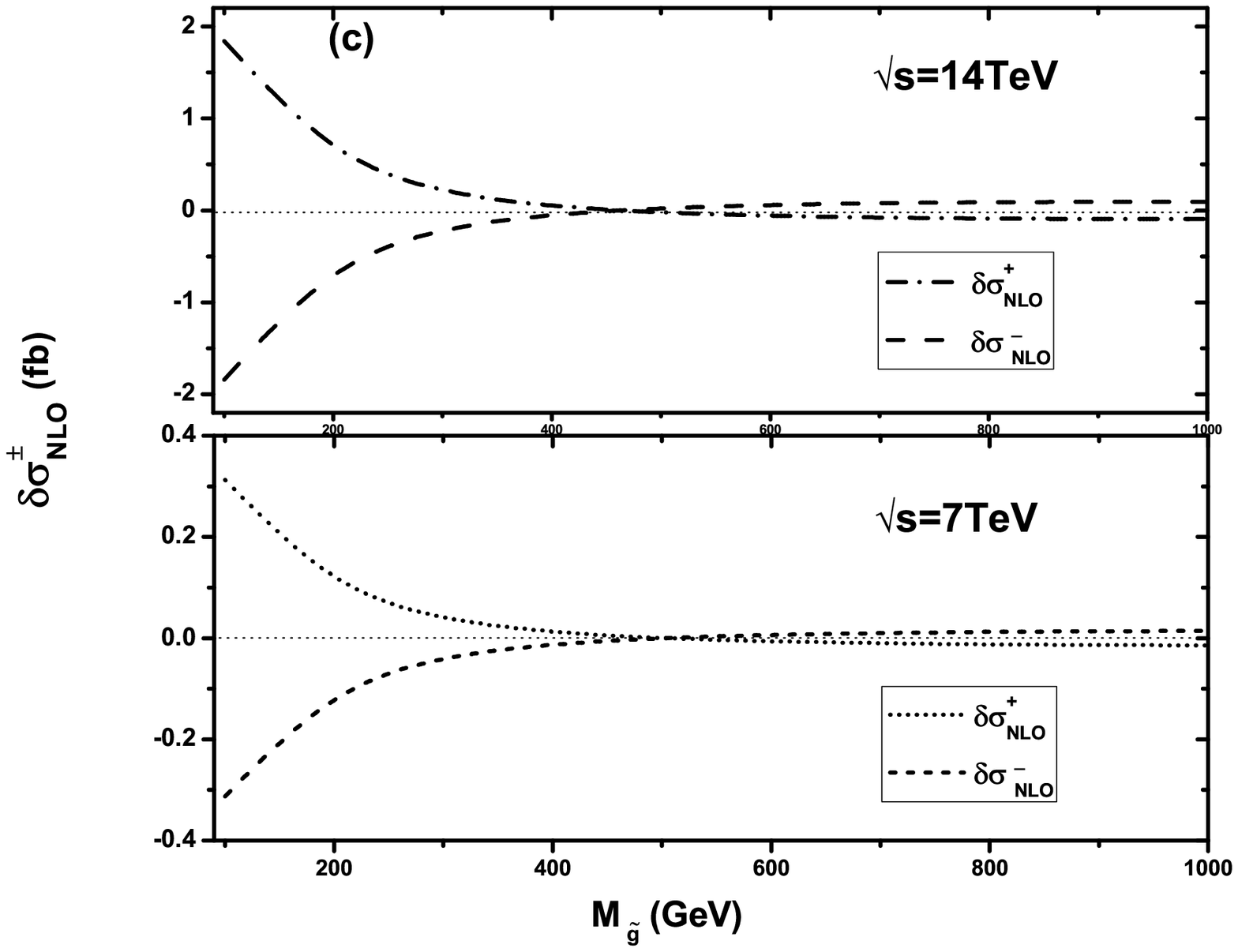}%
\caption{\label{fig9-4} The difference between the NLO QCD corrections
with $A=\pm 1~TeV$ and $A=0~TeV$ versus $m_{\tilde{g}}$
at the $\sqrt{s}=7~TeV$ and $\sqrt{s}=14~TeV$ LHC.
(a) For the $pp\to gd \to t\slep^-_i+X$ process.
(b) For the $pp\to g \bar{d} \to \bar{t}\slep^+_i+X$ process.
(c) For the $pp\to gs \to t\slep^-_i+X$ process.  }
\end{figure}

Figs.\ref{fig8-1} and Figs.\ref{fig9-1} show that if we take $A=0$,
the NLO QCD corrections are still related to $m_{\tilde{t}_1}$
and $m_{\tilde{g}}$. This can be understood from the fact that
the NLO QCD contributions from the counterterms
shown in Eqs.(\ref{eq2-4})--(\ref{eq2-5}) and the loop diagrams
such as Figs.\ref{fig2}(5,6), are relevant to the masses of
$\tilde{t}$ and $\tilde{g}$. We can also see the $\Delta \sigma_{NLO}$
curves versus $m_{\tilde{t}_1}$ in Fig.\ref{fig8-3}(c) for the
$pp\to gs \to t\slep^-_i+X$ process, and the $\Delta \sigma_{NLO}$
curves in the region around $m_{\tilde{g}}\sim 200~GeV$ in
Figs.\ref{fig9-3}(a,b,c) are obviously distorted by the
contributions from the diagrams involving the
$\tilde{t}_j$--$\tilde{d}_k$--$\tilde{l}_i$ coupling with $A=1~TeV$.
In order to understand the contributions from the non-zero $\tilde{t}_j$--$\tilde{d}_k$--
$\tilde{l}_i$ coupling more clearly, we provide the plots of
$\delta \sigma_{NLO}^{\pm} (\equiv \sigma_{NLO}(A=\pm 1~TeV)-\sigma_{NLO}(A=0~TeV))$
versus the $m_{\tilde{g}}$ in Figs.\ref{fig9-4}(a,b,c) for the processes
$pp\to gd \to t\slep^-_i+X$, $pp\to g \bar{d} \to \bar{t}\slep^+_i+X$ and
$pp\to gs \to t\slep^-_i+X$, separately. There we can see that
in the region of $m_{\tilde{g}}<400~GeV$, $\delta \sigma_{NLO}^{+}$
is always positive, while $\delta \sigma_{NLO}^{-}$ remains negative.

\par
\begin{figure}[htbp]
\includegraphics[width=3.2in,height=3.2in]{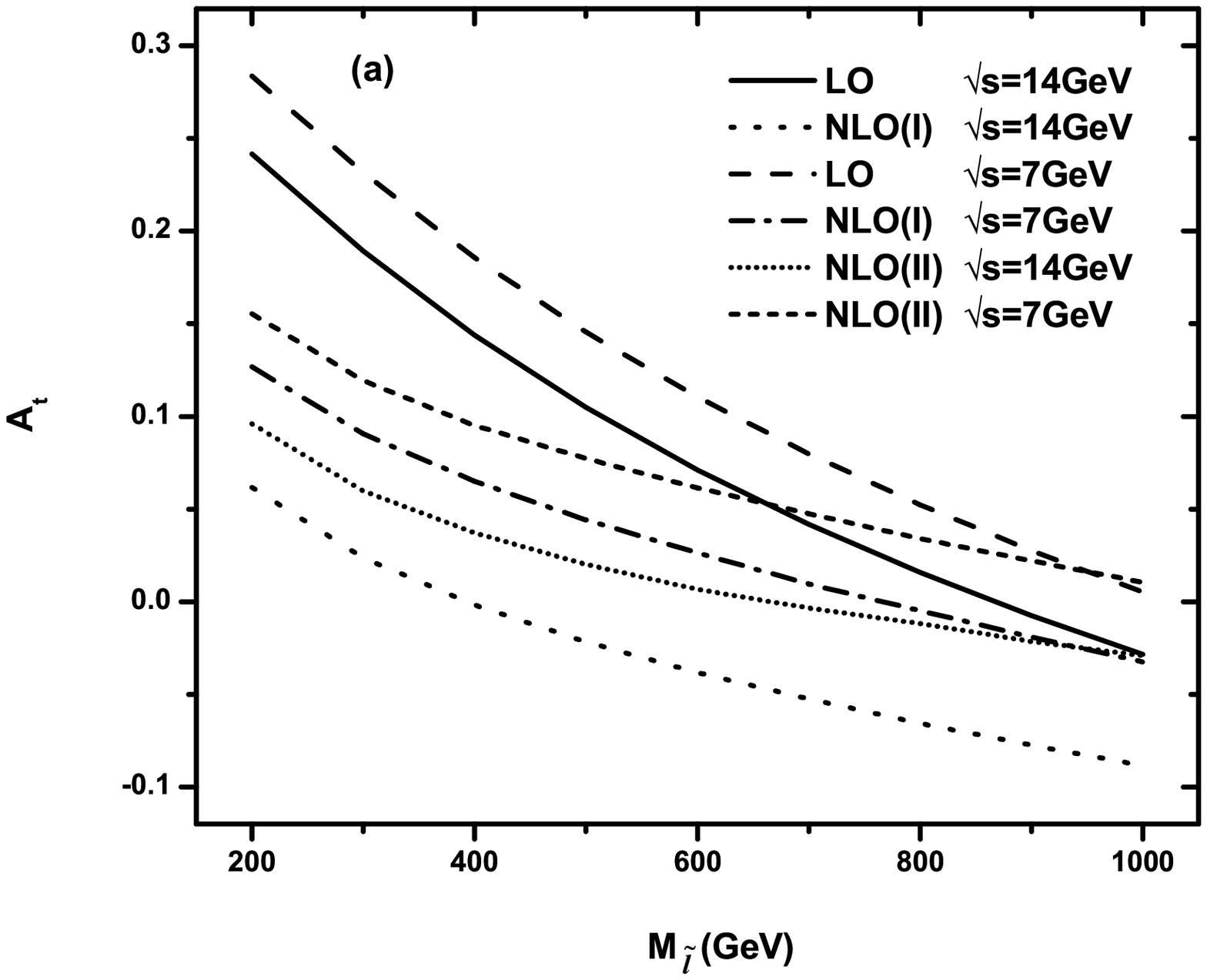}%
\hspace{0in}%
\includegraphics[width=3.2in,height=3.2in]{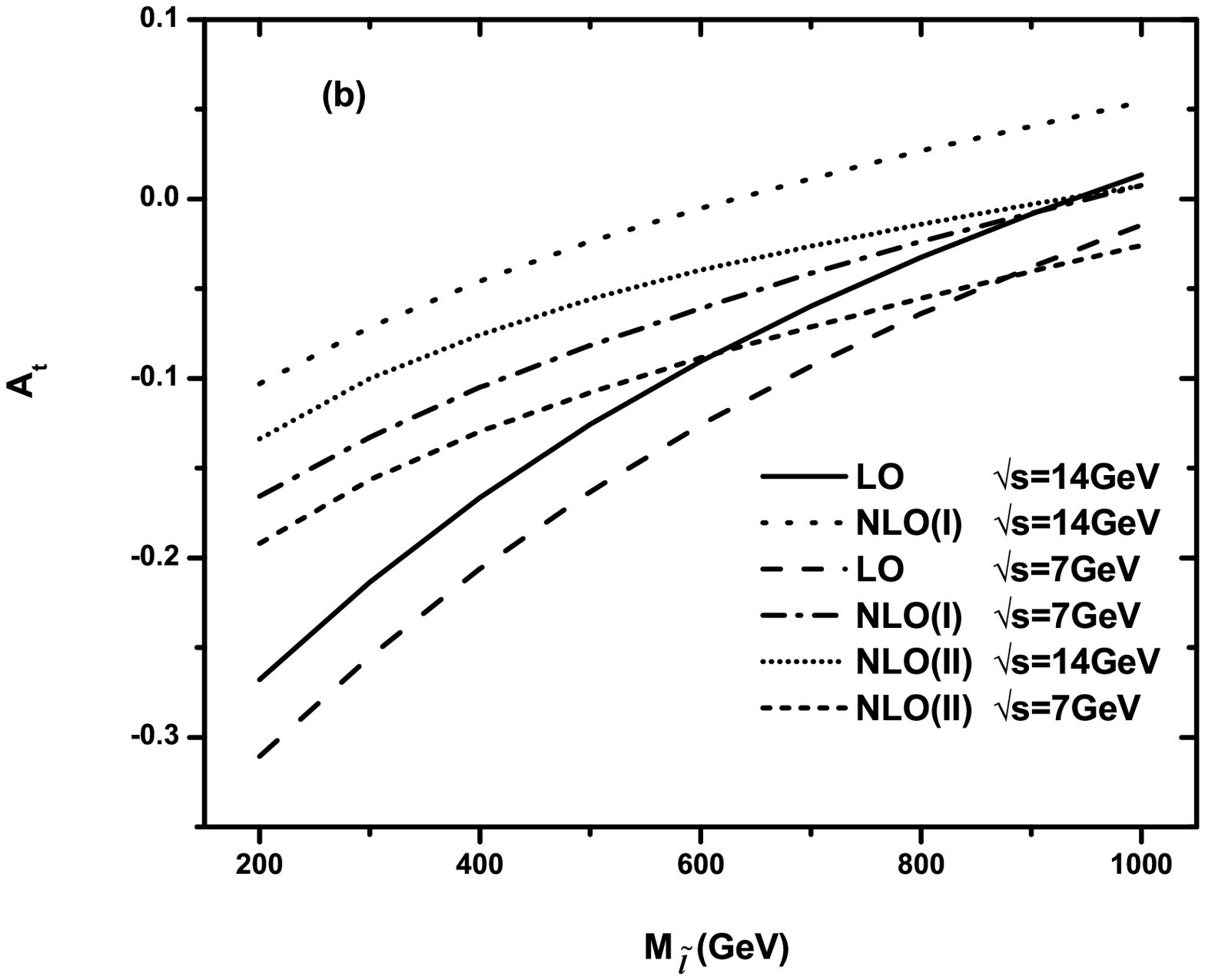}%
\hspace{0in}%
\includegraphics[width=3.2in,height=3.2in]{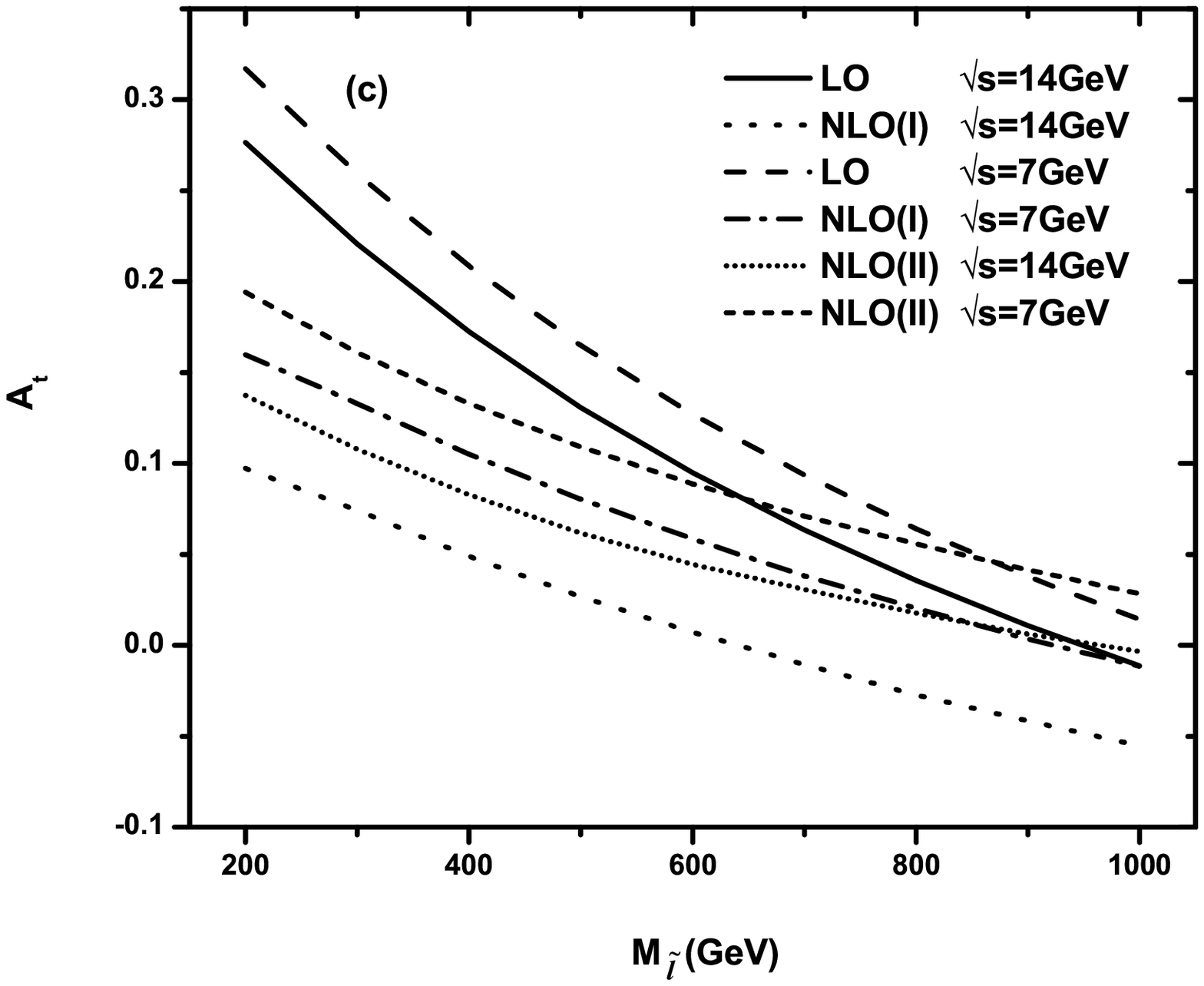}%
\hspace{0in}%
\includegraphics[width=3.2in,height=3.2in]{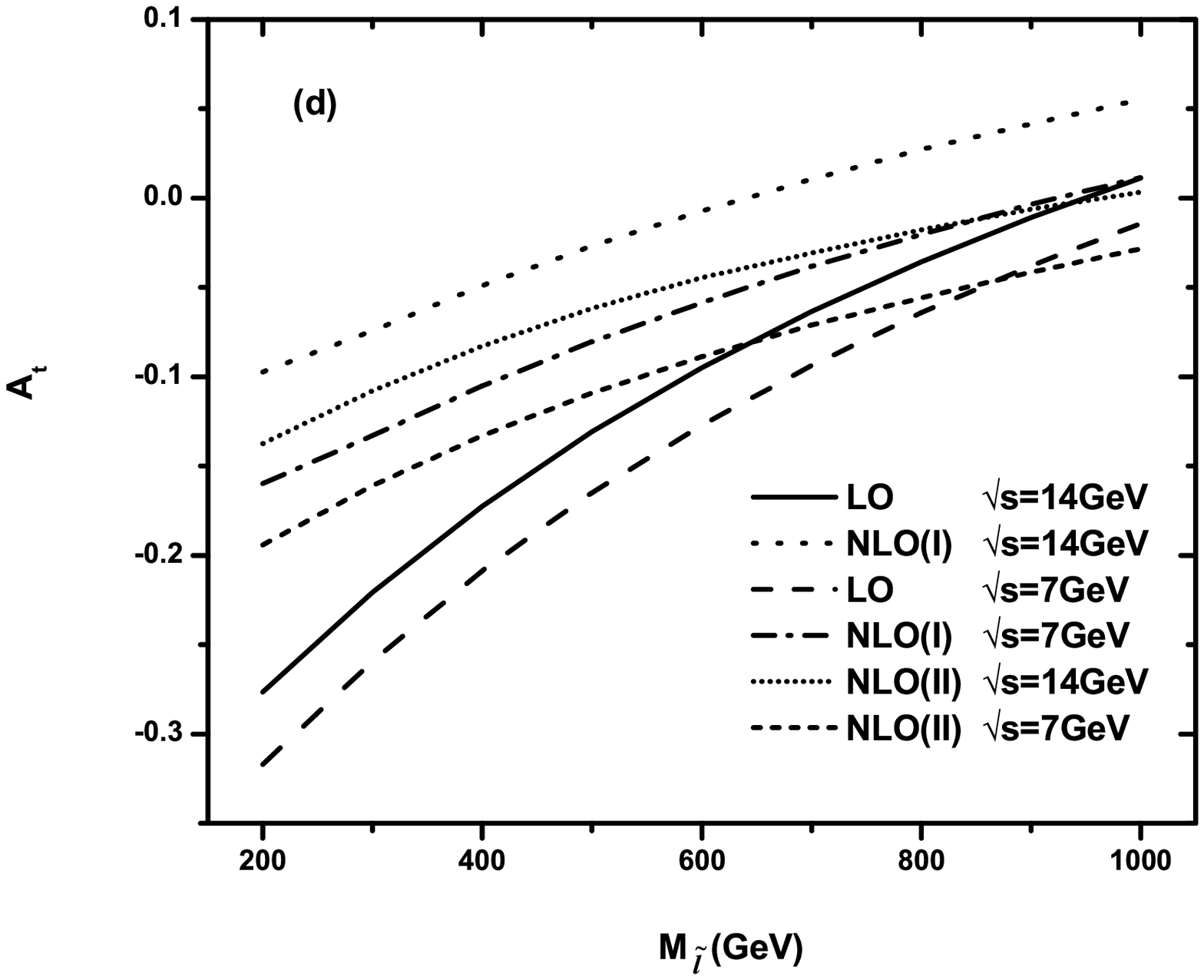}%
\caption{\label{fig10} The LO and NLO QCD corrected polarization
asymmetries ($A_t$) of the (anti)top quark at the early and future
LHC as the functions of $m_{\slep_i}$ with $A=0~TeV$. (a) For the process $pp\to
gd\to t\slep^-_i+X$. (b) For the process $pp\to g\bar d\to \bar
t\slep^+_i+X$. (c) For the process $pp\to gs\to t\slep^-_i+X$. (d)
For the process $pp\to g\bar{s}\to \bar{t}\slep^+_i+X$.}
\end{figure}

In the following calculations we fix the SUSY-breaking parameter $A=0~TeV$.
We show the polarization asymmetries ($A_t$) of the (anti)top quark
as a function of $m_{\slep_i}$ at the LO and NLO QCD for the
processes $pp\to gd_k(g\bar d_k)\to t\slep^-_i(\bar t\slep^+_i)+X$
in Figs.\ref{fig10}(a,b,c,d). $A_t$ is defined as
$A_t=\frac{N_+-N_-}{N_++N_-}$, where $N_+$ and $N_-$ refer to the
numbers of positive and negative helicity (anti)top quarks
respectively. In Ref.\cite{topspin}, the LO $A_t$ at the LHC has
been plotted and it shows that the LO $A_t$ changes signs for a
slepton mass of around 870--900$~GeV$. As we can see in
Figs.\ref{fig10}(a,b,c,d), the polarization degree of the (anti)top
quark has been reduced obviously by the NLO QCD correction. That is
because the NLO QCD radiation corrections destroy the chiral
structure of the interaction of top quark and reduce the LO
polarization asymmetry $A_t$. Although the cross sections for the
processes $pp\to g\bar s \to \bar t\slep^+_i +X$ and $pp \to gs \to
t\slep^-_i +X$ are equal, those two mutually conjugate processes
have opposite $A_t$ values as shown in Figs.\ref{fig10}(c) and
(d).

\par
\begin{figure}[htbp]
\includegraphics[width=3.2in,height=3.2in]{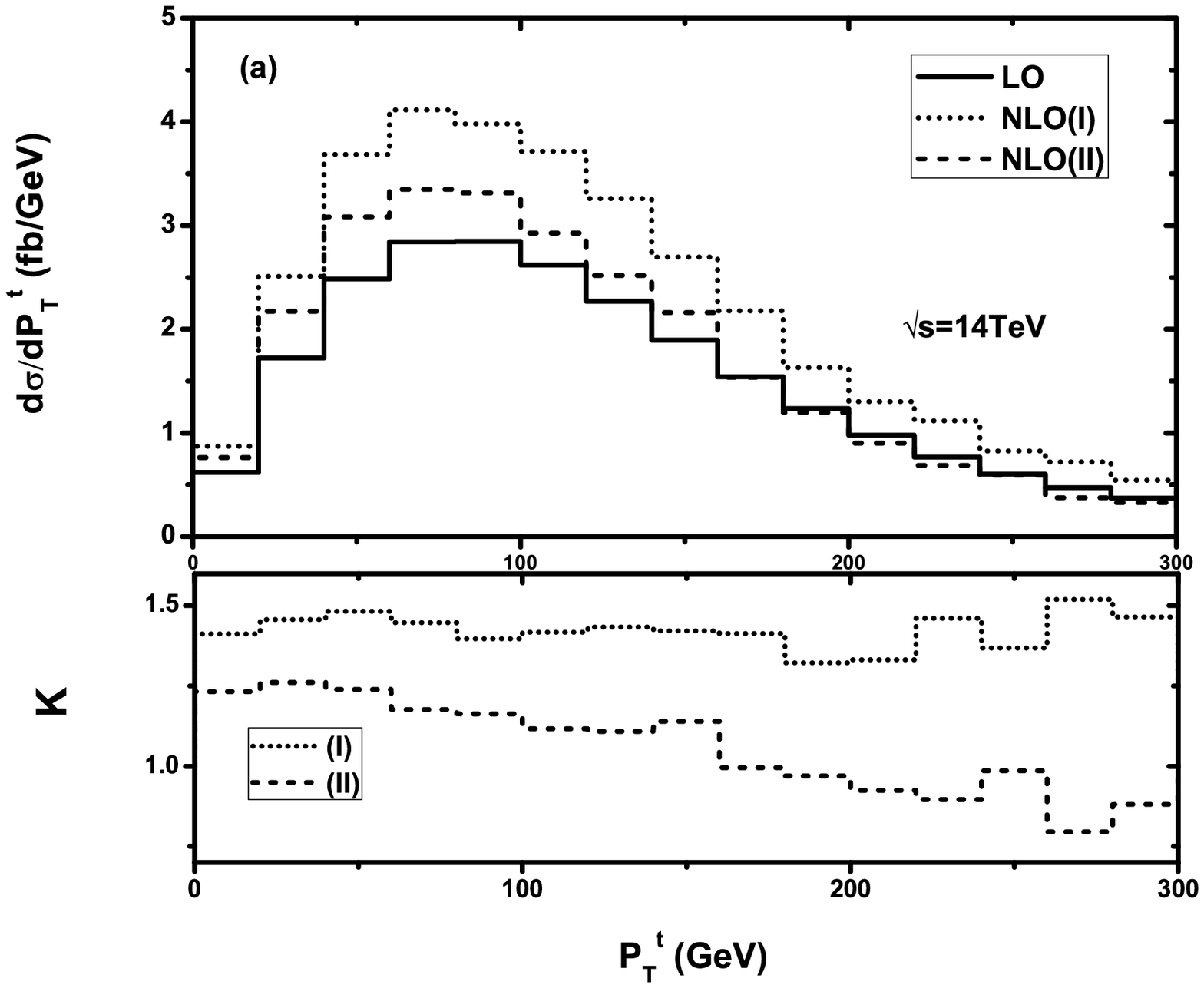}%
\hspace{0in}%
\includegraphics[width=3.2in,height=3.2in]{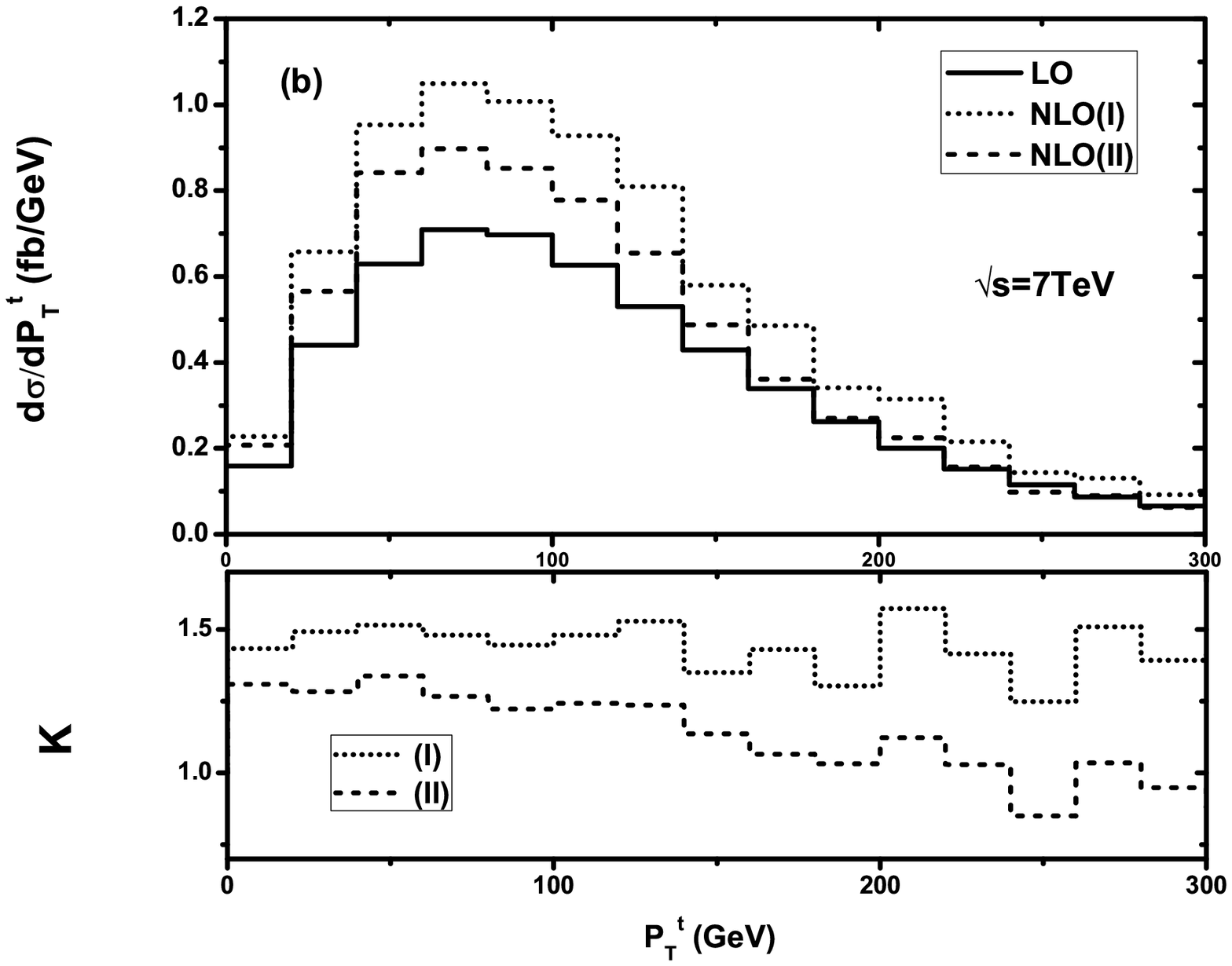}%
\hspace{0in}%
\includegraphics[width=3.2in,height=3.2in]{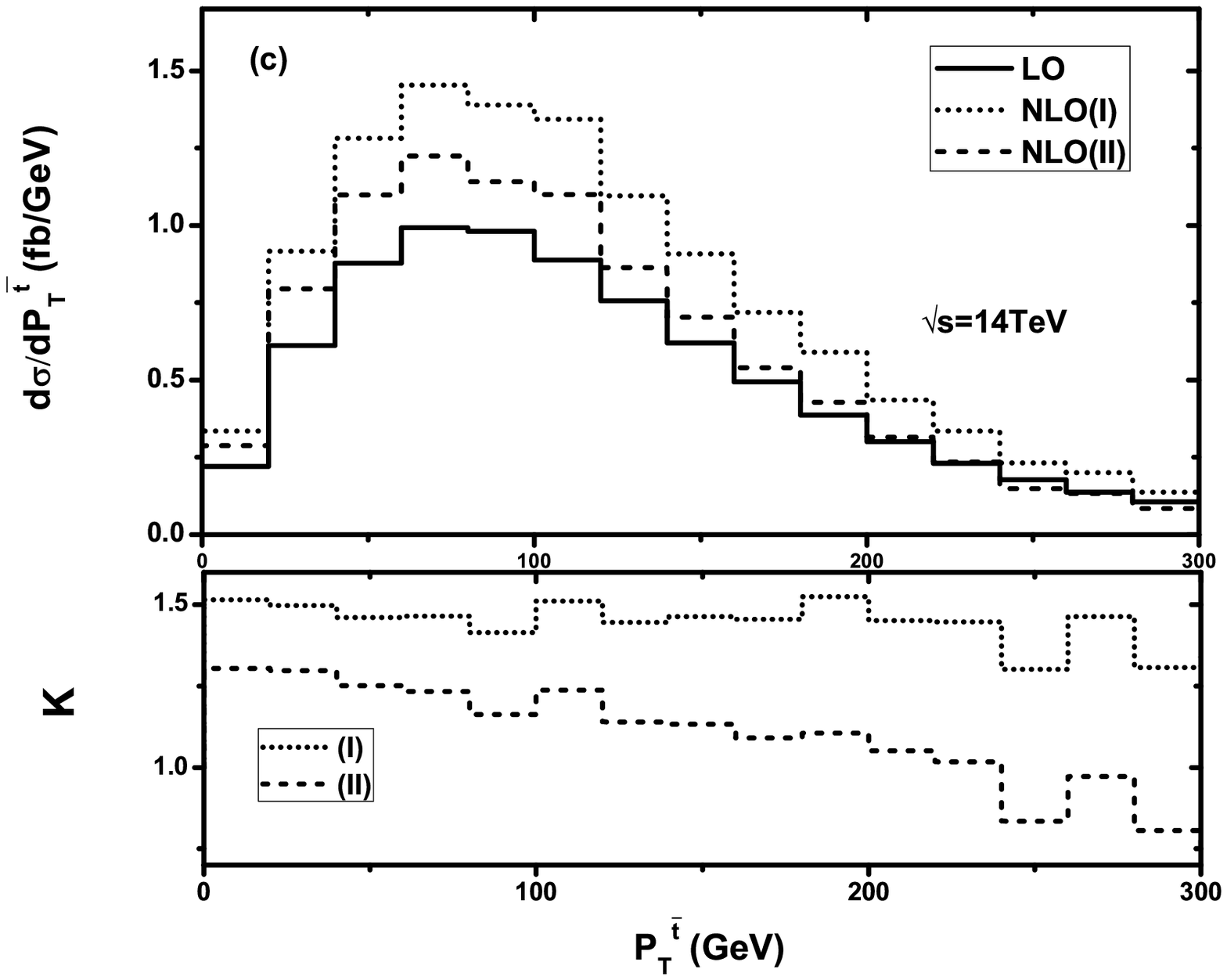}%
\hspace{0in}%
\includegraphics[width=3.2in,height=3.2in]{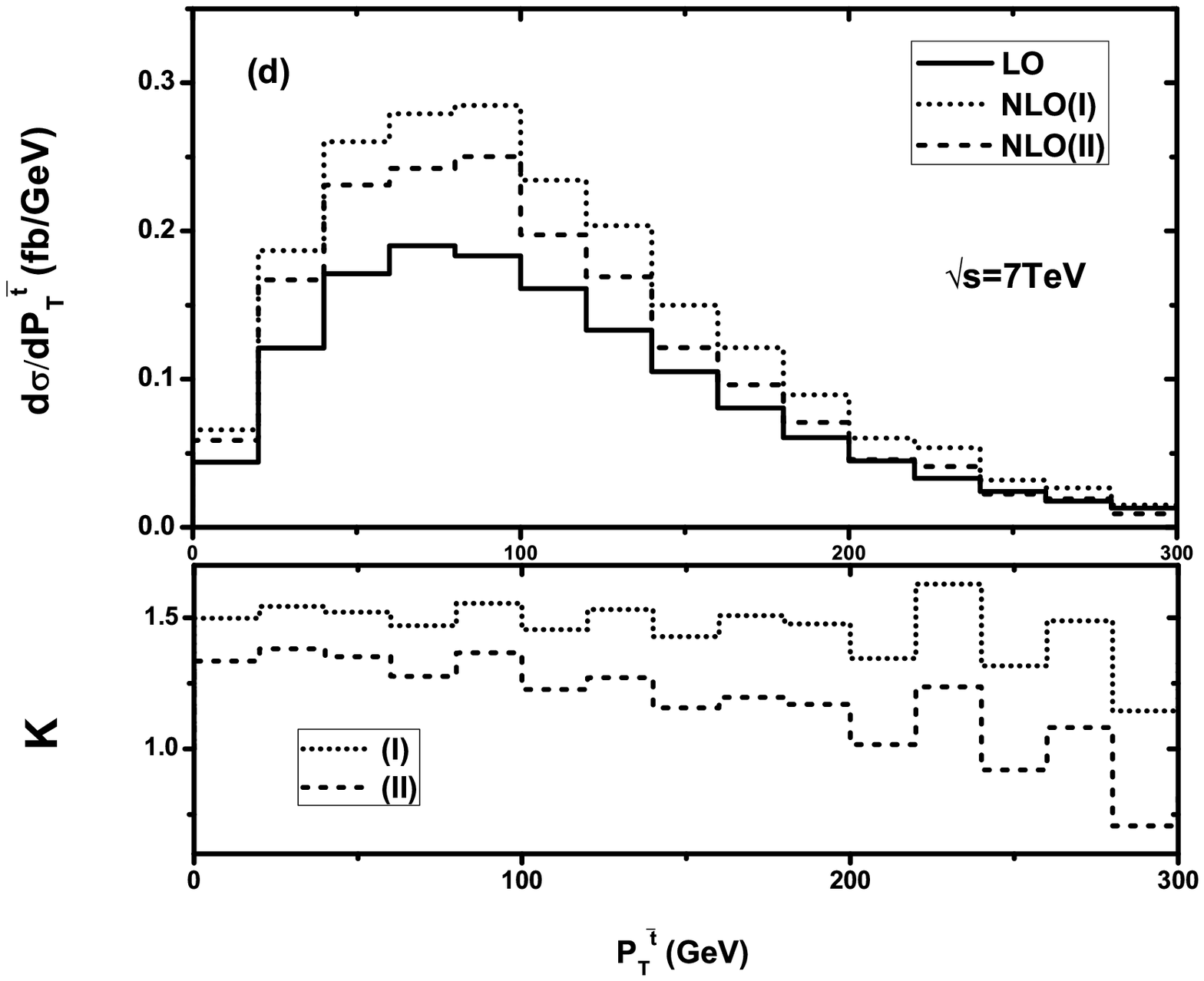}%
\hspace{0in}%
\caption{\label{fig11} The LO, NLO QCD corrected differential cross
sections of the transverse momentum of final (anti)top quark and the
corresponding K-factors
$K(p_T^{t(\bar{t})})=d\sigma_{NLO}(p_T^{t(\bar{t})})/
d\sigma_{LO}(p_T^{t(\bar{t})})$ with $A=0~TeV$. (a) For the process $pp\to gd \to
t\slep^-_i+X$ at the $\sqrt{s} = 14~TeV$ LHC. (b) For the process
$pp\to gd \to t\slep^-_i+X$ at the $\sqrt{s} = 7~TeV$ LHC. (c) For
the process $pp \to g\bar{d} \to \bar{t}\slep^+_i+X$ at the
$\sqrt{s} = 14~TeV$ LHC. (d) For the process $pp \to g\bar{d}\to
\bar{t}\slep^+_i+X$ at the $\sqrt{s} = 7~TeV$ LHC. The curves of
K-factor labeled (I) and (II) correspond to adopting the
inclusive and exclusive scheme, respectively. }
\end{figure}

In further discussion, we consider the case where
$\slep_i^\pm=\tilde{\mu}^\pm$ (i.e., $i=2$). After the production of
the scalar muon, there follows a subsequential decay of
$\tilde{\mu}^{\pm} \to \mu^{\pm}\tilde{\chi}^0_1$ with branch ratio
$92.5\%$ \cite{SPA}, then the final state involves muon, the
lightest neutralino and (anti)top-quark jet
($\mu\tilde{\chi}^0_1t(\bar{t})$). As a demonstration we assume
there exist $m_{\tilde{\mu}} = 189.9~GeV$,
$\lambda^{\prime}_{231}=0.1$ and the other $\lambda^{\prime}=0$; then we
present the LO, NLO QCD corrected transverse momentum distributions
of the final (anti)top quark and the corresponding K-factors
($K(p_T^{t(\bar{t})})=d\sigma_{NLO}(p_T^{t(\bar{t})})/
d\sigma_{LO}(p_T^{t(\bar{t})})$) for the processes of $pp \to gd \to
t\tilde{\mu}^- \to t\mu^{-}\tilde{\chi}^0_1+X$ and $pp \to g\bar{d}
\to \bar{t}\tilde{\mu}^{+} \to \bar{t}\mu^{+}\tilde{\chi}^0_1+X$ at
the LHC in Figs.\ref{fig11}(a,b,c,d). Figs.\ref{fig11}(a) and
(c) are for the $p_T^{t}(p_T^{\bar{t}})$ distributions at the
$\sqrt{s}=14~TeV$ LHC for the processes $pp \to gd \to
t\mu^-\tilde{\chi}^0_1+X$ and $pp \to g\bar{d} \to
\bar{t}\mu^+\tilde{\chi}^0_1+X$, respectively, and
Figs.\ref{fig11}(b) and (d) are at the $\sqrt{s}=7~TeV$ LHC for the
processes $pp \to gd \to t\mu^-\tilde{\chi}^0_1+X$ and $pp \to
g\bar{d} \to \bar{t}\mu^+\tilde{\chi}^0_1+X$, respectively. These
figures show that in Figs.\ref{fig11} there exist peaks located at
the position around $p_T^{t}(p_T^{\bar{t}}) \sim 80~GeV$ at the
early and future LHC, separately. We can see that the LO
differential cross sections are significantly enhanced by the QCD
corrections with inclusive jet selection scheme, while the QCD
correction by using an exclusive jet selection scheme keeps the
convergence of the perturbative series in the plotted
$p_T^{t}(p_T^{\bar{t}})$ range.

\par
\begin{figure}[htbp]
\includegraphics[width=3.2in,height=3.2in]{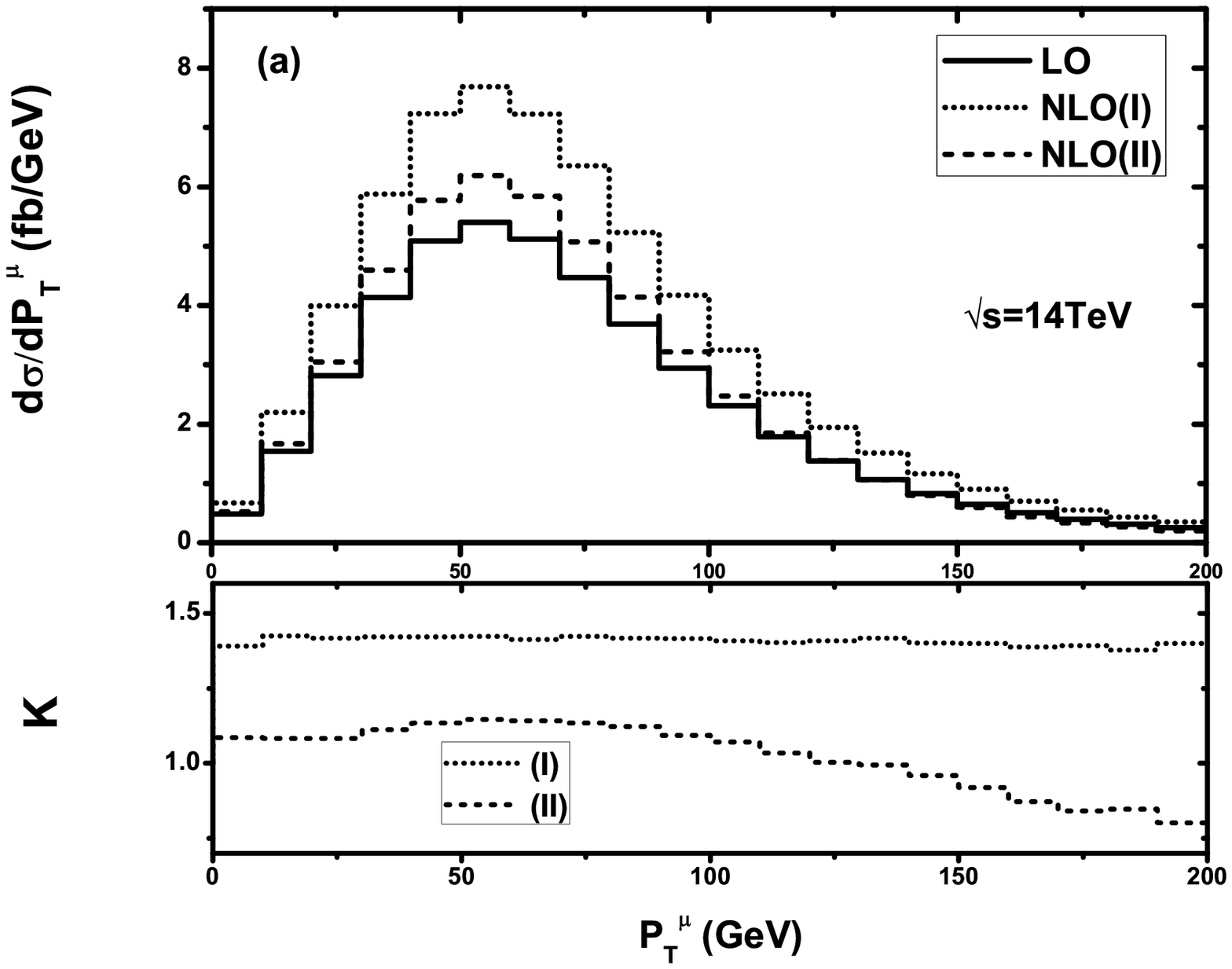}%
\hspace{0in}%
\includegraphics[width=3.2in,height=3.2in]{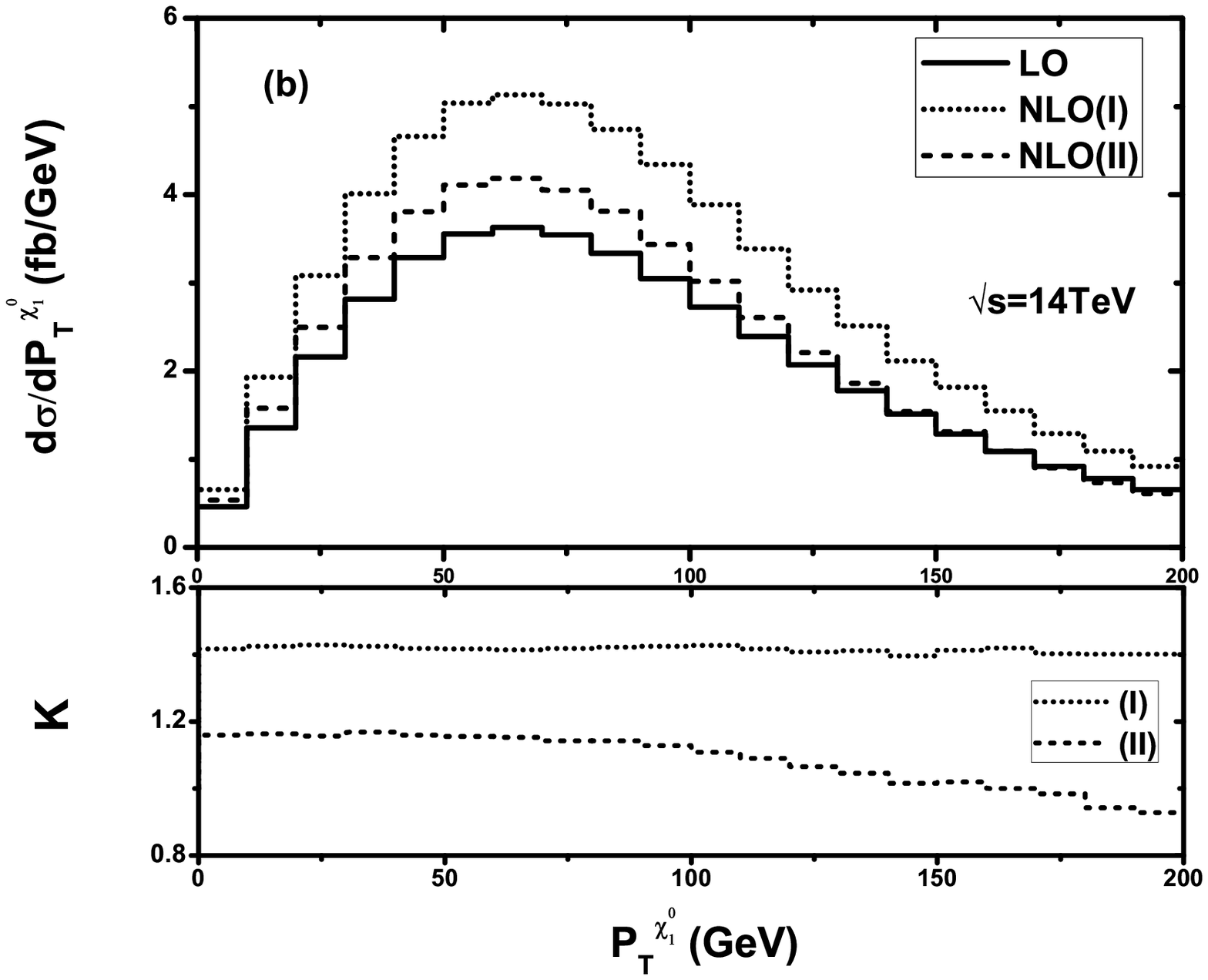}%
\hspace{0in}%
\includegraphics[width=3.2in,height=3.2in]{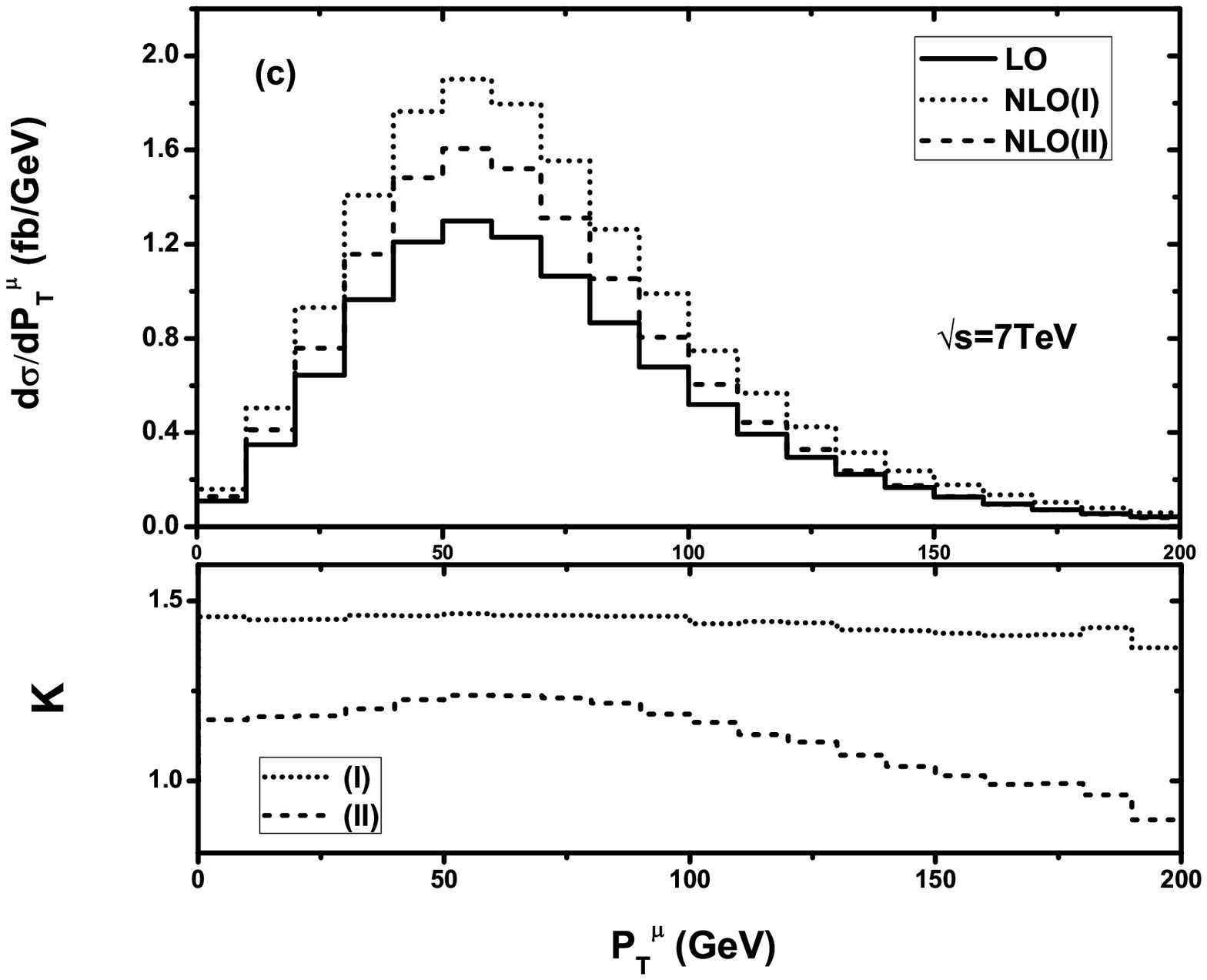}%
\hspace{0in}%
\includegraphics[width=3.2in,height=3.2in]{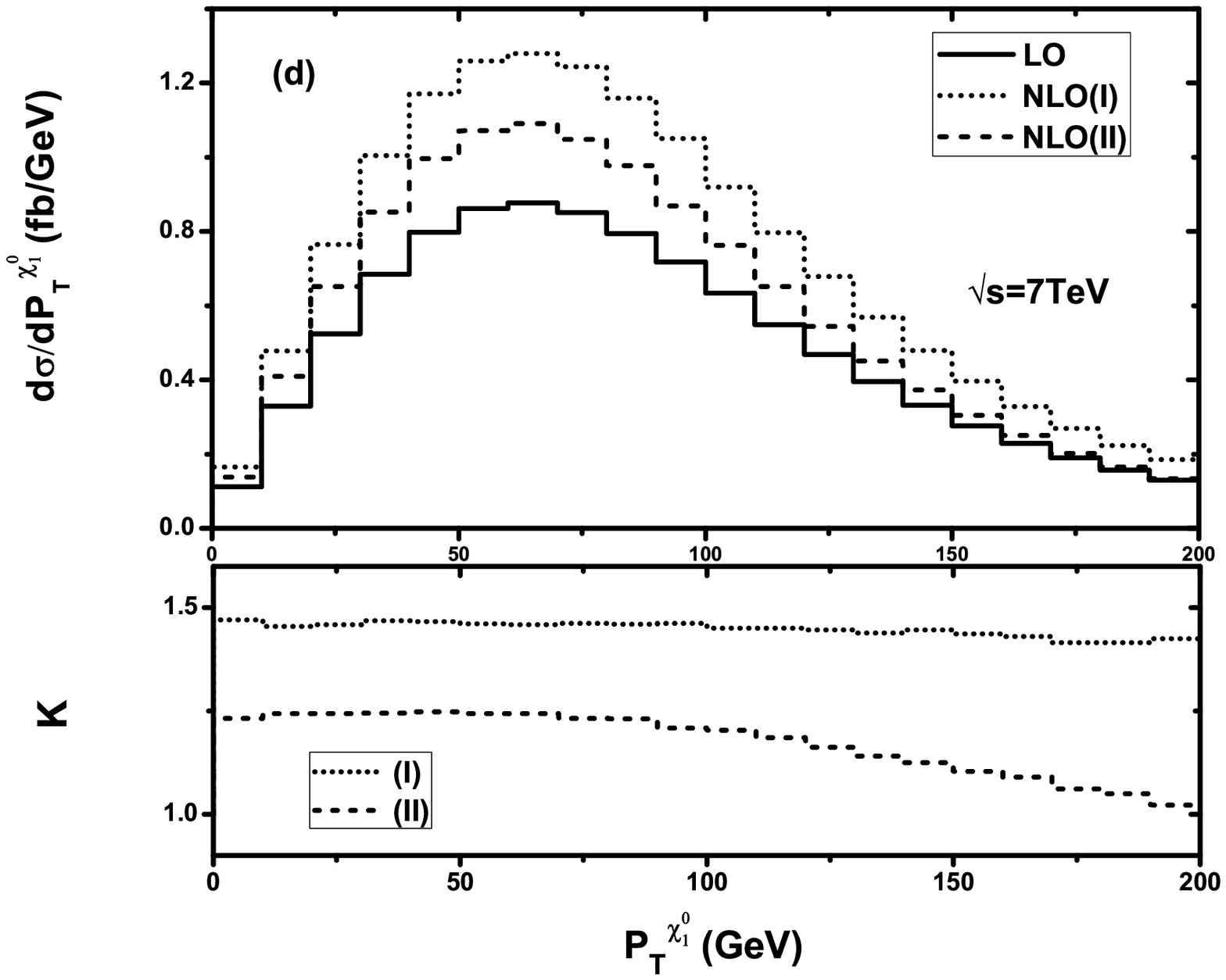}%
\hspace{0in}%
\caption{\label{fig12} The LO, NLO QCD corrected distributions of
the transverse momenta of final muon and neutralino for the process
$pp \to gd \to t\mu^- \tilde{\chi}^0_1+X$, and the corresponding
K-factors $(K(p_T)\equiv \frac{d \sigma_{NLO}}{dp_T}/ \frac{d
\sigma_{LO}}{dp_T})$ with $A=0~TeV$ at the LHC. (a) For the final muon at the
$\sqrt{s}=14~TeV$ LHC. (b) For the final $\tilde{\chi}^0_1$ at the
$\sqrt{s}=14~TeV$ LHC. (c) For the final muon at the
$\sqrt{s}=7~TeV$ LHC. (d) For the final $\tilde{\chi}^0_1$ at the
$\sqrt{s}=7~TeV$ LHC.  }
\end{figure}
\begin{figure}[htbp]
\includegraphics[width=3.2in,height=3.2in]{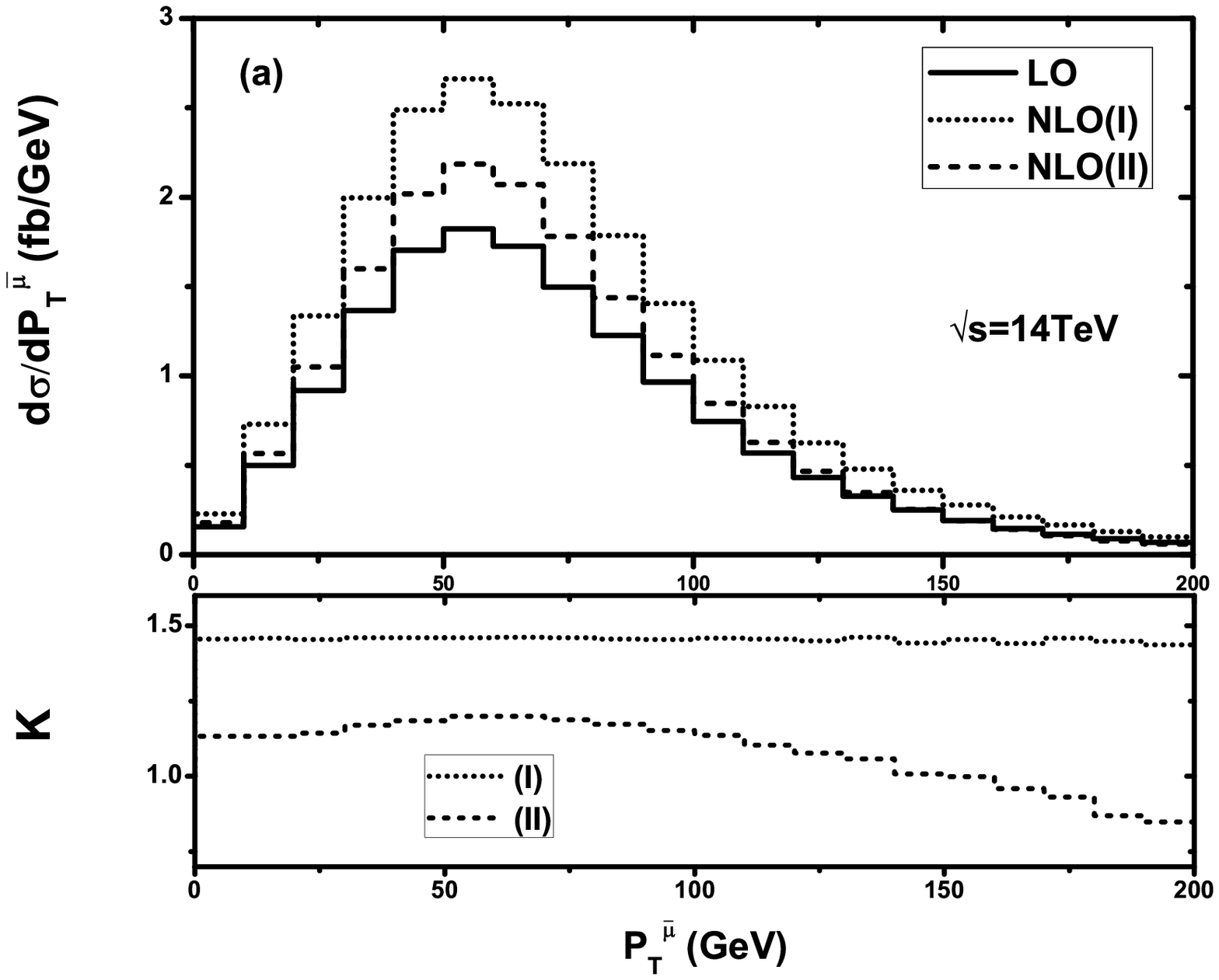}%
\hspace{0in}%
\includegraphics[width=3.2in,height=3.2in]{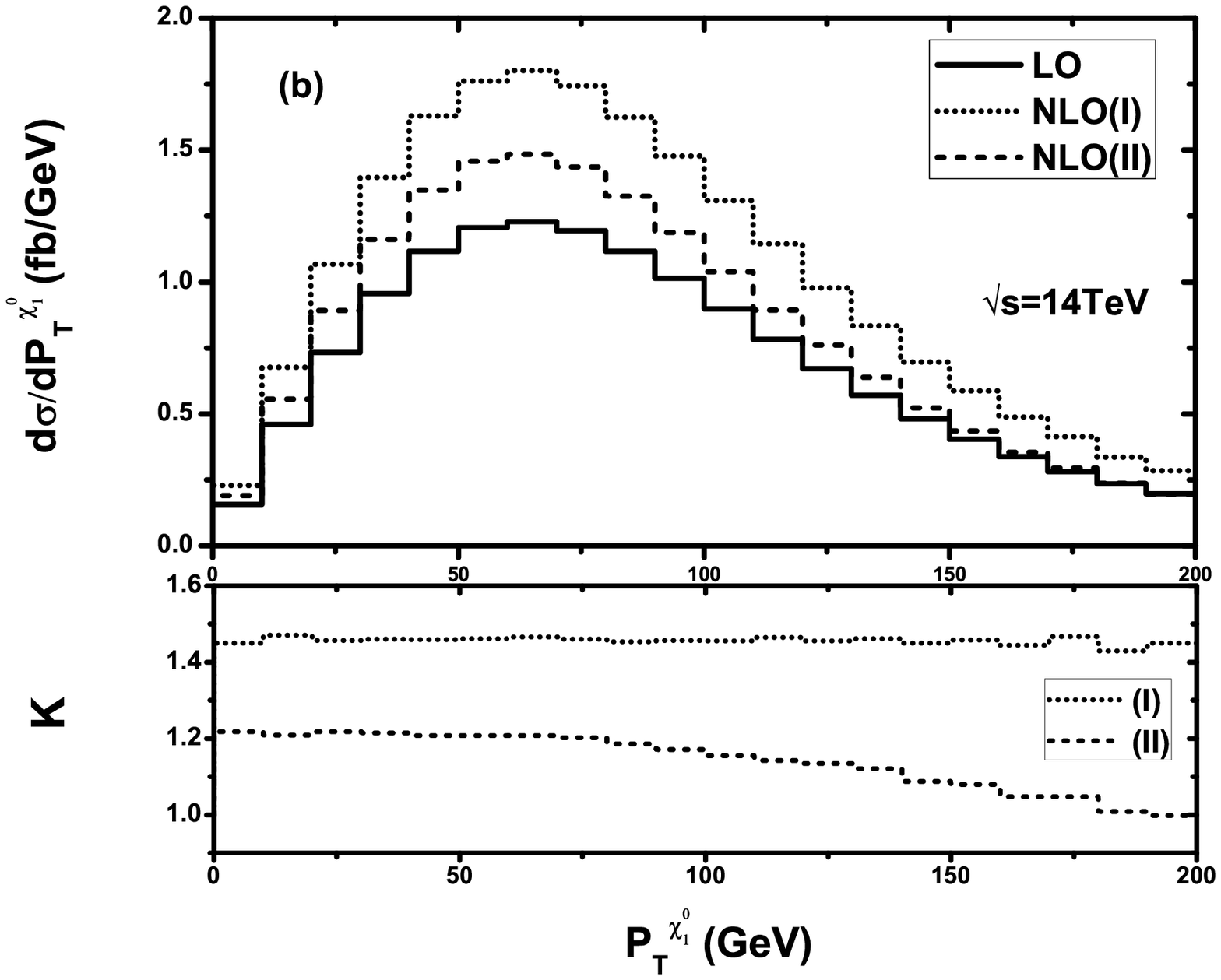}%
\hspace{0in}%
\includegraphics[width=3.2in,height=3.2in]{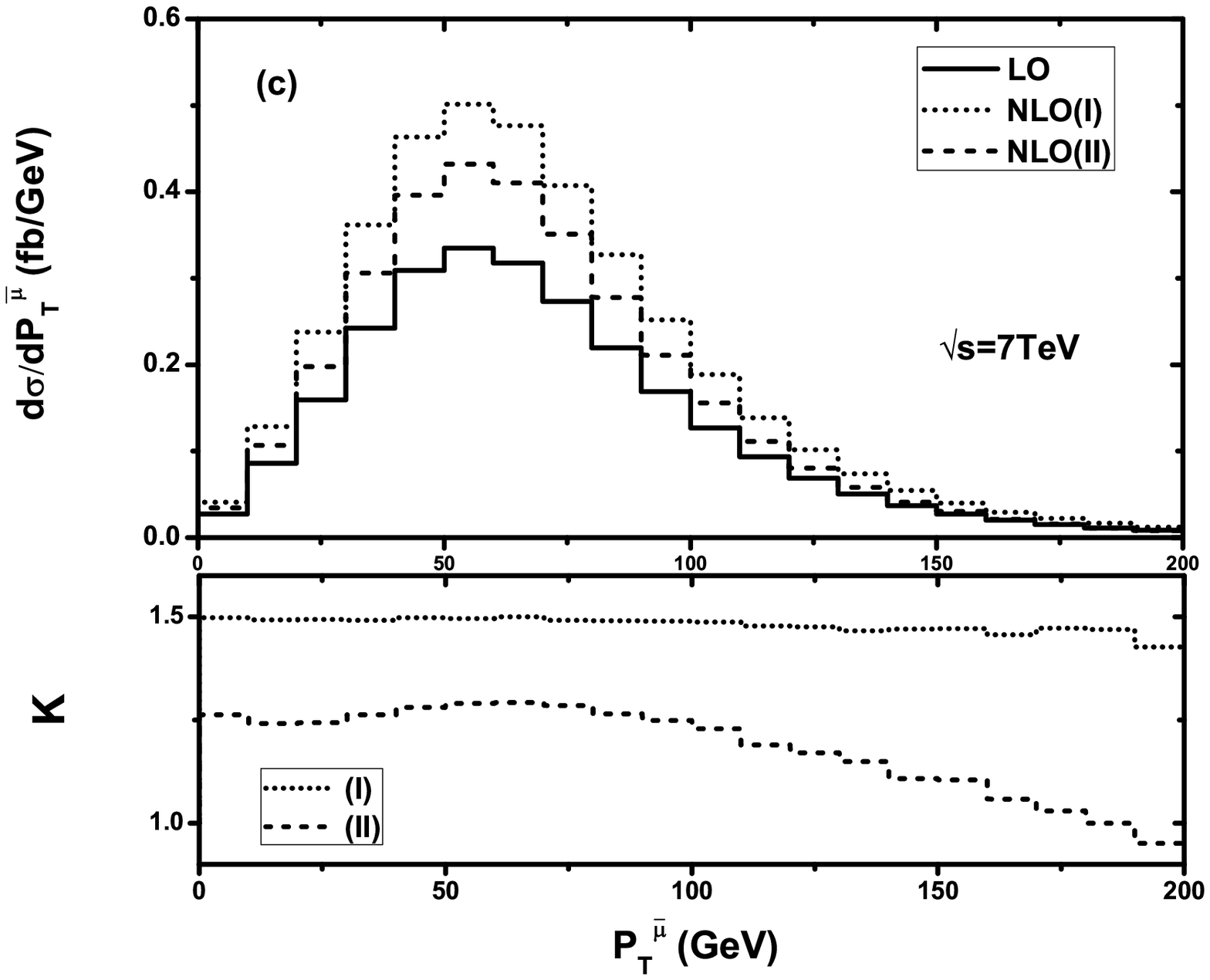}%
\hspace{0in}%
\includegraphics[width=3.2in,height=3.2in]{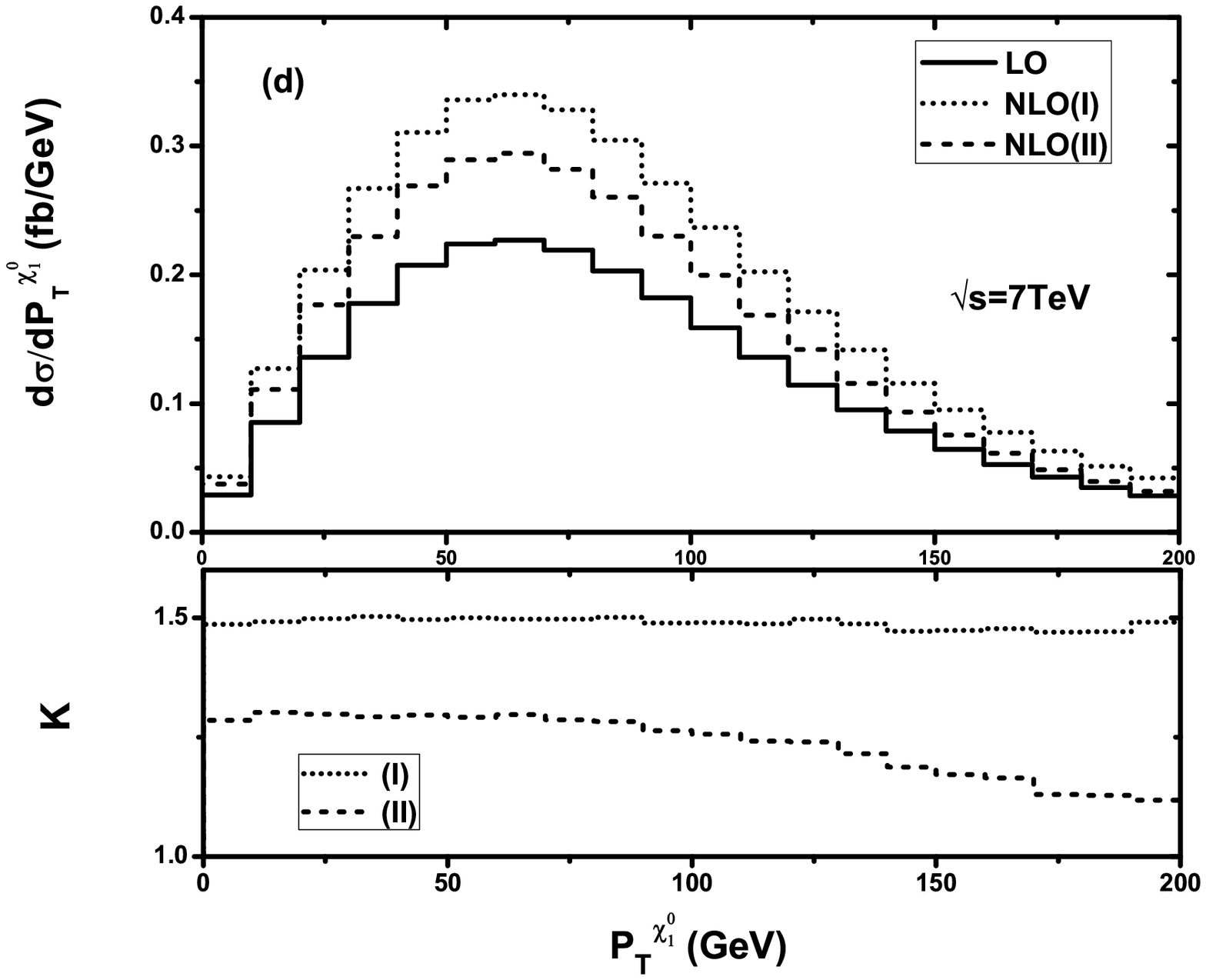}%
\hspace{0in}%
\caption{\label{fig13} The LO, NLO QCD corrected distributions of
the transverse momenta and the corresponding K-factors
$(K(p_T)\equiv \frac{d \sigma_{NLO}}{dp_T}/ \frac{d
\sigma_{LO}}{dp_T})$ of the final particles for the process $pp \to
g\bar{d} \to \bar{t}\mu^+ \tilde{\chi}^0_1+X$ with $A=0~TeV$ at the LHC. (a) For
final muon at the $\sqrt{s}=14~GeV$ LHC. (b) For the final
$\tilde{\chi}^0_1$ at the $\sqrt{s}=14~GeV$ LHC. (c) For the final
muon at the $\sqrt{s}=7~GeV$ LHC. (d) For the final
$\tilde{\chi}^0_1$ at the $\sqrt{s}=7~GeV$ LHC. }
\end{figure}

The transverse momentum distributions of the final muon and the
lightest neutralino, and the corresponding K-factors for the $pp\to
gd \to t\mu^- \tilde{\chi}^0_1+X$ process at the $\sqrt{s}=14~TeV$
LHC are depicted in Figs.\ref{fig12}(a,b), separately, while the
corresponding distributions and K-factors at the $\sqrt{s}=7~TeV$
LHC are depicted in Figs.\ref{fig12}(c,d), respectively. In
Figs.\ref{fig13}(a,b,c,d) we show the transverse momentum
distributions and the corresponding K-factors of the final particles
after the decay of $\tilde{\mu}^+$ for the process $pp \to g\bar{d}
\to \bar{t}\mu^+\tilde{\chi}^0_1 +X$. Figs.\ref{fig13}(a) and (b)
show the $p_T$ distributions of the final $\mu$ and
$\tilde{\chi}^0_1$ and K-factors at the $\sqrt{s}=14~TeV$ LHC,
separately, and Figs.\ref{fig13}(c) and (d) demonstrate the
$p_T^{\mu}$ and $p_T^{\tilde{\chi}^0_1}$ distributions and K-factors
at the $\sqrt{s}=7~TeV$ LHC respectively. In the calculation for
these results we use the narrow-width approximation (NWA) method to
handle the resonant scalar muon effect. Here we assume
$\lambda^{\prime}_{231}=0.1$ and the other $\lambda^{\prime}=0$. The
curves of K-factor labeled (I) and (II) correspond to adopting
the inclusive and exclusive gluon/(anti)quark jet event selection schemes,
respectively. It is clear that with the exclusive jet event selection
scheme we can keep the convergence of the perturbative series, and
the NLO QCD corrections mostly enhance the LO differential cross
sections of the final particles in the plotted $p_T$ range.

\par
In Table \ref{tab-3} we list some of the numerical results for the
LO and NLO QCD corrected total cross sections by adopting the
exclusive gluon/light-(anti)quark jet selection scheme for the
$pp \to gd \to  t\slep_i^-+X$, $pp \to g\bar{d} \to \bar{t}\slep_i^++X$
and $pp \to gs \to t\slep_i^- + X$ processes at the $\sqrt{s}=14~TeV$
and $7~TeV$ LHC. There we take $m_{\slep_i} = 189.9~GeV$,
$\lambda^\prime_{i31}=0.1$ and the other $\lambda^\prime=0$ for the first
two processes, $\lambda^\prime_{i32}=0.1$ and the other $\lambda^\prime=0$
for the last process. We consider the phase space with the restriction of
either $p^{jet}_T < p^{cut}_{T,jet}=35,~50~GeV$ or $\eta^{jet} >
\eta^{cut}_{jet}=2.7,~3$ for gluon/light-(anti)quark jet transverse
momentum.
\begin{table}
\begin{center}
\begin{tabular}{|c|c|c|c|c|c|}
\hline{process} & {$\sigma_{LO}$(fb)} &
\multicolumn{2}{|c|}{$p^{cut}_{T,jet}=50GeV,\eta^{cut}_{jet}=3$}&
\multicolumn{2}{|c|}{$p^{cut}_{T,jet}=35GeV,\eta^{cut}_{jet}=2.7$}\\
\cline{3-6}
&& $\sigma_{NLO}$(fb) & K-factor & $\sigma_{NLO}$ (fb) & K-factor \\
\hline \tabincell{c}{$pp\to gd $\\$\to t\slep^-_i+ X$} &
501.0(1)/114.34(2) & 541.4(4)/134.9(2) & 1.081/1.180 &
491.9(4)/123.6(1) & 0.982/1.081
\\ \hline \tabincell{c}{$pp\to g\bar d $\\$\to \bar t\slep^+_i+ X$} & 163.4(2)/28.39(4) &
184.7(4)/36.08(9) & 1.130/1.271 & 169.7(3)/33.03(8) & 1.038/1.164\\
\hline \tabincell{c}{$pp\to gs$ \\ $\to t\slep^-_i + X$} &
97.78(1)/15.411(2) &
115.9(2)/19.85(9) & 1.185/1.288 & 106.4(2)/18.07(8) & 1.089/1.172 \\
\hline
\end{tabular}
\end{center}
\caption{ \label{tab-3} The numerical results of the LO and NLO QCD
corrected total cross sections by adopting the exclusive selection
scheme for the $pp \to gd \to t\slep^-+X$, $pp \to g\bar{d} \to
\bar t\slep^++X$ and $pp \to gs \to t\slep^-_i + X$ processes with $A=0~TeV$.
The data on the left and right sides of the slash correspond to
the results for $\sqrt{s} = 14~TeV$ and $\sqrt{s}=7~TeV$,
respectively. There we take $m_{\slep_i} = 189.9~GeV$. }
\end{table}

\vskip 5mm
\section{Summary}
\par
In this paper, we calculate the complete NLO QCD corrections to the
single slepton associated with a (anti)top-quark production process
at the early ($\sqrt{s}=7~TeV$) and future ($\sqrt{s}=14~TeV$) LHC.
We investigate the dependence of the LO and the NLO QCD integrated
cross sections on the factorization/renormalization energy scale,
and study the influence of slepton, stop-quark and gluino masses
on the NLO QCD corrected cross sections. We point out that the
uncertainty of the LO cross section due to the introduced energy
scale $\mu$ is apparently improved by including NLO QCD corrections,
and the exclusive jet event selection scheme keeps the convergence
of the perturbative calculations better than the inclusive scheme. We
find also that the SUSY QCD correction generally increases if the 
$\tilde{t}_1$ or $\tilde{g}$ mass is getting large, but the non-zero
$\tilde{t}_j$--$\tilde{d}_k$--$\tilde{l}_i$ coupling with $A=1~TeV$ could
distort that curve tendency in some cases. We present the
LO and the QCD corrected distributions of the transverse momenta of
final products involving (anti)top quark, muon and the lightest
neutralino. Our results show that the NLO QCD corrections suppress
the polarization asymmetries of final (anti)top quark, and mostly
enhance the transverse momentum distributions of final particles.

\vskip 5mm
\par
\noindent{\large\bf Acknowledgments:} This work was supported in
part by the National Natural Science Foundation of China (Contract
No.10875112, No.11075150, No.11005101), and the Specialized Research
Fund for the Doctoral Program of Higher Education (Contract
No.20093402110030).


\vskip 10mm
\begin{thebibliography}{99}

\bibitem{Rv_1} 
P. Fayet, Phys. Lett. B {\bf 69}, 489 (1977)
\bibitem{Rv_2}
G.R. Farrar, P. Fayet, Phys. Lett. B {\bf 76}, 575 (1978)

\bibitem{LBviolation}
L.E. Ibanez, G.G. Ross, Nucl. Phys. B {\bf 368}, 3 (1992)

\bibitem{superpotential_1}
S. Weinberg, Phys. Rev. D {\bf 26}, 287 (1982)
\bibitem{superpotential_2}
N. Sakai, T. Yanagida, Nucl. Phys. B {\bf 197}, 533 (1982)
\bibitem{superpotential_3}
R. Barbier, C. Berat, M. Besancon, M. Chemtob, A. Deandrea, E. Dudas, P. Fayet, S. Lavignac,
G. Moreau, E. Perez, Y. Sirois, Phys. Rept. {\bf 420}, 1-202 (2005).
[arXiv:hep-ph/0406039]

\bibitem{top}
J.J. Cao, Z.X. Heng, L. Wu, J.M. Yang, Phys. Rev. D {\bf 79}, 054003 (2009) 

\bibitem{toppol}
T.M.P. Tait, C.-P. Yuan, Phys. Rev. D {\bf 63}, 014018 (2000)

\bibitem{Barbier}
R. Barbier et al., Phys. Rep. {\bf 420}, 1 (2005).
[arXiv:hep-ph/0406039]

\bibitem{singletopRPV_1}
P. Chiappetta, A. Deandrea, E. Nagy, S. Negroni, G. Polesello, J.M. Virey, Phys. Rev. D {\bf 61}, 115008 (2000).  
[arXiv:hep-ph/9910483]
\bibitem{singletopRPV_2}
F. Borzumati, J. L. Kneur and N. Polonsky, Phys. Rev. D {\bf 60} 115011 (1999).
[arXiv:hep-ph/9905443]
\bibitem{singletopRPV_3}
Z.H. Yu, P. Herbert, W.G. Ma, L. Han, Y. Jiang, Eur. Phys. J. C {\bf16}, 695 (2000).
[arXiv:hep-ph/9910323]
\bibitem{singletopRPV_4}
M. Chaichian, K. Huitu, Z.H. Yu, Phys. Lett. B {\bf 490} 87 (2000).
[arXiv:hep-ph/0007220]
\bibitem{singletopRPV_5}
H. Zhou, W.G. Ma, Y. Jiang, R.Y. Zhang, L.H. Wan, Phys. Rev. D {\bf 64} 095006 (2001)

\bibitem{singletop_1}
R.J. Oakes, K. Whisnant, J.M. Yang, B.-L. Young, X. Zhang, Phys. Rev. D {\bf 57}, 534-540 (1998)

\bibitem{singletop_2}
A. Datta, J.M. Yang, B.-L. Young, X. Zhang, Phys. Rev. D {\bf 56}, 3107-3113 (1997)

\bibitem{topspin}
M. Arai, K. Huitu, S.K. Rai, K. Rao, J. High Energy Phys. {\bf1008}, 082 (2010)

\bibitem{LO}
M.A. Bernhardt, H.K. Dreiner, S. Grab and P. Richardson, Phys.
Rev. D {\bf 78}, 015016 (2008)

\bibitem{feynarts}
T. Hahn, Comput. Phys. Commun. {\bf140}, 418 (2001)

\bibitem{formcalc}
T. Hahn, M. Perez-Victoria, Comput. Phys. Commun. {\bf118}, 153 (1999)

\bibitem{bounds}
F. Borzumati, J.S. Lee, Phys. Rev. D {\bf 66}, 115012 (2002)

\bibitem{counterterm}
A. Denner, Fortschr. Phys. {\bf 41}, 307 (1993)

\bibitem{MSbar_1}
W.J. Marciano, Phys. Rev. D {\bf 29}, 580 (1984)
\bibitem{MSbar_2}
W.J. Marciano, Phys. Rev. D {\bf 31}, 213 (1984) (E)

\bibitem{KLN_1}
T. Kinoshita, J. Math. Phys. {\bf 3}, 650 (1962) 
\bibitem{KLN_2}
T.D. Lee, M. Nauenberg, Phys. Rev. B {\bf 133}, 1549 (1964)

\bibitem{TCPSS}
B.W. Harris, J.F. Owens, Phys. Rev. D {\bf 65}, 094032 (2002).
hep-ph/0102128

\bibitem{cteq_1}
J. Pumplin, et al., J. High Energy Phys. {\bf 0207}, 012 (2002)
\bibitem{cteq_2}
D. Stump, et al., J. High Energy Phys. {\bf 0310}, 046 (2003)

\bibitem{Amsler}
C. Amsler, et al., Phys. Lett. B {\bf 667}, 1 (2008)

\bibitem{SPA}
J.A. Aguilar-Saavedra, et al., Eur. Phys. J. C {\bf 46}, 43 (2006)

\bibitem{isajet}
http://www.phys.ufl.edu/$\sim$jblender/isajet/isajet.html

\bibitem{Dreiner}
H.K. Dreiner, S. Grab, M. Kramer, M.K. Trenkel, Phys. Rev. D {\bf 75}, 035003 (2007).
[arXiv:hep-ph/0611195]


\end{thebibliography}
\end{document}